\documentclass[11pt]{article}
\pdfoutput=1
\usepackage{amsmath,amssymb,jheppub,graphicx}

    \usepackage{etoolbox}
    \makeatletter
    \patchcmd{\maketitle}{\@fpheader}{}
    \makeatother
    
\vspace*{0.0 mm}

\usepackage{verbatim}
\usepackage{bm} 
\usepackage{slashed}
\usepackage{mathdots}
\usepackage{color}%

\usepackage{hyperref}
\def\hhref#1{\href{http://arxiv.org/abs/#1}{arXiv:#1}} 

\allowdisplaybreaks[1]

\usepackage[all,cmtip]{xy}

\newcommand{\cpn}{${\mathbb C}{\mathbb P}^{N-1}$\,}
\newcommand{\cpone}{${\mathbb C}{\mathbb P}^{1}$}
\newcommand{\rtwo}{${\mathbb R}^2$}
\newcommand{\rone}{${\mathbb R}^1$}
\newcommand{\sone}{${\mathbb S}^1$}

\def\Z{\mathbb{Z}} 
\def\R{\mathbb{R}} 
\def\C{\mathbb{C}} 
\def\P{\mathbb{P}} 
\def\K{{\cal{K}}}
\def\A{{\cal{A}}}

\def\tn{\widetilde{n}}
 
\def\tO{\widetilde{\Omega}}
\def\be{\begin{equation}}
\def\ee{\end{equation}}
\def\cI{{\cal I}}

\def\Dslash{{\rlap{\raise 1pt \hbox{$\>/$}}D}}

\def\Z{\mathbb{Z}} 
\def\R{\mathbb{R}} 
\def\C{\mathbb{C}} 
\def\B{\mathbb{B}}

\def\Im{\text{Im}}
\def\Re{\text{Re}}
\def\tr{\text{tr}}

\def\v{{\vee}}

\def\til{\widetilde}
\def\hat{\widehat}
\def\bar{\overline}

\def\cA{{\mathcal A}}
\def\cB{{\mathcal B}}

\def\cO{{\mathcal O}}

\def\a{{\alpha}}
\def\ba{{\bar\a}}
\def\b{{\beta}}

\def\G{{\Gamma}}
\def\d{{\delta}}

\def\e{{\epsilon}}

\def\th{{\theta}}

\def\l{{\lambda}}
\def\L{{\Lambda}}
\def\m{{\mu}}

\def\s{{\sigma}}

\def\f{{\varphi}}

\title{Resurgence  and Trans-series  in Quantum Field Theory:   The \cpn  Model}

\author[1]{Gerald~V.~Dunne}
\author[2]{and Mithat \"Unsal}
\affiliation[1]{Department of Physics, University of Connecticut, Storrs CT 06269, USA}
\affiliation[2]{Department of Physics and Astronomy, SFSU, San Francisco, CA 94132}
\emailAdd{gerald.dunne@uconn.edu}
\emailAdd{unsal.mithat@gmail.com}
\abstract{This work is a step towards  a non-perturbative continuum definition of quantum field theory (QFT), beginning with asymptotically free  two dimensional non-linear sigma-models,  using recent ideas from mathematics and QFT.  The ideas from mathematics are resurgence theory, the trans-series framework, and  Borel-\'Ecalle resummation.  The ideas from QFT use continuity on  \rone$\times \mathbb S^1_L$, i.e, the absence of any phase transition  as $N \rightarrow \infty$  or rapid-crossovers for  finite-$N$, 
  and the small-$L$ weak coupling limit 
   to render the semi-classical sector  well-defined and  calculable. 
We classify semi-classical configurations with actions $1/N$ (kink-instantons), $2/N$ (bions and bi-kinks),  in units where the 2d instanton action is normalized to one.  
Perturbation  theory possesses  the  IR-renormalon ambiguity that arises due to non-Borel summability of the  large-orders perturbation series (of Gevrey-1 type),  for which a microscopic cancellation mechanism was unknown.  
This divergence must be  present  because the corresponding expansion is on a singular Stokes ray   in the complexified coupling constant plane, and the sum exhibits the Stokes phenomenon crossing the ray. 
 We show that there  is also a  non-perturbative ambiguity 
 inherent to certain neutral topological molecules  (neutral bions and  bion-anti-bions)   in the semiclassical expansion.   We find a set of ``confluence equations'' that encode the {\it exact cancellation} of the two different type of ambiguities.  
   There exists a resurgent behavior in the semi-classical trans-series  analysis of the QFT,  whereby subleading orders of exponential terms mix in a systematic way, canceling all ambiguities.  We  show that a new notion of 
  ``graded resurgence triangle''  is necessary to capture the path integral approach to resurgence, and  
   that graded resurgence  underlies a potentially rigorous definition of general QFTs.  The mass gap and the $\Theta$ angle dependence of vacuum energy  are calculated from first principles, and are in accord with large-$N$ and lattice results. 
}

\begin{document}
\maketitle

\vfill

\eject

\allowdisplaybreaks


\section{General idea of resurgence in QFT }
\label{intro}
\begin{quote}
``Series don't diverge for  no reason; it is not a capricious thing. The divergence of a series must reflect its cause."
\end{quote}
\vspace{-0.3cm}
 \hspace{4cm}    M. V. Berry,  {\it Stokes and the Rainbow}, Newton Institute Lecture, 2003 \\
 
This work  aims to give a non-perturbative continuum definition of  quantum field theory,  specifically here for two-dimensional non-linear  sigma models,  using two recent developments in mathematics and  quantum field theory (QFT). The ideas from mathematics come from the beautiful and powerful notions of  resurgent functions and trans-series  which go beyond  conventional (Poincar\'e) asymptotic analysis \cite{Dingle:1973,Ecalle:1981,BerryHowls,s07,delabaere,Costin:2009,Sternin:1996}. The new insights from QFT are the  semi-classically consistent  compactifications \cite{Unsal:2007vu, Unsal:2007jx,Anber:2011de} and deformations   \cite{Unsal:2008ch, Shifman:2008ja}  of quantum field theories, to control their infrared behavior,  rendering them  well-defined and calculable. 
 
Since these quantum field theories possess a non-Borel-summable asymptotic perturbative expansion around  {\it any} background,  perturbation theory  on its own is ambiguous, and 
 does not define the QFT. 
 This is one of the major difficulties  why many mathematicians  would  say  QFT is still non-rigorous, as recently emphasized in Ref.\cite{Douglas}.  
  A  lesser known  fact is that the non-perturbative semi-classical expansion on its own is also ambiguous  (in the context of QFT, see  \cite{Argyres:2012vv,Argyres:2012ka} and the present work), and also does not define the QFT. However, there exists a mechanism   to cancel the  non-perturbative ambiguities of  perturbation  theory with the  ambiguities of the semi-classical expansion   within   resurgence theory,   and one obtains {\it unique}, ambiguity-free, answers for physical quantities. This is a provocative and ambitious goal which has been explored in some detail   in certain quantum mechanical systems with degenerate vacua \cite{Bogomolny:1977ty,Bogomolny:1980ur,Brezin:1977gk,Stone:1977au,Balian:1978et,ZinnJustin:1981dx,zinn-book,Balitsky:1985in,h2plus,silverstone,Achuthan:1988wh,LeGuillou:1990nq,Jentschura:2004jg}, 
 in connection with the  pioneering work of Bender and Wu \cite{Bender:1969si}.
 Here and in joint work  with P. Argyres,   \cite{Argyres:2012vv,Argyres:2012ka},  we take the first steps
in applying these ideas to  quantum field theory, where new effects appear, such as  asymptotic freedom and renormalons \cite{Beneke:1998ui}.

Philosophically, the idea of resurgence is to combine perturbation theory (with small parameter $\lambda$, or $\lambda \hbar$ if one restores $\hbar$) and non-perturbative analysis (with small parameter $e^{-A/\lambda}$, with $A>0$) into a unified ``trans-series'' representation of a physical observable, in which the trans-series encodes much more [and potentially all] information about the function, rather than being viewed simply as a perturbative approximation or as an asymptotic approximation\footnote{In general a trans-series also includes a sum over powers of logarithms \cite{Ecalle:1981,Costin:2009,ZinnJustin:1981dx,zinn-book}, which are associated with quasi-zero-modes. We will comment on such terms later. After (\ref{trans}), we generally set $\hbar=1$.}:
 \begin{eqnarray}
 f(\lambda \hbar )\sim  \sum_{k=0}^{\infty}   c_{(0, k)} \, (\lambda \hbar) ^{k}  + \sum_{n=1}^\infty \, (\lambda\hbar)^{-\beta_n} \, e^{-n\, A/(\lambda \hbar)} \sum_{k=0}^{\infty}  c_{(n, k)} \, (\lambda\hbar)^{k} 
 \label{trans}
 \end{eqnarray}
The main point of resurgence  is that the perturbative and non-perturbative sectors can be   related in a systematic and mutually consistent manner, and unified as a trans-series. Paraphrasing the perspective of Dingle \cite{Dingle:1973} and Berry and Howls \cite{BerryHowls}, a trans-series may be a coded version of the exact function, which requires decoding in a systematic manner. Resurgence is the statement that the coefficients $c_{(n, k)}$ in (\ref{trans}) of the series expansion in powers $\lambda^k$ at large order $k$, for some $n$, are directly related to the coefficients at low order in $k$, for some other $n$. In other words, the perturbative expansion about some semiclassical configuration [a multi-instanton-anti-instanton] is directly related to the perturbative expansion about some other semiclassical configuration with different action.
 In Section \ref{subsec:examples} we give some elementary but illustrative examples where such a trans-series expansion (rather than a perturbative {\it or} a non-perturbative expansion)  is obviously necessary in order to give a complete description of the function as the phase of the expansion parameter $\lambda$ is varied.

A natural and powerful approach to this kind of problem is known as 
{\it 
Borel-\'Ecalle resummation},   or   {\it generalized Borel summation} \cite{Ecalle:1981, s07, Costin:2009, Sternin:1996}, a
technique for extracting mathematical and physical information from a
divergent series. In quantum mechanics, this method may be applied both to
the divergent perturbation series representing an energy eigenvalue 
\cite{Bogomolny:1980ur, Brezin:1977gk,   Stone:1977au, 
Balian:1978et,
ZinnJustin:1981dx,
zinn-book, 
Balitsky:1985in,
h2plus,
silverstone,
Achuthan:1988wh,  
LeGuillou:1990nq, 
Jentschura:2004jg},
 as well as to the divergent asymptotic series representation of
the semiclassical wavefunction  \cite{Dingle:1973, Ecalle:1981, s07, Costin:2009, Sternin:1996}. For certain polynomial
oscillator-type potentials the relation to exact semiclassical
quantization rules has been explored in detail 
\cite{voros,ddp}. As a direct
physical application, resurgence yields trans-series representations of
both eigenvalues and wavefunctions that, in principle, encode {\it all}
information about the solution. Earlier, Dingle and others developed
resurgent forms of WKB expansions for quantum mechanical problems and
special functions. These ideas also underlie improved {\it
hyper-asymptotic} approximations in which the remainder tails left after
optimal truncation of a divergent series are repeatedly Borel resummed
\cite{Dingle:1973,BerryHowls},  revealing interesting universal structures.

It is not immediately clear that we can extend these quantum mechanical
results to quantum field theories with renormalization. 
 We present here some evidence that resurgence can be applied to QFT, in the  two-dimensional \cpn model, one of the simplest non-trivial quantum field theories which possesses features analogous to QCD such as asymptotic freedom, confinement and instantons \cite{Novikov:1984ac}. The resurgence perspective allows us to  identify certain semi-classical objects in the \cpn model with the elusive  infrared (IR) renormalons, so that ambiguities in the perturbative and non-perturbative sector cancel one another.
 This builds on, but goes much further than, the fundamental results of Lipatov connecting instantons in QFT path integrals and the divergence of perturbation theory \cite{Lipatov:1976ny}.

In fact,  resurgence theory in QFTs, or in certain quantum mechanics problems with degenerate classical vacua, offers an extra feature:
when we consider the effect of a topological theta angle we find that the semi-classical exponential factors may also acquire phases, and sectors with different phases cannot mix in perturbation theory, by the simple fact that perturbation theory is independent of $\Theta$. 
Thus, the $\Theta$ dependence serves as a simple and useful guide that ``grades'' the distinct {\it resurgent sectors}, linking those that talk to one another and cure one anothers' ambiguities.  
Using the abbreviations $\frac{\widetilde \Theta_k}{N} \equiv   \frac{ \Theta + 2 \pi k }{N}, A= 4\pi $,  and for the 't Hooft coupling 
$\lambda  \equiv  g^2N$ in the bosonic \cpn model, the general structure that  emerges out of the path integral formulation   can be summarized symbolically in the following ``graded resurgence triangle'':
\begin{eqnarray}
&f_{(0,0)} &\nonumber\\ 
\cr
  e^{-\frac{A}{\lambda}+i  \frac{\widetilde \Theta_k  }{N}} f_{(1, 1)}  \hskip -20pt
&& \hskip -20pt  \quad
e^{-\frac{A}{\lambda} -i  \frac{\widetilde \Theta_k  }{N} } f_{(1, -1)} \nonumber\\ \cr
e^{-\frac{2A}{\lambda}+2i  \frac{\widetilde \Theta_k  }{N}  } f_{(2, 2)}
\quad \quad & 
e^{-\frac{2A}{\lambda}} f_{(2, 0)}  &
\quad \quad
e^{-\frac{2A}{\lambda}-2i  \frac{\widetilde \Theta_k  }{N}} f_{(2, -2)}
\nonumber\\   \cr
e^{-\frac{3A}{\lambda}+3i  \frac{\widetilde \Theta_k  }{N}} f_{(3,3)}
\quad \quad
e^{-\frac{3A}{\lambda}+i  \frac{\widetilde \Theta_k  }{N}} f_{(3,1)}
 \hskip -20pt
&& 
e^{-\frac{3A}{\lambda}-i  \frac{\widetilde \Theta_k  }{N}} f_{(3,-1)}
\quad \quad
e^{-\frac{3A}{\lambda}-3i  \frac{\widetilde \Theta_k  }{N}} f_{(3, -3)} \nonumber\\ \cr
e^{-\frac{4A}{\lambda}+4i  \frac{\widetilde \Theta_k  }{N}} f_{(4,4)}
\quad \quad
e^{-\frac{4A}{\lambda}+2i  \frac{\widetilde \Theta_k  }{N}} f_{(4,2)} \quad \quad 
&   e^{-\frac{4A}{\lambda} } f_{(4,0)}
  & \quad \quad 
e^{-\frac{4A}{\lambda}-2i  \frac{\widetilde \Theta_k  }{N}} f_{(4,-2)}
\quad \quad
e^{-\frac{4A}{\lambda}-4i  \frac{\widetilde \Theta_k  }{N}} f_{(4, -4)} \nonumber\\ \cr
\iddots \qquad \qquad \qquad \qquad\qquad &\vdots & \qquad \qquad\qquad \qquad\qquad  \ddots 
\label{triangle}
\end{eqnarray}
which represents a general expansion of some observable. The rows correspond to a given instanton number $n$, with associated perturbative loop expansions times an  instanton prefactor $f_{(n, k)}(\lambda) \equiv (\lambda)^{-\beta_n} \, \sum_{k=0}^{\infty}  c_{(n, k)} \, (\lambda)^{k}  $, and with the topological phases specified. Only columns of this triangle with matching $\Theta$ dependence can possibly mix via resurgence\footnote{For notational simplicity we suppress log terms that generally also appear in the prefactor sums.}.  
For example, in the Bogomolny-Zinn-Justin (BZJ) approach to the periodic potential problem  \cite{zinn-book,delabaere-periodic}, which has degenerate vacua and a topological theta angle,
the ambiguity in the perturbative contribution $f_{(0, 0)}(\lambda)$ to the ground state energy is cured by an ambiguity in the  instanton-anti-instanton amplitude, 
at order $e^{-\frac{2A}{\lambda}} f_{(2, 0)}(\lambda)$. 
 This is in fact a general phenomenon that extends throughout the triangle:  the $\Theta$-sectors are correlated with instanton sectors, which gives another tool for probing the mixing of the different terms in the trans-series representation. Our main conjecture is that each column is a resurgent function of $\lambda$.
 This graded resurgence structure provides an interesting new perspective on instanton calculus and is born in  a natural implementation of  the theory of resurgence in the path integral formalism. 

Our long-term goal in  applying  resurgence to QFT  is rather ambitious:  
We aim  to  give  a non-perturbative continuum definition of quantum field theory, and  provide  a mathematically rigorous foundation. We also would like that such such a definition  should be of practical value (not a formal tool) whose results  can be compared with the numerical analysis of lattice field  theory. In other words,  generalizing the title of  't Hooft's seminal Erice lectures \cite{'tHooft:1977am},  we want to make sense out of general QFTs in the continuum.

We emphasize that  our immediate goal is {\it not} to provide theorems;  rather we would like to reveal  {\it structure underlying QFT}, a framework in which we can define QFT in a self-consistent manner without running  into internal  inconsistencies.  We hope that whatever framework emerges along these lines 
may form the foundation of a rigorous definition. This point of view is close in spirit to  Refs.~\cite{Douglas,Thurston}. However, we ultimately hope that we will be able to use resurgence theory to provide exact and rigorous  results for general QFTs, at least in the semi-classical domain. Our optimism stems from the work of Pham et.al. where they proved that the semi-classical expansions in certain non-trivial quantum mechanical systems are resummable to finite exact and unambiguous answers \cite{ddp,delabaere-periodic}.

 Similar ideas that appear in the current work 
 and in  \cite{Argyres:2012vv, Argyres:2012ka} also appeared in a 
 recent talk by Kontsevich \cite{Kontsevich}. Kontsevich also examines resurgence from the path integral perspective,  with the intention of establishing a non-perturbative definition of certain special  QFTs and quantum mechanics, directly  from the path integral. The notion of analytic continuation of paths in field space, which is also crucial in our analysis of  quasi-zero mode integrals, plays an important role in his discussion. 
 We also note that some progress has recently been made in applying the ideas of resurgence to matrix models, and certain string theories and quantum field theories (which do not require renormalization)   in the recent works of Schiappa et al \cite{schiappa}, Mari\~no et al \cite{marino,Marino:2012zq}, and Costin et al \cite{CG}. 
Our works   differ from the above in the sense that we   study  both   realistic QFTs, with asymptotic freedom and  renormalons, and comment  upon theories with extended supersymmetry. The study of realistic QFTs requires, 
 apart from new mathematical inputs, new physical inputs  as well, 
 rendering them  calculable; this QFT program began with new compactifications  \cite{Unsal:2007vu, Unsal:2007jx} and  deformations  of gauge theories \cite{Unsal:2008ch, Shifman:2008ja}.\footnote{
 \label{Marinocomment}The  ability to use the weak coupling limit in realistic QFTs is a newly developing program, and it is what makes the current detailed analysis possible. In non-supersymmetric theories,  it  is also to  a certain extent an unexpected, but quite welcome aspect. The state of the art concerning the large orders in perturbation theory versus non-perturbative effects  in the framework of resurgent functions and trans-series in ODEs, integral equations, quantum mechanics and a sub-class of special QFTs is explained clearly in a recent review by  Mari\~no \cite{Marino:2012zq}, where he asserts
\begin{quote}
 ``In realistic QFTs, perturbation theory is so wild that it is not feasible to pursue this program [theory of resurgence], but in some special QFTs -- namely, those without renormalons, like Chern-Simons theory in 3d or ${\cal N}=4$ Yang-Mills theory -- there are some partial results."
\end{quote}
We  provide ample evidence in this work for \cpn,  and in 
\cite{Argyres:2012vv, Argyres:2012ka} for QCD(adj), 
that the resurgence formalism can in fact be applied to more realistic QFTs, even those with asymptotic freedom and renormalons.}

\subsection{Problems with semi-classical analysis on $\R^2$ }
\label{problems}

To motivate our application of resurgence we recall that the  $\C\P^{\rm N-1}$ model in two dimensions is an asymptotically free non-linear sigma model with many features in common with four dimensional Yang-Mills theory \cite{Novikov:1984ac,Coleman}. Despite some progress in this class of theories, especially in the large-$N$ limit,  there are several significant long-standing open problems:

{\bf Problem 1.  Invalidity of  the semi-classical dilute instanton gas  approximation on $\R^2$:  }
  The  theory on $\R^2$ has instanton solutions, but it does not admit a reliable 
 semi-classical analysis because of the existence of the instanton size modulus, which implies that
 instantons of all sizes come with no extra action cost.   Therefore,  a self-consistent dilute instanton gas approximation, which relies on the assumption that  the typical separation between instanton events is much larger than the instanton size, does not exist for  the $\C\P^{\rm N-1}$ model on $\R^2$.    This is a variant of the  long-standing ``infrared   embarrassment" problem \cite{Coleman}.

{\bf Problem 2.   Meaning of the infra-red  renormalon singularities:} 
Another serious (and actually, we claim, related)   problem  is that if one works out the large orders of perturbation theory in sigma models, there are singularities (poles or branch points)   on the positive real axis  $\R^{+}$ of the complex Borel plane located at order  $N$ times closer to the origin than the {\it leading} 2d instanton-anti-instanton $[\cal I  \overline{ \cal I}]$  singularity.  These are called  {\it infrared (IR) renormalon} singularities, and it was previously  unknown if there exists a semi-classical (or non-semiclassical) field configuration which may cure this disease of perturbation theory \cite{'tHooft:1977am,David:1980gi}. Thus, perturbation theory is ill-defined for \cpn on $\R^2$.

{\bf Problem 3. Relation between large-$N$ results and instantons:} 
\cpn in the large-$N$ limit admits a solution with non-perturbative mass gap 
\begin{align}
m_g = \Lambda  = \mu e^{-S_I/N} \equiv  \mu e^{-4 \pi/ (g^2N)} \ ,   
\label{gap-largeN}
\end{align}
 where $\mu$ and  
$\Lambda$ are, respectively,  the renormalization and the strong scale, and $S_I$ is the 2d-instanton action. There is no known semi-classical (or non-semiclassical)  configuration which leads to this mass gap. This has been studied, for example, in several fundamental works \cite{D'Adda:1978kp,Jevicki:1979db,Affleck:1980mb,Munster:1982sd,Aguado:2010ex}, but so far no fully consistent semi-classical analysis exists on ${\mathbb R}^2$.
\\

We emphasize that none of these  problems is formal.  They indicate that 
some crucial ingredients are  lacking in our current understanding of QFT.   
We show here that it is possible to solve these problems  on $\R^1 \times$\sone, and in fact, the solutions of all three problems  are deeply related. We also argue that this solution may be continuously connected to the situation on $\R^2$, leading to a quite radical resolution of the problems listed above. Our proposal, valid for both supersymmetric and non-supersymmetric theories,  is also consistent with arguments based on mirror symmetry \cite{Hori:2000kt} and appropriate generalization of the  quantum chiral ring relations  \cite{Cachazo:2002ry} in the minimal supersymmetric case, as discussed below in Section \ref{sec:mirror}.
  
\subsection{Good and bad for semi-classics in compactified theories}
\label{subsec:goodbad}

Traditionally,  an infrared cut-off  such as  thermal compactification  is used   to tame the problem of the size of an instanton in asymptotically free theories \cite{Gross:1980br}.  However,  this thermal compactification approach, as opposed to the spatial  (non-thermal) compactification that we propose here, does not provide  a 
semi-classical analysis of the confined phase.
 As reviewed in Section \ref{thermal1},  at finite temperature, the \cpn model has two phases (regimes):   a   deconfined phase  $\beta < \beta_c \approx \Lambda^{-1}$ (where $\beta_c$ is the deconfinement temperature and $\Lambda$ is the strong scale), and a confined phase,  $\beta > \beta_c$. \footnote{Strictly speaking, there is no phase transition, but a rapid crossover in the behavior of the theory at the strong scale. At $N=\infty$ thermodynamic limit, this becomes a sharp phase transition; at finite-$N$, the deconfined regime has $O(N)$-free energy, and the confined regime has $O(1)$ free energy.}

 \begin{list}{$\bullet$}{\itemsep=0pt \parsep=0pt \topsep=2pt}
\item The theory in the  weak coupling   small--$\mathbb S^1_\beta$ (high-temperature)  regime is in a deconfined phase, governed by trivial holonomy (\ref{minthermal}).  The instanton size problem can indeed be tamed in this phase, since 
\begin{equation}
\rho \lesssim \beta < \beta_c \approx \Lambda^{-1} 
\label{scales-th1}
\end{equation} 
i.e., the instanton size  $\rho$  is cut-off by the box size $\beta$, {\it and} the box size is smaller than the strong scale. However, the  information that one acquires in this  regime does not apply to the confined-phase and is not continuously connected to $\R^2$.  Despite the fact that one can make sense of a dilute instanton gas  for $\beta \ll \Lambda^{-1}$   \cite{D'Adda:1978kp,Jevicki:1979db,Affleck:1980mb,Munster:1982sd,Aguado:2010ex},  whatever is learned in this phase  is not obviously directly relevant for  understanding the   physics below the deconfinement temperature, at large--$\mathbb S^1_\beta \times \R $,  or on $\R^2$. 

\item   If the theory is in the confined phase, i.e., $\beta > \beta_c\approx \Lambda^{-1}$, it is governed by a non-trivial holonomy for the line operator at strong coupling. In this regime,  there is no weakly coupled description of  the long-distance physics.  The dilute instanton gas approximation breaks down whenever the size modulus is of order the strong scale, $\rho \sim \Lambda^{-1}$.    Therefore, the existence of a box whose size is larger than $ \beta_c $ is  not  helpful for a semi-classical dilute gas approximation. Formally, it is still true that the size modulus of the ``classical instanton solution"  is cut-off at the scale $\beta$. However,  for 
\begin{equation}
 \Lambda^{-1} \lesssim \rho   \lesssim \beta
 \label{scales-th2}
\end{equation} 
the notion of a semi-classical instanton is not meaningful, due to strong coupling. The semi-classical approximation is simply inapplicable in this regime.
 \end{list}

\noindent  In \cpn   on $ \R^1 \times \mathbb S^1$,  as well as in gauge theories on  $\R^3 \times \mathbb S^1$, there are  three   types of  gauge holonomies. (In \cpn, this is the gauge  holonomy associated with the $\sigma$-model connection, defined below in (\ref{hol}), and it plays the same role as the Wilson line in gauge theory). In studying calorons, Yi and Lee \cite{Lee:1997vp}, and
van Baal et al \cite{Kraan:1998pm}, analyzed the difference between trivial  (degenerate eigenvalue distribution) and non-trivial holonomy (maximally non-degenerate distribution), and realized  the importance of non-trivial holonomy for topological configurations with fractional topological charges. However, a further refinement is still needed in order to  find a quantitative semi-classical   theory describing the dynamics.  The types of holonomies are: 
 \begin{list}{$\bullet$}{\itemsep=0pt \parsep=0pt \topsep=2pt}
\item[{\bf 1)}]   {\bf   Weak coupling trivial holonomy:}  Semi-classical analysis applies, but this regime  is separated from strong coupling non-trivial  holonomy by a rapid cross-over or a phase transition. 
\item[{\bf 2)}]   {\bf   Weak coupling non-trivial holonomy:}  Semi-classical analysis applies, and for a large class of theories, this regime  may be
 continuously connected to   strong  coupling non-trivial holonomy regime. 
  \item[{\bf 3)}]  {\bf   Strong coupling non-trivial holonomy:} Weak-coupling semi-classical analysis is not applicable. 
\end{list}
The appreciation of the existence and significance of the second  type of holonomy is  relatively new in gauge theory  \cite{Unsal:2007vu, Unsal:2008ch}, and for \cpn it is presented in this work. \footnote{The associated 
``free-energy" in both strong and weak coupling non-trivial holonomy regime is $O(1)$, whereas the free-energy of the weak coupling trivial holonomy regime is $O(N)$.  As $N \rightarrow \infty$ thermodynamic limit, there is   a sharp phase transition between 1) and 3), but the limit is smooth between 2) and 3).  This is what we mean by continuity on  $ \R^1 \times \mathbb S^1$. }

 \begin{figure}[htb]
\centering{\includegraphics[scale=0.3]{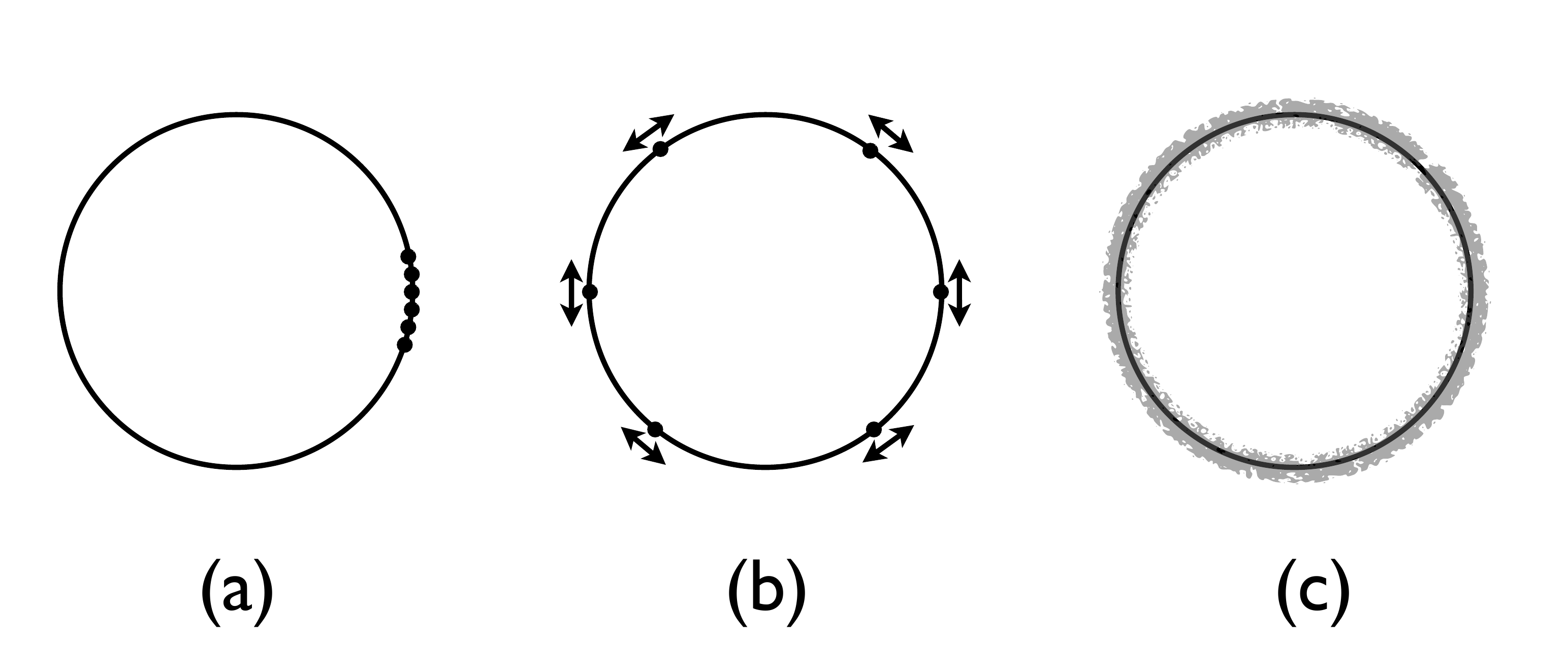}
\\
\includegraphics[scale=0.27]{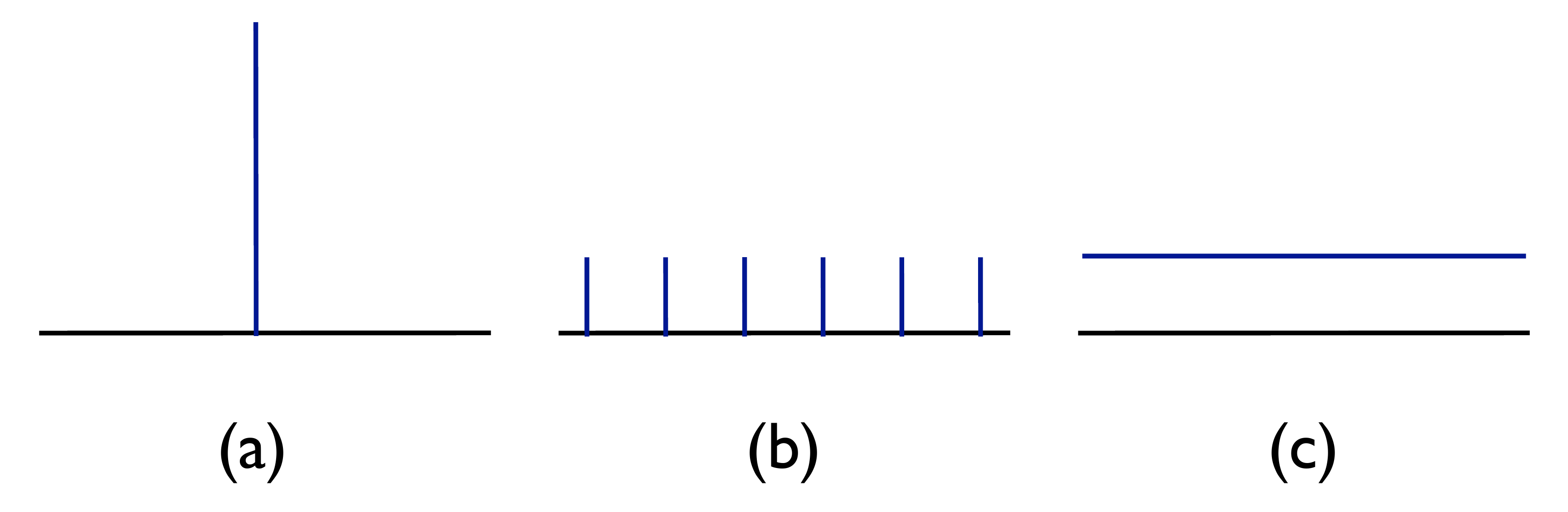}}
\caption{Three types of gauge  holonomy in non-abelian gauge theories and analogously $\s$-connection holonomy (defined in   (\ref{hol}))  in non-linear sigma models, and their classification according to their eigenvalue distribution.  
a) is the weak coupling trivial holonomy. b) and c) are, respectively,  the weak and strong  coupling non-trivial holonomy, and they are  continuously connected in the sense of gauge invariant order parameters. 
In b), the eigenvalues of holonomy  are located at the roots of unity and their fluctuations are small.  In c), the positions of the eigenvalues are uniform and randomized.  
In gauge theory on $\R^3 \times S^1$, b) is  the counterpart of the weak coupling  adjoint Higgs regime.  There is no weak coupling or even a potential description for c). The difference between  (b) and (c) in gauge theory is discussed in  Ref.~\cite{Unsal:2008ch}.} 
\label{fig:holonomy}
\end{figure}

Specifically,  we propose  a  method to  understand   the dynamics of the \cpn    theory on  $\R^2$  using continuity and weak coupling methods.  We  work with  a  spatial (non-thermal) compactification with twisted boundary conditions.    In this case, as we will show, there are  \cpn theories without {\it any } phase transition (the associated free energy always remains order one, as opposed to order $N^1$ as in the deconfined regime of the thermal theory)  and  whatever is learned at weak coupling is expected to smoothly interpolate to strong coupling.  We study a class of theories related to $\C\P^{\rm N-1}$: 
the bosonic  model,  the  supersymmetric $ {\cal N}= (2,2)$ theory with $N_f=1$ Dirac fermionic flavor, as well as non-supersymmetric multi-flavor theory with $N_f >1$.   Our methods  
apply equally well both to supersymmetric and non-supersymmetric theories, and the role that supersymmetry plays in  $N_f=1$  is in fact not particularly significant.

\subsection{Perturbation theory and spatial  twisted boundary conditions}
\label{subsec:spatial-twists}
 
In the classical theory, we show that the twisted boundary conditions can be recast into a background field, a $U(1)^N$ $\sigma$-connection.  It is realized as a background field for the line operator,  defined below in (\ref{hol}).  This is analogous  to the background Wilson line in gauge theory on $\R^3 \times \mathbb S^1$. Determining the stability of a given background in quantum theory 
 requires a perturbative one-loop Coleman-Weinberg type analysis at small-$L$, similar to gauge theory.  The idea of twisted  boundary conditions has  recently been used   to study the substructure of calorons   in \cpn at finite temperature \cite{Bruckmann:2007zh,Brendel:2009mp,Harland:2009mf}, with earlier relevant work by Sutcliffe \cite{Sutcliffe:1993yy}, and we are in part inspired by these  results. However, our physical interpretation is  different, as we use spatially twisted boundary conditions, and we further address the crucial issue of the stability of the twisted background in the quantum theory, not addressed in earlier works. 

The main result of the perturbative one-loop analysis is parallel to Yang-Mills theory with adjoint fermions,   QCD(adj),   on $\R^3 \times \mathbb S^1$. We find that on a temporal circle, 
 i.e., thermal case, twisted boundary conditions are  unstable for any   $N_f  \geq 0 $ in the small-$\mathbb S_\beta$  regime. In contradistinction, upon spatial compactification on a 
 cylinder $\R^1 \times \mathbb S^1_L$,  ${\mathbb Z}_N$ twisted boundary conditions are  {\it stable} for $N_f \geq 1$, and  unstable  for  $N_f  =0 $.  However, even in the ($N_f=0$) bosonic case, 
a  ${\mathbb Z}_N$ symmetric twist can be achieved, by exploiting the  fermion induced stabilization by taking fermions heavy with respect to the strong scale, but light with respect to $1/(LN)$. 
This   theory at distances larger then $m^{-1} $ emulates the bosonic theory   and has a stable   ${\mathbb Z}_N$-twisted background. We will refer to the bosonic theory obtained in this manner or obtained by adding an explicit   center-stabilizing potential as  ``deformed-\cpn".  A   stable 
${\mathbb Z}_N$ twist  is the first step   towards a resolution of the problems listed in Section \ref{problems}.

 \subsection{Topological defects and molecules} 
 \label{subsec:defects}
 
The stability of the spatially-twisted-background in the weak coupling regime allows us to investigate 
topological defects.  The leading  topological defects  and molecules  in the \cpn model on \rone $\times  \mathbb S^1_L$   are:

 \indent (i) Kinks (or more rigorously kink-instantons) $\K_i$.  \\ 
\indent (ii) Charged   bions (correlated kink-anti-kink events)   $\cB_{ij} = [\K_i\bar\K_j]$, \\
\indent (iii) Neutral bions   $\cB_{ii} = [\K_i\bar\K_i]$,  and \\ 
\indent(iv)  Neutral bion-anti-bion molecular events such as   $[\cB_{ij}  \cB_{ji}],  [\cB_{ij} \cB_{jk}   \cB_{ki}]$  etc.  

\noindent Note that the 2d instanton $\cal I$   does not appear in this  list of  leading topological configurations.  Apart from the fact that it causes the anomaly in the classical $U(1)_A$  chiral symmetry  for theories with multiple massless fermions [reducing it to a discrete chiral $\Z_{2N}$], the role of the 2d instanton in the semi-classical expansion is actually {\it insignificant} for the resolution of the problems listed in Section \ref{problems}. We emphasize this, as it is surprising and goes against the general  philosophy of many works. Also note that our classification of topological defects in \cpn  is  identical to QCD(adj) compactified on $\R^3 \times \mathbb S^1$, with the role of the  monopole-instantons of \cite{Argyres:2012vv} being played by the kink-instantons in \cpn.

Before going into a description of the topological defects, we note the important hierarchy of length (or energy) scales, 
\begin{eqnarray}\label{scale-hierarchy}
\begin{matrix}
r_{\rm k}&\ll&r_{\rm b}&\ll&d_{\rm k-k}&\ll&d_{\rm b-b}, 
\end{matrix}
\end{eqnarray}
where   $r_\text{k}$  is the size of  of  kink-instanton, $r_\text{b}$ is the size of a bion, $d_\text{k-k}$  and  $d_\text{b-b}$ are the typical  inter-kink and  inter-bion separations. This  
justifies the use of of a dilute gas of kinks, bions, etc. This hierarchy  is also inherent to the hierarchy of successive terms in the semi-classical trans-series expansion. 

The kink-instantons are self-dual configurations with topological charge $1/N$, as discussed in detail in Section \ref{sec:selfdual}.  The kink-instantons  come in $N$ types,  associated with the (extended) root system of the  $SU(N)$ global symmetry of the \cpn theory.     The first $(N-1)$ types are  1d kink-instantons associated with the simple roots $\alpha_i$,   and the remaining one is the  twisted (affine)  kink-instanton, associated with the affine root $\alpha_0= -  \sum_i\alpha_i$. We give an explicit construction of this affine-kink-instanton.   In a theory with massless fermions, these defects carry fermion zero modes. In the bosonic theory, they have a dependence on the topological $\Theta$-angle. Both of these ingredients will be crucial in the  construction of the trans-series and resurgent analysis. 

At second order in the semi-classical expansion, we  have  two  types of topological configurations.  These are $[\K_i \K_j]$ type correlated kink-kink events, or bi-kinks,  with topological charge and action $2/N$ of a 2d instanton.  However, these events turn out to be not very important.  This is easy to understand in the $N_f \geq 1$ theory, where $[\K_i \K_j]$  amplitudes carry twice as many fermion zero modes and so are much less important for IR physics. It is a bit more subtle to understand this in the deformed bosonic theory, but it is nevertheless true. 

The more interesting events at second order are correlated kink-anti-kink  instanton events   $\cB_{ij}$, which exist for all non-vanishing entries of the extended Cartan matrix $\hat A_{ij} \neq 0$ of $SU(N)$. These defects have a typical size much larger than the kink size, but exponentially  smaller than the typical inter-kink separation. Therefore, these objects may be seen as  molecular  instanton events. 

For  $\hat A_{ij} < 0$,  there exists a {\it charged   bion}.  Charge $\mu_{\cB_{ij}}  =  \alpha_i -  \alpha_j$ stands for the fact that the root associated with this correlated tunneling event is non-zero. 
For $N_f \geq 1$ theories,  the  
interaction between  $\K_i $ and $\bar\K_j$ has a repulsive component due to boson exchange, in addition to an attractive fermion zero mode exchange induced interaction, leading to a bound (correlated) event, with characteristic size   $r_\text{b}$.  The charged bion is the counter-part of the magnetic bion in  QCD(adj) on $\R^3 \times \mathbb S^1$  \cite{Unsal:2007jx}.    The $N_f=0$ case requires more care and is discussed in Section \ref{sec:2defects}.  

For  $\hat A_{ii} >0  $,  there exists a neutral  bion. These are the correlated tunneling events beginning and ending at the same position in the landscape of vacua, hence the name neutral,   $\mu_{\cB_{ii}}  =  \alpha_i -  \alpha_i=0$. In this case, the interaction between 
 $\K_i $ and $\bar \K_i$ has two attractive components: one  due to boson exchange and another due to  fermion zero mode exchange.  Consequently, and naively, it does not make sense to talk about such molecules, because such an object would have a size as  small as  the kink-instanton size,  where it is meaningless to talk about a   kink-anti-kink molecule. In fact, this problem in 2d field theory  compactified down to quantum mechanics,  reduces to  a variant of a readily solved problems in quantum mechanics. The  $N_f=0$ case reduces to a problem addressed by  Bogomolny and Zinn-Justin in the context of bosonic quantum mechanics,  
\cite{Bogomolny:1977ty,Bogomolny:1980ur,Brezin:1977gk,Stone:1977au,ZinnJustin:1981dx},  while the  $N_f=1$ case reduces to  supersymmetric quantum mechanics addressed by Balitsky and Yung \cite{Balitsky:1985in},  and the $N_f >1$ case is new. 
The  Bogomolny--Zinn-Justin  (BZJ) prescription tells us how to make sense of such kink-anti-kink configurations in quantum mechanics.

\subsection{Classification  of non-perturbative ambiguities and ``confluence equations''}
\label{subsec:classification}

 According to the BZJ-prescription, applied now to  the   $N_f=0$ deformed--\cpn model,  the instanton--anti-instanton   $[\K_i\bar\K_i]$ amplitude is two-fold ambiguous. This is the first of many  non-perturbative ambiguities in   non-perturbative  amplitudes,  which  will appear  repeatedly. In fact, the amplitude exhibits a jump exactly on the real positive $g^2$ axis. Slightly above and below the real axis  $g^2 =  |g^2 |  e^{ i \theta},  \;\;  \theta= 0 \pm \epsilon $,  we define the left and right  bion amplitude as  
 \begin{equation}
 \label{dcp-bion}
     [\K_{i}  \bar \K_{i} ]_{\theta=0^{\pm}} = \Re   [\K_{i}  \bar \K_{i} ]   \pm i g^{-r_1} e^{-2S_0}, 
      \end{equation}
Whenever there are massless fermions in the theory,  the neutral bion event   $\cB_{ii}$ is ambiguity free.   
For \cpn  theories with massless  fermions, the first non-perturbative ambiguity appears at {\it fourth order} in the semi-classical expansion, and is due to the  bion-anti-bion amplitude:
  \begin{equation}
 \label{cp-bion}
     [\cB_{ij}  \bar \cB_{ij} ]_{\theta=0^{\pm}} = \Re   [\cB_{ij}  \bar \cB_{ij} ]   \pm ig^{-r_2} e^{-4S_0}\quad .
      \end{equation}
This is a new result: namely that there is an ambiguity associated with neutral  topological molecules  whose constituents carry a fraction of the  2d instanton action,  $1/N$ or $ 2/N$, respectively.
On  the face of it,  this is a  disaster, rendering the semi-classical expansion  meaningless.  
After all, we are calculating a  real  physical observable, say, a mass gap or vacuum energy density, 
in a system without an instability. What does an imaginary part mean?

 \begin{figure}[htb]
\centering{\includegraphics[scale=0.3]{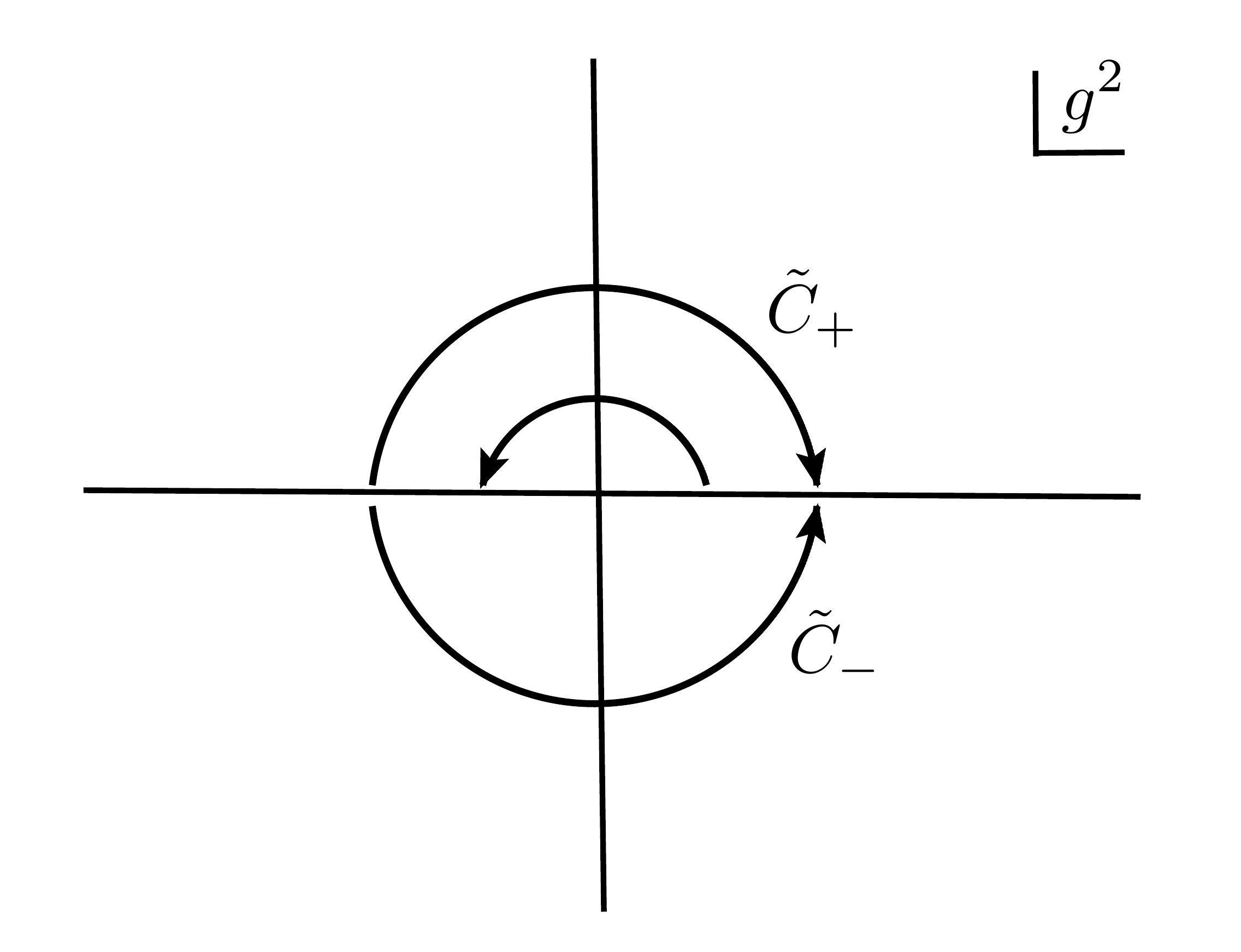}}
\caption{ This figure depicts one of the  main ideas of confluence equations and this work:  \newline  {\bf i)}  For real positive $g^2$, perturbation theory  is non-Borel summable, i.e., ill-defined.   Continue to negative $g^2$ where the perturbation theory becomes Borel summable. Then, continue back to $ |g^2| \pm i \epsilon$, where one obtains left(right) Borel sums,  $\B_{0, \th=0^\pm}$. The absence of a smooth $\theta =0$ limit  means non-Borel summability, i.e.,  perturbation theory does not define the theory. 
\newline {\bf ii)}  For real positive $g^2$,   the {\it neutral} bion amplitude is also ill-defined.  Continue to negative $g^2$ where it  is well-defined.  Then, continue it back, via $\tilde C_{\pm}$ to 
$ |g^2| \pm i \epsilon$.  Upon continuation, one obtains a left(right) neutral bion amplitude  $[\cB_{ii}]_{\th=0^\pm}$, with an imaginary discontinuity between the two. This means  an ambiguity at $\theta=0$. 
We demonstrate, analytically,  the {\it exact cancellation} of these two ambiguities at order $e^{-2S_0}\equiv  e^{-2S_I/N}$.   This is the first of many such cancellation encoded into confluence equations.  This means quantum field theory is non-perturbatively well-defined  in continuum up to ambiguities at order   $e^{-4S_0}$, which can further be improved systematically. 
\newline {\bf iii)}The mathematical  reason  behind this phenomenon is
that  $\R^{+}$ in the complex $g^2$-plane is a Stokes ray.  The jump
in the resummed perturbation theory is the Stokes jump, which is
(remarkably) mirrored by a jump in the neutral bion amplitude,  to
render observables meaningful even along the Stokes ray.  
  } 
\label{fig:continuation}
\end{figure}

In fact, what looks like a disaster turns out to be a blessing in disguise.  In asymptotically-free  confining field theories,  perturbation theory on  $\R^2$ is factorially divergent, a Gevrey-1 series (defined below in Section \ref{sec:large}). It is often non-Borel 
summable; that is to say, the sum is ambiguous.  One can define a left and right Borel sum, 
$\B_{0, \th=0^\pm}$, the imaginary part of which is the ambiguity. 
The ambiguity of perturbation theory on  $\R^2$  cannot be cured solely with instanton-anti-instanton amplitudes (which were sufficient to cure the problem in quantum mechanics \cite{Bogomolny:1980ur,ZinnJustin:1981dx,zinn-book}).  The reason is that perturbation theory on $\R^2$ develops singularities in the Borel plane, called  ``infrared renormalons", \cite{'tHooft:1977am, Beneke:1998ui} located at  approximately $1/N$ of the 2d   $[\cal I \bar {\cal I}]$ singularity. But there is  no known field configuration on $\R^2$ to cancel the  non-perturbative renormalon  ambiguity of the perturbation theory. However, on small  $\R^1 \times \mathbb S^1$ this is precisely what our neutral molecules  (\ref{dcp-bion}) and (\ref{cp-bion}) do. 

In  the semi-classical domain of the \cpn model on small $\R  \times \mathbb S^1$,  as in gauge theory on $\R^3 \times \mathbb S^1$,   we encounter two  classes  of non-perturbative ambiguities: 
 \begin{list}{$\bullet$}{\itemsep=0pt \parsep=0pt \topsep=2pt}
\item {\bf Ambiguities in the Borel resummation of perturbation theory} either 
 around the perturbative vacuum or in the background of instantons or kinks. 
\item {\bf   Ambiguities  in the  definition of the non-perturbative amplitudes}  (\ref{dcp-bion}), (\ref{cp-bion}) associated with neutral  topological molecules, or molecules which include neutral sub-components. 
\end{list}
In order for the  \cpn model to have a meaningful,  semi-classical  non-perturbative definition in the continuum,  these two class of ambiguities must cancel.  Denote the left (right) Borel resummation of perturbation theory, as described above, by $\B_{0, \th=0^\pm}$.  The fact that   
$\B_{0, \th}  $ exhibits a jump at $\theta=0$ is the statement that the expansion is  on a Stokes ray and the jump is the Stokes discontinuity. In this work,  we  will show explicitly 
the cancellation between the leading ambiguity in perturbation theory with the leading ambiguity in the non-perturbative neutral bion amplitude:  
  \begin{align}\label{borelbion0}
\Im\,\B_{0, \th=0^\pm}  + \Im\, [\cB_{ii}]_{\th=0^\pm}  
=0 \;  ,\quad {\rm up \; to}  \; e^{-4S_0}
\end{align}
This is the first entry in a hierarchy of such cancellation  equations: see Section \ref{confluence}  and  equations (\ref{rBZJ2}). These equations  hold the key to the possibility of defining QFTs in the continuum using the resurgence framework. For this reason we give these equations a name: {\it perturbative--non-perturbative confluence equations}, or {\it confluence equations} for short. 
Our main conjecture is that all columns, i.e., all independent $\Theta$ sectors  shown in the resurgence triangle (\ref{triangle}) are resurgent functions of the parameter $\lambda \hbar \equiv \lambda$.

The first class of ambiguities is  well-known in perturbation theory, both in quantum mechanics 
and QFT \cite{LeGuillou:1990nq}. The latter class is perhaps less well known, but has been  studied in the context of quantum mechanics  \cite{Bogomolny:1977ty,Bogomolny:1980ur,Brezin:1977gk,Stone:1977au,Balian:1978et,ZinnJustin:1981dx,zinn-book,
Balitsky:1985in,h2plus,silverstone,Achuthan:1988wh}.
Our main new result here is the connection with IR renormalons in QFT for this second class of ambiguities, in non-perturbative amplitudes for topological molecules in QFTs. This effect   is primarily explored in this work and  its companion on $\R^3 \times \mathbb S^1$ \cite{Argyres:2012vv}. 
In the case of \cpn, compared to Yang-Mills theories on $\R^3 \times \mathbb S^1$, the situation simplifies 
relatively  because  the small-compactified circle limit reduces directly to quantum mechanics, and certain important technical results already exist  in the literature \cite{Bogomolny:1980ur, ZinnJustin:1981dx,zinn-book,Balitsky:1985in}.\footnote{The same can be done for QFTs on $\R^4$ as well, by an unconventional compactification  on $T^3 \times \R $. This will be explored elsewhere.}

By using the resurgence formalism, we calculate the mass gap, a solely non-perturbative quantity  and the $\Theta$ angle dependence of the vacuum energy density at arbitrary $N$.  Both results are nontrivial, and are in accord with large-$N$ and lattice results.  Unlike the large-$N$ considerations, which provide little microscopic insight, our  derivation also makes explicit the microscopic origin of the mass gap and $\Theta$-dependence  in the \cpn-model. To our knowledge, this is the first time that these non-perturbative quantities are analytically derived from first principles.

\subsection{Interlude: Prototypes of trans-series expansions}
\label{subsec:examples}

The cancellation of ambiguities between perturbative and non-perturbative  expansions is an example of the resurgent behavior of trans-series expressions, in which the full expansion of a physical observable should be viewed as unifying the perturbative and non-perturbative parts, as in (\ref{trans}). Given the physical significance of such behavior, we sketch the main mathematical ideas of resurgence with some elementary examples. We refer the interested reader to \cite{Dingle:1973,Ecalle:1981,BerryHowls,s07,Costin:2009,Sternin:1996} for excellent introductory reviews. Let us state clearly at the beginning that a major advantage of this formalism is that it enables us to keep track of the relation between perturbative and non-perturbative expansions as we analytically continue in the phase of the coupling constant. The familiar relation between the single instanton sector and the large-order growth of perturbative coefficients, the Bender-Wu relations in quantum mechanics \cite{Bender:1969si,arkady} and the Lipatov analysis in QFT \cite{Lipatov:1976ny}, are the simplest examples, but they are the proverbial ``tip of the iceberg''. Resurgence produces a whole series of such relations between different non-perturbative sectors, and these are needed to demonstrate the full consistency of QFT.

A simple illustrative class of examples of trans-series and resurgence arises for the asymptotic behavior of functions satisfying a second-order differential equation [such as Bessel functions, Airy functions, parabolic cylinder functions, etc, or indeed for general Schr\"odinger equations], for which there are just {\it two} non-perturbative  exponential terms in the trans-series expansion (\ref{trans}). Consider the following integral, related to the modified Bessel function $K_0$:
\begin{eqnarray}
Z_1(\lambda)&=& \int_{-\infty}^\infty dx\, e^{-\frac{1}{2\lambda} \sinh^2(\sqrt{\lambda}\, x)} 
\label{z11}\\
&=& \frac{1}{\sqrt{\lambda}}\, e^{\frac{1}{4\lambda}}\, K_0\left(\frac{1}{4\lambda}\right)
\label{z12} \\
&\sim & \sqrt{\frac{\pi}{2}}\sum_{n=0}^\infty (-1)^n (2\lambda)^n\frac{\Gamma(n+\frac{1}{2})^2}{n!\,\Gamma\left(\frac{1}{2}\right)^2} \quad, \quad \lambda\to 0^+
\label{z13}
\end{eqnarray}
This is the ``partition function'', $Z_1(\lambda)={\rm tr}\, e^{-V_1}$, for the 0-dimensional field theory with potential $V_1(x)=\frac{1}{2\lambda} \sinh^2(\sqrt{\lambda}\, x)=\frac{1}{2}x^2+\frac{\lambda}{6}x^4+\dots$. The perturbative series in (\ref{z13}) is factorially divergent (Gevrey class 1; see Section \ref{sec:large}), but has coefficients alternating in sign, and is Borel summable (see (\ref{z1-borel})  below). On the other hand, for the periodic potential, $V_2(x)=\frac{1}{2\lambda} \sin^2(\sqrt{\lambda}\, x)=\frac{1}{2}x^2-\frac{\lambda}{6}x^4+\dots$, obtained formally by $\lambda\to -\lambda$, we have the partition function, $Z_2(\lambda)={\rm tr}\, e^{-V_2}$,
related to the modified Bessel function $I_0$:
\begin{eqnarray}
Z_2(\lambda)&=& \int_0^{\pi/\sqrt{\lambda}}dx\, e^{-\frac{1}{2\lambda} \sin^2(\sqrt{\lambda}\, x)} 
\label{z21}\\
&=& \frac{\pi}{\sqrt{\lambda}}\, e^{-\frac{1}{4\lambda}}\, I_0\left(\frac{1}{4\lambda}\right)
\label{z22} \\
&\sim & \sqrt{\frac{\pi}{2}}\sum_{n=0}^\infty  (2\lambda)^n\frac{\Gamma(n+\frac{1}{2})^2}{n!\,\Gamma\left(\frac{1}{2}\right)^2} \quad, \quad \lambda\to 0^+
\label{z23}
\end{eqnarray}
Formally, it is tempting to conclude from the perturbative expansions in (\ref{z13}) and (\ref{z23}) that 
$Z_1(-\lambda)=Z_2(\lambda)$, but this is \emph {not true}, as it misses important non-perturbative contributions [see (\ref{contin}) below]. 

The first clear sign of a problem is that the perturbative expansion in (\ref{z23}) is a {\it non-alternating} divergent series and is not Borel summable. There is therefore the possibility of a non-perturbative imaginary ambiguity in $Z_2(\lambda)$. However, we know that the periodic potential system is stable, so there should be no imaginary part, with $Z_2(\lambda)$ being real.

We can resolve this problem using Borel-\'Ecalle summation. From (\ref{z13}) we see that the Borel sum of the perturbative series for $Z_1(\lambda)$ can be expressed in terms  of a hypergeometric function:
\begin{eqnarray}
Z_1(\lambda)=  \sqrt{\frac{\pi}{2}}\, \frac{1}{2\lambda}\int_0^\infty dt\, e^{- \frac{t}{2\lambda}} ~_2F_1\left(\frac{1}{2},\frac{1}{2}, 1; - t\right)
\label{z1-borel}
\end{eqnarray}
For $\lambda>0$, this Borel integral is well defined, as the hypergeometric function has a cut $(-\infty, -1)$ only along the negative $t$ axis. On the other hand, a formal Borel expression for $Z_2(\lambda)$ has a cut on the contour of integration. So we must define $Z_2(\lambda)$ by analytic continuation from $Z_1(\lambda)$. 
Consider rotating the phase of $\lambda$ in the $Z_1(\lambda)$ Borel expression (\ref{z1-borel}) so that $\lambda\to |\lambda | e^{i\theta}$. Then the direction of the branch cut rotates, and when $\theta$ approaches $\pm \pi$, the branch cut approaches the contour of integration, either from below or above. In this limit, when $\theta=\pm \pi$, the alternating asymptotic series in $Z_1(\lambda)$ in (\ref{z13}) becomes non-alternating. There is however an ambiguity, because we can rotate either clockwise or anti-clockwise. The difference between these two results is the difference of the two so-called ``lateral Borel sums'', ${ \B}_\pm(\lambda)$, a measure of the ambiguity in summing the non-alternating series, and can be written as a Borel integral above and below a cut:
\begin{eqnarray}
Z_1(e^{i\pi}\,\lambda)-Z_1(e^{-i\pi}\,\lambda)&=& \sqrt{\frac{\pi}{2}}\, \frac{1}{2\lambda}\
\int_1^\infty dt\, e^{-\frac{t}{2\lambda}}\left[~_2F_1\left(\frac{1}{2},\frac{1}{2}, 1, t-i\varepsilon\right)-~_2F_1\left(\frac{1}{2},\frac{1}{2}, 1, t+i\varepsilon\right)\right]
\nonumber\\
&=& -(2i)\, \sqrt{\frac{\pi}{2}}\, \frac{1}{2\lambda}\, e^{-\frac{1}{2\lambda}}  \int_0^\infty dt\, e^{-\frac{t}{2\lambda}}\, ~_2F_1\left(\frac{1}{2},\frac{1}{2}, 1, -t\right) 
\label{disc21}\\
&=&-2i\, e^{-\frac{1}{2\lambda}} \, Z_1(\lambda)
\label{disc22}
\end{eqnarray}
To obtain (\ref{disc21})  we used the known discontinuity property of the hypergeometric function \cite{nist}:
\begin{eqnarray}
~_2F_1\left(\frac{1}{2},\frac{1}{2}, 1, t+i\varepsilon\right)-~_2F_1\left(\frac{1}{2},\frac{1}{2}, 1, t-i\varepsilon\right)= 2i ~_2F_1\left(\frac{1}{2},\frac{1}{2}, 1, 1-t\right)
\label{hyper}
\end{eqnarray}
Notice the amazing fact that in the discontinuity of $Z_1(\lambda)$ in (\ref{disc22}) we recover an exponential factor, $e^{-2(\frac{1}{4\lambda})} $, multiplying the original function $Z_1(\lambda)$. This is not an accident -- it is a sign of resurgence at work. 

In fact,  the modified Bessel functions $K_0(z)$ and $I_0(z)$ are related under analytic continuation of their argument by the following  connection formula:
\begin{eqnarray}
K_0(e^{\pm i\pi}\, |z|) =K_0(|z|)\mp i\,\pi\, I_0(|z|)
\label{conn}
\end{eqnarray}
Therefore, we deduce that
\begin{eqnarray}
Z_1(e^{\pm i \pi}\, \lambda)&=& \frac{\pi}{\sqrt{\lambda}}\, e^{-\frac{1}{4\lambda}}\, I_0\left(\frac{1}{4\lambda}\right) \mp \frac{i}{\sqrt{\lambda}} \, e^{-\frac{1}{4\lambda}}\, K_0\left(\frac{1}{4\lambda}\right) \\
&=& Z_2(\lambda) \mp i \, e^{-\frac{1}{2\lambda}}\, Z_1(\lambda)
\label{contin}
\end{eqnarray}
which is consistent with (\ref{disc22}).

Another way to understand this is that the naive asymptotic expansions of $K_0(z)$ and $I_0(z)$ for $z\to +\infty$ are not consistent with the connection formula (\ref{conn}), reflecting the fact that these asymptotic expansions do not fully define the functions. The asymptotic expansion, as $z\to +\infty$,  of $K_0(z)$ is easily obtained from the following integral representation:
\begin{eqnarray}
K_0(z)&=&\frac{1}{2}\int_0^\infty \frac{dt}{t}\, e^{-\frac{z}{2}\left(t+\frac{1}{t}\right)}
\sim  \sqrt{\frac{\pi}{2z}}\, e^{-z}\sum_{n=0}^\infty \frac{(-1)^n}{(2z)^n}\frac{\Gamma\left(n+\frac{1}{2}\right)^2}{n!\, \Gamma\left(\frac{1}{2}\right)^2}
\label{k0-series}
\end{eqnarray}
The integral has saddle points at $t=\pm 1$, but only the saddle at $t=+1$ contributes.
The resulting series is asymptotic but Borel summable. The other modified Bessel function, $I_0(z)$, has a quite different integral representation
\begin{eqnarray}
I_0(z)&=&\frac{1}{2\pi i}\int_C \frac{dt}{t}\, e^{-\frac{z}{2}\left(t+\frac{1}{t}\right)}
\label{i0-int}
\end{eqnarray}
where $C$ is the contour of the anti-clockwise unit circle. Now both saddle points, at $t=\pm 1$, contribute, and the full resurgent asymptotic expansion reads:
\begin{eqnarray}
I_0(z)&=& \sqrt{\frac{1}{2\pi z}}\left[e^{z}\sum_{n=0}^\infty \frac{1}{(2z)^n}\frac{\Gamma\left(n+\frac{1}{2}\right)^2}{n!\, \Gamma\left(\frac{1}{2}\right)^2}+
\left\{\begin{matrix}
i\cr -i\cr 0
\end{matrix}
 \right\}
  e^{-z}\sum_{n=0}^\infty \frac{(-1)^n}{(2z)^n}\frac{\Gamma\left(n+\frac{1}{2}\right)^2}{n!\, \Gamma\left(\frac{1}{2}\right)^2}\right]
\label{i0-series}
\end{eqnarray}
where the three cases are for $0< {\rm arg}\, z<\pi$, $-\pi< {\rm arg}\, z<0$, and ${\rm arg}\, z=0$, respectively. This is a two-term trans-series expansion, with two different non-perturbative (in $z=\frac{1}{4\lambda}$) exponential terms, $e^{\pm z}$, one associated with each saddle point. The second term is exponentially sub-leading when ${\rm Re}(z)>0$, and so is often neglected. But as the phase of $z$ approaches $\pm \pi/2$, both terms contribute, accounting for the real oscillatory nature of the ordinary Bessel functions $J_0(z)$ and $Y_0(z)$. More importantly, the coefficients in the asymptotic expansions  multiplying each exponential term are related to one another in a  particular  way,  differing simply by a factor of $(-1)^n$. This correspondence is a direct consequence of Darboux's theorem: the high orders of the asymptotic expansion about one saddle involve high derivatives, and these are determined by the behavior of the function in the vicinity of the nearest singularity, which is the other saddle.  This is also manifest in our transseries expansion: The value of the  ``action"  $S(z) = \frac{z}{2}\left(t+\frac{1}{t}\right)$ at the two saddle points  $t=\pm 1$ are   $S_{+} = z$ and $S_{-}=-z$, and the two exponential terms in the trans-series are respectively $e^{-S_{+}}=e^{-z}$ and $e^{-S_{-}}=e^{z}$.   Dingle defines the difference of the two ``actions" as  the {\it singulant}  \cite{Dingle:1973,BerryHowls}
\begin{align} 
\label{singulant}
\Delta S_{+-}=S_{+} - S_{-} \; , 
\end{align}
 and notes that the large order behavior of the  asymptotic series is controlled by the singulant. The trans-series  is roughly (ignoring inessential details)
\begin{eqnarray}
I_0(z) \sim  \left[e^{-S_{-}}\sum_{n=0}^\infty \frac{n!}{(\Delta S_{+-})^n} +
\left\{\begin{matrix}
i\cr -i\cr 0
\end{matrix}
 \right\}
  e^{-S_{+}}\sum_{n=0}^\infty \frac{(-1)^n n!}{(\Delta S_{+-})^n}
\label{i0-series-2} \right]
\end{eqnarray}
This feature is generic, as was discovered by Dingle in the context of WKB analysis \cite{Dingle:1973}, and elaborated by Berry and Howls \cite{BerryHowls} for more general integrals with multiple saddles.  
Let  $t_i, i=1,2,\ldots$ denote the set of saddle points with actions $S(t_i) =S_i$. Ref.\cite{BerryHowls} defines a  singulant  $S_{ij} = S_i -S_j$ for each adjacent saddle $j$.    The leading $n!$ divergence of the asymptotic series is controlled by the saddle(s) $j$  for which $|S_{ij}| <   |S_{ij'}|$ for all $j'$.  We refer the reader   to  Ref.\cite{BerryHowls} for a discussion of the intricate  topological structure that arises in an integral with multiple saddles, characterizing which saddles affect one another. This is one aspect of resurgence -- the expansions around different saddles 
 [i.e., the expansions about different instantons] 
  are necessarily related to one another  in a very precise way, as encoded in Darboux's theorem. \footnote{ 
Clearly, the form given in  (\ref{i0-series-2})  is very suggestive for quantum field theory. Indeed, 
we  will identify 
a very similar structure in  quantum field theory (albeit with infinitely many exponential factors, as in the sub-sequent $\Gamma$ function example)  in the perturbative expansion around the perturbative vacuum or different topological  sectors.  In particular, the singulants  will be identified with  neutral topological composites (like neutral bion),  as opposed to single instantons. }

Other important examples of resurgence, more like what we expect to encounter  in quantum field theory, involve an {\it infinite} series of non-perturbative exponential terms. The simplest example of this type  is given by Stirling's formula for the gamma function. In fact, it is easier to describe in terms of the digamma function $\psi(x)=\Gamma^\prime(z)/\Gamma(z)$. Consider the divergent but Borel summable ``perturbative'' expansion for $z\to +\infty$ (so $\frac{1}{z}$ is the small perturbative parameter):
\begin{eqnarray}
\psi\left(\frac{1}{2}+z\right)
&= & \ln z-\int_0^\infty dt\, e^{-2 z t}\left(\frac{1}{\sinh t}-\frac{1}{t}\right)
\label{psi1}\\
&\sim & \ln z-2\sum_{n=0}^\infty \frac{(-1)^n(2n+1)!}{(2\pi z)^{2n+2}}\left(\frac{1-2^{2n+1}}{2^{2n+1}}\right)\zeta(2n+2)
\label{psi2}
\end{eqnarray}
The series expansion is sufficient to satisfy the basic recurrence relation, $\psi\left(\frac{1}{2}+z\right)=\psi\left(-\frac{1}{2}+z\right)+1/(z-1/2)$, derived from the   gamma function relation $z\,\Gamma(z)=\Gamma(z+1)$. Now imagine rotating $z$ to $e^{\pm i \pi/2} z$, so that the series becomes non-alternating. The series representation (\ref{psi2}) suggests that the difference between these two rotations might be: $\psi\left(\frac{1}{2}+i z\right)-\psi\left(\frac{1}{2}-i z\right)\sim i \pi$, coming from the log term. However, we also know the ({\it global}) reflection formula for the gamma function, $\Gamma(z)\Gamma(1-z)=\pi/\sin(\pi z)$, which implies that
\begin{eqnarray}
\psi\left(\frac{1}{2}+i z\right)-\psi\left(\frac{1}{2}-i z\right)=
i\, \pi\, \tanh(\pi\, z) =
i\, \pi\left(1+2 \sum_{k=1}^\infty (-1)^k\, e^{-2 \pi k z}\right)
\label{psi-bridge}
\end{eqnarray}
We see that in addition to the expected  $i\,\pi$ term, there is an infinite series of exponentially small non-perturbative terms (in $e^{-2\pi z}=e^{-2\pi/\lambda}$). These are neglected in the series representation (\ref{psi2}) but are needed in order to satisfy the reflection formula. Moreover, we obtain the following expression for the real part:
\begin{eqnarray}
\hskip -.5cm {\rm Re}\, \psi\left(\frac{1}{2}+i\, z\right)\sim 
\ln z+2\sum_{n=0}^\infty \frac{(2n+1)!}{(2\pi z)^{2n+2}}\left(\frac{1-2^{2n+1}}{2^{2n+1}}\right)\zeta(2n+2) -i\pi \sum_{k=1}^\infty (-1)^k\, e^{-2 \pi k z}
\end{eqnarray}
This formula looks strange since the LHS is obviously real.  The formal first sum on the RHS looks real, but this is deceptive. It also looks like there is an imaginary contribution on the RHS, and this is also deceptive. Upon Borel resummation of the formal series, 
 the imaginary non-perturbative part on the RHS cancels the imaginary part coming from the Borel summation of the non-alternating divergent series.  This is how the RHS is (not so obviously) real. 

The reflection formula is derived from an analytic continuation of the gamma function, so it encodes {\it  global} information about the function, more than just the basic recursion formula 
$z\,\Gamma(z)=\Gamma(z+1)$  
(which may be viewed as the local Schwinger-Dyson equation derived from the integral representation of the gamma function) 
and its series solution in (\ref{psi2}). In other words, a global resurgent trans-series expression for the digamma function at large $|z|$ must include both perturbative powers of $1/z$ and also exponentially small non-perturbative terms $e^{-2\pi k z}$, for all positive integer $k$. Furthermore, the perturbative series part is not independent of the non-perturbative exponential part, as they are connected by the reflection formula.
\footnote{In QFT, the counter-part of this global relation is the confluence equations  discussed in 
Section \ref{confluence}.}  
 Equivalently, we can deduce the exponential terms from directed Borel integrals that go around the poles of the Borel integrand at $t=i k \pi$ in  (\ref{psi1}), which can in turn be identified with saddles of an integral representation, which interact with one another according to Darboux's theorem. An analogous analysis is possible for the Barnes multiple gamma functions and for the Hurwitz zeta function, the basic functions known to underly quantum field theoretic effective actions in constant gravitational and electromagnetic background fields \cite{Das:2006wg}.

At first sight one might think that these examples are unrealistically special, since we know extra information about these functions [such as a differential equation or a functional relation]. However, 
the first example of the modified Bessel functions captures much of the physics of the perturbative and non-perturbative ambiguities arising in the quantum mechanics of periodic potentials such as the Sine-Gordon model \cite{Stone:1977au,zinn-book,delabaere-periodic}, along the lines of the Bogomolny-Zinn-Justin analysis \cite{Bogomolny:1977ty,Bogomolny:1980ur,ZinnJustin:1981dx}, and is central to our \cpn analysis in this paper (see Section \ref{sec:resurgence-cpn}). The second example of the digamma function underlies the non-perturbative ambiguities found in quantum field theoretic effective actions in constant electromagnetic and gravitational backgrounds such as found in the analytic continuation between de Sitter and anti-de Sitter spaces \cite{Das:2006wg}, and in certain highly symmetric Chern-Simons models \cite{schiappa,marino,Marino:2012zq,Russo:2012kj}. 

Beyond these examples, the essence of \'Ecalle's theory of resurgence is that by studying  the analytic structure of Borel integrals in the vicinity of {\it all} the poles and cuts, one can in principle reconstruct all information about the function starting from a series representation \cite{Ecalle:1981}. One of \'Ecalle's main results is that the set of ``analyzable functions'', for which these operations apply, forms a basis of trans-series expansions which appears to be sufficient to capture all the required information for the types of functions that appear in physics and mathematics. This is a profound and provocative viewpoint, which has motivated our study here of the ${\mathbb C}{\mathbb P}^{N-1}$  model, to see to what extent this bridge between perturbative and non-perturbative physics is actually realizable in quantum field theory.  

\section{\cpn Models}
\label{section-model-a}

We study the \cpn model defined on  two-dimensional Euclidean space-time, \rone$\times$\sone, with the topology of a cylinder, using coordinates $(x_1, x_2)$, where $x_1$ is the non-compact Euclidean time direction, and $x_2$ is the compactified spatial dimension of length $L=2\pi R$. The classical action of the \cpn model is
\begin{eqnarray}
S=\frac{2}{g^2}\int d^2 x \left(D_\mu n\right)^\dagger D_\mu n 
\label{action}
\end{eqnarray}
where $n=(n_1, n_2, \dots n_N)^T$ is a complex  $N$-component column vector satisfying $n^\dagger n=1$. The
covariant derivative operator is $D_\mu=\partial_\mu+i A_\mu$, and the abelian gauge field $A_\mu$ is determined by its equation of motion to be
\begin{eqnarray}
A_\mu=\frac{i}{2}\left(n^\dagger \partial_\mu n- \partial_\mu n^\dagger n\right)
\label{gauge}
\end{eqnarray}
The \cpn model has a local $U(1)$ gauge redundancy,
\begin{eqnarray}
n(x)\to e^{i\alpha(x)}n(x)\quad, \quad A_\mu(x)\to A_\mu(x)  -\partial_\mu \alpha(x)
\label{u1}
\end{eqnarray}
and also a global $U(N)$ symmetry 
\begin{eqnarray}
n(x)\to U\, n(x)\quad,\quad U\in U(N)
\label{global}
\end{eqnarray}
The   $n$-field  parametrizes    the   coset space 
 \begin{equation}
 {\cal M}_{N, 1} \equiv \mathbb C  \mathbb P ^{N-1} =   \frac{U(N)}{U(N-1) \times U(1)} 
 \label{coset}
 \end{equation}
and is   characterized by $N^2-1-(N-1)^2=2(N-1)$ real fields, which  are massless to all orders in  perturbation theory. 

It will be useful to add a topological  theta-term  to the action:
\begin{eqnarray}
S_\Theta= i   \Theta\,  Q \label{theta}  
\end{eqnarray}
where $\Theta$  is an angular parameter with period $2\pi$, and  $Q$ is the topological charge:
\begin{eqnarray}
Q  =  -\frac{i }{2\pi} \int d^2 x \, \epsilon_{\mu\nu} (D_\mu n)^\dagger D_\nu n
= -\frac{i }{2\pi} \int d^2 x \, \epsilon_{\mu\nu}\partial_\mu\left(n^\dagger \partial_\nu n\right) 
=\frac{1}{2\pi} \int d^2 x \, \epsilon_{\mu\nu}\partial_\mu A_\nu
\label{charge}  
\end{eqnarray}
In bosonic and deformed \cpn theories,  physical observables exhibit $\Theta$ dependence.  
The  $\Theta$ term will be particularly useful in analyzing the semi-classical trans-series 
expansion.  We will also consider  the $\C\P^{\rm N-1}$ model with $N_f$ Dirac fermions: \begin{equation}
S_{\rm fermion}= \frac{2}{g^2} \int d^2 x\left[  - i \bar \psi \gamma_\mu D_{\mu} \psi + \textstyle \frac{1}{4} \left( (\bar \psi \psi)^2  + (\bar \psi \gamma_3 \psi)^2   -  (\bar \psi \gamma_\mu \psi)^2 \right)\right]
\label{fermion}
\end{equation} 
where a sum over the fermion flavor index  is assumed. The $N_f=1$  theory  is the  ${\cal N} =(2,2)$ supersymmetric  model \cite{D'Adda:1980ie}, while for $N_f>1$ the theory is non-supersymmetric.

All $\C\P^{\rm N-1}$ models are asymptotically free, because the first order $\beta$-function is independent of $N_f$.  Whether this class of theories is confining or conformal in the IR depends on $N_f$. 
The coupling $g^2(\m)$ is a function of energy scale $\m$. At one loop order, we have
\begin{equation}
\label{dimtrans}
\L^{\beta_0} =  \mu^{\b_0}  e^{ - \frac{ {4\pi}}{ g^{2}(\m)}} , \qquad {\b_0} =N \qquad  { \rm or } \qquad 
\L  =  \mu \, e^{ -  S_I / \b_0 }  = \mu \, e^{ - \frac{ {4\pi}}{ Ng^{2}(\m)}} ,
\end{equation}
 where $\L$ is the strong coupling scale, $\b_0$ is the coefficient of the 1-loop beta function, 
and  $S_I$ is the instanton action.  $\L$ is the natural scale of the mass gap of the theory. 
  Note the simple but important fact that  the relation between $\L $ and   $\mu $, via dimensional transmutation, is {\it not} determined by the instanton factor, but by the instanton factor divided by $N$. Thus, on $\R^2$, a physical quantity like the mass gap cannot have a semi-classical origin or description. On the other hand, in the weakly coupled semi-classical domain on $\R^1 \times \mathbb S^1$, we will show that the semi-classical expansion is in $ e^{ -  S_I / \b_0 } \equiv e^{ - \frac{ {4\pi}}{ Ng^{2}(\m)}} $, allowing us to identify the configurations which lead to a mass gap. 

The $N_f=1$ model has a classical $U(1)_V \times U(1)_A$ symmetry, acting on the 2d 
Dirac fermion $\psi= (\psi_{+}, \bar \psi_{-})^T  $ as 
\begin{equation}
V : \psi  \longrightarrow e^{ i \delta } \psi , \qquad  A : \psi 
 \longrightarrow e^{ i \sigma_3 \beta} \psi   
\end{equation} 
where $\psi_{+}$ and  $\psi_{-}$ are the left and right-mover modes.
The $U(1)_A$ symmetry is anomalous due to instanton effects. In the background of an instanton,  chirality is violated by  $2N$ units, and 
 the associated 2d instanton amplitude is given by 
\begin{equation}
I_{\rm 2d} =e^{-S_I}   (\psi_{-}   \psi_{+ })^N     
\end{equation} 
Quantum mechanically, the exact  axial symmetry of the theory is   $\Z_{2N}$.  On $\R^2$, there is compelling evidence that  the  $\Z_{2N}$ discrete chiral symmetry is dynamically broken down to  $\Z_{2}$ by a fermion-bilinear condensate 
 \begin{equation}
  \label{cc1}
 \langle  \psi_{-}   \psi_{+ }  \rangle = N \Lambda\, e^{i \frac{2 \pi k}{N}}, \qquad  k=0,1, \ldots, N-1 
 \end{equation}
  leading to $N$ isolated vacua. The supersymmetric index for this theory is also $N$.
  We will give a new derivation of the chiral condensate and supersymmetric index in Section \ref{sec:susy}.

  In the multi-flavor  theory $N_f>1$ theory, the classical global symmetry is  $U(N_f)_V \times U(N_f)_A$. Quantum mechanically, due to instanton effects, this symmetry is reduced to 
  \begin{equation}
  \left [ U(1)_V \times SU(N_f)_V \times SU(N_f)_A \times \Z_{2N\, N_f} \right] / {\Z_2 \times \Z_{N_f}} 
  \label{symm}
  \end{equation}
  The terms in the denominator are there to prevent double counting. $\Z_2$ is fermion number mod two, which lives  both in $U(1)_V$  and  $\Z_{2N\, N_f}$.  $\Z_{N_f}$ lives in both the 
 diagonal subgroup of the continuous chiral symmetries as well as in  $\Z_{2N\, N_f}$. The true discrete chiral symmetry of the multi-flavor theory is also just $\Z_N$, as in the case of single flavor theory. The 2d instanton amplitude in the $N_f$ flavor theory is given by  
\begin{equation}
I_{\rm 2d} =e^{-S_I}  \left[ \det_{ab} (\psi^{a}_{-}   \psi^{b}_{+ }) \right]^N    \sim 
e^{-S_I}  \left[(\psi^{n_1}_{-}   \psi^{n_1}_{+ })     (\psi^{n_2}_{-}   \psi^{n_2}_{+ })  
\ldots (\psi^{N_f}_{-}   \psi^{N_f}_{+ })  + \; {\rm permutations} \;  \right]^N  
\end{equation} 
This is of course a singlet under continuous chiral symmetries, but reduces the discrete chiral symmetry down to   $\Z_{2N\, N_f}$. In two-dimensions, the continuous chiral symmetry cannot be spontaneously broken, by the Coleman-Mermin-Wagner theorem \cite{Coleman}.  However, the discrete chiral symmetry may be broken. The order parameter associated with the discrete chiral symmetry is 
 \begin{equation}
 \label{cc2}
 \langle  \det_{ab} (\psi^{a}_{-}  \psi^{b}_{+ })  \rangle   \sim 
  \Lambda^{N_f} e^{i \frac{2 \pi k}{N}}, \qquad  k=0,1, \ldots, N-1 
 \end{equation}
  leading to $N$ isolated vacua, as in the case of the supersymmetric theory.  The chiral condensates  (\ref{cc1}) and    (\ref{cc2})  on $\R^2$, just like the mass gap of the bosonic theory, cannot have a semi-classical description.    Since the  instanton size is a modulus parameter, as noted in the Introduction, there is  no well-defined semi-classical dilute instanton gas description on $\R^2$. 

\subsection{Twisted boundary conditions vs.  background in classical theory}
\label{section-model-b}

Twisted boundary conditions can be used to probe more finely the structure of the \cpn 
model. We first show the  equivalence at the  {\it classical} level between the theory with 
twisted   boundary conditions  and  action (\ref{action}),  and the theory with untwisted (periodic)  boundary conditions in the presence of  a  $U(1)^N$  background  ``gauge" field. In the next section (\ref{section-model-c}), we show that the background field is associated with a gauge invariant line operator that we construct from the \cpn\, fields.  The quantum mechanical stability of the background is discussed in  Section \ref{section-oneloop-a}. As stated in the Introduction, twisted boundary conditions  on their own do not necessarily imply semi-classical calculability. What is crucial is that a certain type of background must be stable, in order for a semi-classical analysis  to apply usefully  to \cpn. 

Consider  twisted boundary conditions for the bosonic  and fermionic fields in \cpn:
\begin{eqnarray}
n (x_1, x_2+L)&=&  \Omega_0 \,n (x_1, x_2)\qquad , \qquad  
\; \; n_i(x_1, x_2+L)  = e^{2\pi i \mu_i}  \;  n_i(x_1, x_2)
  \cr 
\psi (x_1, x_2+L)&=& \pm \Omega_0 \, \psi (x_1, x_2)  \qquad, \qquad  \psi_i(x_1, x_2+L)  =  \pm e^{2\pi i \mu_i}  \;  \psi_i(x_1, x_2)
\label{twist}
\end{eqnarray}
where $\Omega_0 \in U(N)$ is the twist matrix
\begin{eqnarray}
\Omega_0&=&\begin{pmatrix}
e^{2\pi i\mu_1} &0&\dots &0 \cr
0& e^{2\pi i\mu_2}&\dots &0\cr
\vdots\cr
0&0&\dots & e^{2\pi i\mu_N}
\end{pmatrix}
\quad, \quad 0\leq \mu_1\leq \mu_2 \leq \dots \leq\mu_N<1
\label{twist2}
\end{eqnarray}
In other words, the fields are periodic only up to a global $\Omega_0$ rotation, with $\Omega_0 \in U(N)$.  The twist matrix $\Omega_0$ can always be brought to this diagonal form by a similarity transformation. 

A general   twisted boundary condition  with non-degenerate twist explicitly breaks  the global 
$U(N)$ symmetry  down to $U(1)^N$.  A special role will be  played by the maximally symmetric twist, or $\Z_N$-symmetric twist: 
\begin{eqnarray}
(\mu_1, \mu_2, \dots, \mu_N)=\frac{1}{N}(0, 1, 2, \dots, N-1)
\label{max}
\end{eqnarray}
for which  the eigenvalues of $\Omega_0$ are equidistant.   

The twisted   boundary condition  for the theory with action (\ref{action}) is equivalent to the theory with  periodic  boundary conditions and a  twisted $U(1)^N$  background field. To see this, define  periodic fields ${\tn}_j$, for $j=1, \dots N$:
   \begin{equation}
 {\tn}_j (x_1, x_2)  = e^{- i \frac{2\pi  \mu_j x_2}{L}}  n_j(x_1, x_2)  \qquad, \qquad   {\tn}_j (x_1, x_2+L)  =  
 {\tn }_j (x_1, x_2),
\label{aperiodic}
\end{equation}
The gauge field $A_{\mu}$ in (\ref{gauge}) can be re-expressed in terms of the periodic fields   ${\tn}(x_1, x_2)$ as
\begin{eqnarray}
A_{\mu}(n) &&= A_{\mu} (\tn)     -  \frac{2 \pi}{L}  \delta_{\mu 2}  \sum_{j=1}^N {\tn}^\dagger_j \,  \mu_j\,{\tn}_j
\label{gauge2}
\end{eqnarray}
Then the bosonic action (\ref{action}) can be expressed in terms of the periodic field as 
\begin{eqnarray}
S= \frac{2}{g^2} \int d^2 x\left(  |D_{\mu}^{\rm b} {\widetilde n}_j |^2  -  |{\tilde n}^*_j D_{\mu}^{\rm b} {\tilde n}_j |^2 \right)  \; ,  \cr 
D_{\mu}^{\rm b} \cdot =  \ \partial_\mu  \;   \cdot  +   i  A_\mu + i \; \delta_{\mu 2} \;   \frac{2\pi}{L} \;  \begin{pmatrix}
\mu_1 &0&\dots &0 \cr
0& \mu_2 & \dots &0\cr
\vdots\cr
0&0&\dots &\mu_N
\end{pmatrix} \; \cdot
\label{action2}
\end{eqnarray} 
 The change in the fermonic action (\ref{fermion}), under similar manipulations, is also equivalent to the replacement 
$D_{\mu} \rightarrow D_{\mu}^{\rm b}$.
This corresponds to the action of the ${\tn }$-particles coupled to a  
``background $U(1)^N$ gauge holonomy" $\Omega_0$.  The gauge connection 
that we refer to here is "the sigma model connection"  that   will be defined in the next section. 
We comment that the twisted boundary conditions introduced here have some relations to, but also some significant differences from, twisted mass terms in sigma models \cite{Hanany:1997vm,Dorey:1998yh,Gorsky:2005ac,Cui:2010si,Bolokhov:2011mp}. A closer, but also not exact, analogy is with 
symmetry breaking terms for studying sphalerons in sigma models \cite{Mottola:1988ff,Snippe:1994fk,Vachaspati:2011ad}.

\subsection{Parametrization of \cpn manifold  and  $\s$- connections}
\label{section-model-c}

The local splitting of the $n$ field into a modulus and phase  $n_i=  e^{i \varphi_i}  |n_i|$ 
 provides a useful parametrization to study the  dynamics of the theory.  Despite the fact that the splitting is local, it will help us to build new line operators,  which are useful in the study of the phases, and to make  connections with the  4d gauge theory. 

{\bf Definiton 1:  Point-wise modulus and phase splitting  } The  point-wise splitting amounts to rewriting the \cpn field in {\it complexified hyperspherical coordinates}, involving $2(N-1)$ angular fields:
\begin{equation} 
  \begin{pmatrix}
n_1 \cr
n_2 \cr
n_3 \cr
\vdots\cr
n_N \cr
\end{pmatrix}    = 
\left( \begin{array}{l}
e^{i \varphi_1}  \cos \frac{\theta_1}{2} \\
e^{i \varphi_2} \sin\frac{\theta_1}{2}\, \cos \frac{\theta_2}{2} \\
e^{i \varphi_3} \sin \frac{\theta_1}{2} \, \sin\frac{\theta_2}{2}\, \cos\frac{\theta_3}{2} \\ 
\vdots \cr
e^{i \varphi_N} \sin \frac{\theta_1}{2} \, \sin \frac{\theta_2}{2} \, \sin \frac{\theta_3}{2} \,\dots \sin \frac{\theta_{N-1}}{2}
\end{array} \right) 
\qquad  \; \theta_i \in [0, \pi],  \qquad   \varphi_i \in [0, 2\pi).
\label{chs}
\end{equation}
The angular fields $\{\theta_1, \ldots, \theta_{N-1}\}$ are independent, and one out of  $\{\varphi_1, \ldots,  \varphi_N\}$ can be gauged away. Hence,  there are $2(N-1)$ microscopic degrees of freedom, as noted in the coset construction (\ref{coset}).  
We set the  gauge choice to  
\begin{align}
\sum_{i=1}^{N} \varphi_i =0  \qquad ({\rm mod}\,\, 2\pi)
\label{constraint}
\end{align} 
 by using the   local $U(1)$ gauge redundancy (\ref{u1}):  
\begin{equation}
|n_i(x_1, x_2)|  \rightarrow |n_i(x_1, x_2)| \qquad \varphi_i(x_1, x_2)  \rightarrow  \varphi_i(x_1, x_2)   +\alpha(x_1, x_2)
\label{gt}
\end{equation}
The modulus is gauge invariant, whereas each phase transforms as a  ``gauge" connection.  This splitting is crucial in the construction of a refined line-operator probing the structure of the theory. 

{\bf Definiton 2:    Sigma-model connection or $\s$-connection}   
The   derivative of each  phase, $-\partial_\mu \varphi_i, i=1, \ldots, N$, transforms under (\ref{u1}) like a gauge connection,   modulo the constraint (\ref{constraint}).  Therefore, and due to reasons that will follow, it is useful to define   $ - \partial_\mu \varphi_i  \equiv {\cal A}_{\mu,i} $ as  the ``sigma-model connection", or ``$\sigma$-connection" for short.    Under (\ref{gt}),  it rotates as 
\begin{equation}
- \partial_\mu \varphi_i  \equiv {\cal A}_{\mu,i}  \longrightarrow - \partial_\mu \varphi_i   +  \partial_\mu 
\alpha \equiv {\cal A}_{\mu,i}  +   \partial_\mu  \alpha
\label{gt2}
\end{equation}
This $N$-component  (minus the constraint) $\s$-connection is crucial to build new line operators. 
 
The relation between various useful representations of the \cpn fields are 
\begin{eqnarray}  
&&n (x_1, x_2)  =  \Omega(x_1, x_2)    {\cal R}( x_1,x_2), \qquad  \tn(x_1, x_2)= \tO(x_1, x_2)    {\cal R}(x_1,x_2)   
\end{eqnarray}
where  ${\cal R} \equiv {\rm Diag}(|n_i|)$, with  $0 \leq   |n_i|   \leq 1$, according to (\ref{chs}) and 
\begin{eqnarray} 
\Omega(x_1, x_2) &=& \tO  (x_1, x_2) T(x_2) \,,   \qquad \qquad   \text{in components} \,,  \cr\cr
  \begin{pmatrix}
e^{i \varphi_1 } &0&\dots &0 \cr
0& e^{i \varphi_2}  & \dots &0 \cr
\vdots\cr
0&0&\dots &e^{i  \varphi_N} 
\end{pmatrix} 
&=  &
 \begin{pmatrix}
e^{i \phi_1 } &0&\dots &0 \cr
0& e^{i \phi_2}  & \dots &0 \cr
\vdots\cr
0&0&\dots &e^{i  \phi_N} 
\end{pmatrix} 
 \begin{pmatrix}
e^{i \frac{2 \pi \mu_1 x_2}{L}} &0&\dots &0 \cr
0& e^{i \frac{2 \pi \mu_2 x_2}{L}} & \dots &0 \cr
\vdots\cr
0&0&\dots &e^{i \frac{2 \pi \mu_N x_2}{L}} 
\end{pmatrix}    
\end{eqnarray}
The $\varphi_i (x_1,x_2)$ field is not periodic, as per  (\ref{twist}), $\varphi_i (x_1,x_2+L) = \varphi_i (x_1, x_2) + 2 \pi \mu_i $, while the $\phi(x_1,x_2)$ fields are periodic, and $T(x_2)$ is a twist matrix depending only on the compactified coordinate $x_2$.

{\bf Line operators:} 
The local (point-wise)  decomposition of fields  can be used to construct line operators, which are useful to probe the phases of the theory in a way which provides {\it more information} than the usual 
Wilson line associated with  the auxiliary gauge field (\ref{gauge}). In particular,  
\begin{equation}
A_{\mu} = - \sum_{i=1}^{N} |n_i|^2  \partial_\mu \phi_i =  \sum_{i=1}^{N} |n_i|^2  {\cal A}_{\mu,i}
\label{gauge22}
\end{equation}
and obeys   (\ref{u1}). 
 On $\R^2$,  one can define an  open Wilson line 
\begin{equation}
W ({\bf a} , {\bf b}) =   e^{i \int_{{\bf a} }^{{\bf b}}  d x_{\mu}  A_{\mu}  } =  e^{ -i  \sum_{i=1}^{N}   |n_i|^2  \int_{{\bf a} }^{{\bf b}}    dx_{\mu} \partial_\mu \phi_i} = \prod_{i=1}^{N} 
 e^{ -i   |n_i|^2  \int_{{\bf a} }^{{\bf b}}    dx_{\mu} \partial_\mu \phi_i}
\end{equation}
which transforms covariantly.    Instead of this  conventional   operator, we define a more refined version, because there is more information in the phases than there  is in their  weighted-sum (\ref{gauge22}). 

{\bf Definition 3:  $\s$-connection holonomy }
We define a line operator associated with the  $\s$-connection holonomy: 
\begin{equation}
 (^L\Omega)_{ij} =   ({^L}\Omega)_i \delta_{ij}, \qquad 
^L\Omega_i({\bf a} , {\bf b}) =  e^{i \int_{{\bf a} }^{{\bf b}}  dx_{\mu}  {\cal A}_{\mu,i}  } 
=  e^{i ( \varphi_i( {\bf a}) - \varphi_i( {\bf b}) )}     
\label{line}
\end{equation}
as an $N\times N$ matrix.  Note that the line operator, because of the way that it is constructed, can be written as a product of two-point operators at the end-points.  This is because the $\sigma$-connection ${\cal A}_{\mu,i}$ is a total derivative. 
 Under a $U(1)$ gauge rotation, the line operator   transform covariantly: 
 \begin{equation}
^L\Omega({\bf a} , {\bf b})  = \Omega({\bf a})  \Omega^{\dagger}({\bf b}) 
  \rightarrow  e^{i  \alpha( {\bf a}) }  \;\;  {^L}\Omega({\bf a} , {\bf b}) \;\; e^{-i \alpha( {\bf b}) }     
\end{equation}
The traced version of this  line operator making a topologically non-trivial loop on the $\mathbb S^1$ circle 
 will play a role parallel to the Wilson line  (or Polyakov loop) in $SU(N)$ gauge theory. 

\subsection{Center symmetry}
\label{sec:center}
When compactified on $\R^1 \times \mathbb S^1$, the  \cpn model   has a global symmetry, called  {\it center symmetry}, which acts non-trivially on certain line operators. In order to see this, recall that the local gauge invariance does not require  the gauge rotations to be strictly periodic.  
Aperiodicity   up to a {\it global} element of the center-group is both permitted and useful: 
\begin{equation} 
e^{i \alpha(x_1, x_2+L) } =  e^{-i \xi}  e^{i \alpha(x_1, x_2) } \qquad, \qquad  e^{-i \xi}  \in U(1) \; . 
\label{global2} 
\end{equation}
A canonical gauge invariant order parameter  on the cylinder \rone$\times$\sone\, is given by the Wilson loop
\begin{eqnarray}
W(x_1)=\exp\left[i \int_0^L A_2(x_1, x_2)\, dx_2\right] 
\label{wilson1}
\end{eqnarray}
In order to  extract more information about the structure of the theory, we introduce a  refined order parameter. 
Consider the $\s$-connection holonomy  (\ref{line})  which makes a circuit around the 
 compact direction, namely
 \begin{eqnarray}
&& (^L\Omega)_j (x_1) = \exp \left[ i \int_{0}^{L}  dx_2 \;  {\cal 
A}_{2,j} \right] = \exp \left[ i (\varphi_j (x_1, 0) -\varphi_j(x_1 , L)) \right]
\cr \cr \cr
&&^L\Omega(x_1) =   \begin{pmatrix}
e^{i [\varphi_1( x_1, 0) - \varphi_1(x_1, L)]} &0&\dots &0 \cr
0& e^{i [\varphi_2(x_1, 0) - \varphi_2(x_1, L)]}  & \dots &0\cr
\vdots\cr
0&0&\dots &e^{i [\varphi_N(x_1,0) - \varphi_N(x_1, L)]}
\end{pmatrix}  
\label{hol}
\end{eqnarray} 
Under an aperiodic global  gauge rotation, $e^{i [\varphi_1( x_1, 0) - \varphi_1(x_1, L)]}  \rightarrow e^{i \xi} 
e^{i [\varphi_1( x_1, 0) - \varphi_1(x_1, L)]} $. The constraint (\ref{constraint}) implies
\begin{align}
\det  \; \left( ^L\Omega \right) =1
\label{constraint2}
\end{align} 
which is same as   $e^{i N \xi} = 1$ or equivalently, 
\begin{align}
e^{i \xi} = e^{i \frac{2 \pi k}{N}},  \qquad k=1, \ldots, N 
\label{gphase}
\end{align} 
Therefore, the center symmetry of the theory is $\Z_N$ and   (\ref{hol}) is its order parameter. Under center-symmetry, it rotates by the  global $\Z_N$- phase   (\ref{gphase}): 
\begin{equation} \Z_N \;:\; 
   ^L\Omega  \longrightarrow    e^{i \frac{2 \pi k}{N}}   \;  \;  ^L \Omega \; .  \end{equation}
This gauge invariant  operator plays the same role as a Polyakov loop or a  Wilson line does in   non-abelian gauge theories.  

Note that the twisted boundary conditions used in (\ref{twist})  and (\ref{twist2}) can be viewed as 
the background field associated with the line operator (\ref{line}).  In quantum field theory,  we would 
associate $2\pi\mu_i \leftrightarrow  \langle[\varphi_i(0) - \varphi_i(L)]  (x_1)\rangle $ with   the vev of the  dynamical field, and we would interpret  $[\phi_i(0) - \phi_i(L)] (x_1)$ as the quantum fluctuations around the vev. In quantum mechanics,  (\ref{twist2})  will be the configuration that {\it minimizes} the one-loop potential,  the background around which we can make a Born-Oppenheimer approximation:
 \be
 ^L\Omega =  \underbrace{ \begin{pmatrix}
e^{2\pi i\mu_1} &0&\dots &0 \cr
0& e^{2\pi i\mu_2}&\dots &0\cr
\vdots\cr
0&0&\dots & e^{2\pi i\mu_N}
\end{pmatrix}  }_{\rm background}
\underbrace{  \begin{pmatrix}
e^{i [\phi_1(0) - \phi_1(L)]} &0&\dots &0 \cr
0& e^{i [\phi_2(0) - \phi_2(L)]}  & \dots &0\cr
\vdots\cr
0&0&\dots &e^{i [\phi_N(0) - \phi_N(L)]}
\end{pmatrix}  }_{\rm fluctuations}
\label{hol2} \qquad  \qquad 
\ee 
and $[\phi_i(0) - \phi_i(L)] (x_1)$ denotes the fluctuation around it.  

This is  an important point, distinguishing a classical analysis from  a  quantum one.  In the quantum theory, the twisted boundary condition (which is equivalent to a background $U(1)^N$ 
$\sigma$-connection)   is actually  not a  choice;  rather it is determined by  dynamics.   In the weak coupling regime, this background is determined  by a Coleman-Weinberg type one-loop analysis. As we show in the next section, not all values of $\mu_i$ are actually stable under quantum fluctuations. 

\section{Perturbative one-loop analysis for $\s$-connection holonomy}
\label{section-oneloop-a}
The realization of center-symmetry in the \cpn model in the small-$\mathbb S^1$ regime can be determined through a one-loop calculation. Because of  asymptotic freedom,  at sufficiently small $\mathbb S^1$ (with respect to the length scale set by the inverse of the strong scale $\Lambda$), the Kaluza-Klein modes of the theory are weakly coupled and can be integrated out perturbatively.  For the general \cpn theory with $N_f$ fermions,  the one-loop effective potential for the gauge invariant line operator $^L\Omega$   can be obtained by standard methods  \cite{Gross:1980br}.  The one-loop  analysis picks out which ``background" appearing in the covariant derivative (\ref{action2})  and (\ref{hol2})  is preferred by thermal or quantum fluctuations.  The use of twisted boundary conditions, and hence a twisted background, in the classical theory  {\it does not} imply the stability of the given background in the quantum theory.

In order to check the stability of the twisted background, under thermal vs. spatial compactification, use
\begin{eqnarray}
\tilde n (x_1, x_2+L)&=&  \tilde n (x_1, x_2), \qquad    \tilde \psi (x_1, x_2+L)=  - \tilde  \psi (x_1, x_2)   \qquad (\rm thermal) \cr
\tilde n (x_1, x_2+L)&=&  \tilde n (x_1, x_2), \qquad    \tilde \psi (x_1, x_2+L)=  + \tilde  \psi (x_1, x_2)   \qquad (\rm spatial)
\label{bc}
\end{eqnarray}
The choice of  anti-periodic (thermal)  vs. periodic (spatial) boundary conditions for fermions in the path integral formalism, correspond, in the operator formalism to studying the theory by using either 
the thermal partition function (response to heating) or twisted (signed)  partition function (response to spatial squeezing). 
\begin{eqnarray}
\label{pf}
&&   Z (\beta)  \equiv \tr [ e^{-\beta H} ]   \equiv Z_{\cal B} 
 + Z_{\cal F}    \qquad \qquad  \;\;\;  (\rm thermal)  \\
&& \widetilde Z (L)  \equiv \tr [ e^{-LH} (-1)^F]   \equiv Z_{\cal B} 
 - Z_{\cal F}   \qquad (\rm spatial) 
\label{tpf}
\end{eqnarray}
where $\beta$ and $L$ are  the size  of the thermal and spatial  \sone,  respectively. 
Here    $(-1)^F$ is fermion number modulo two, grading the Hilbert space  ${\cal H} = {\cal B} \oplus {\cal F}$, where states in the 
bosonic sub-space ${\cal B}$ contribute with a plus sign and states in the fermionic subspace ${\cal F}$ contribute with a minus sign.   In the supersymmetric theory, the $N_f=1$ case, $\widetilde Z (L) $  is the supersymmetric Witten index, and in the non-supersymmetric theory, with  $N_f>1$,  as already stated,   it probes the phase structure as a function of spatial size 
\cite{Unsal:2007vu}. 

The main result of the one-loop analysis of the effective potential  for  the line operator (\ref{hol})  of the $\sigma$-connection holonomy,  is 
   \begin{eqnarray}
&&V_{-} [ {^L}\Omega] =   
\label{tpotential}  \frac{2}{ \pi \beta^2}  \sum_{n=1}^{\infty} \frac{1}{n^2}  (-1+ (-1)^n N_f)  (|\tr \, {^L}\Omega^n |-1)  \qquad \;\;\; (\rm thermal)  \\
&&V_{+} [ {^L}\Omega]  =   (N_f -1) \frac{2}{ \pi L^2}  \sum_{n=1}^{\infty} \frac{1}{n^2}   
 (|\tr \, {^L}\Omega^n |-1)  \qquad \qquad  \qquad  (\rm spatial) 
 \label{spotential}
\end{eqnarray} 
This result is remarkable because it indicates a sharp quantitative difference between  thermal and spatial compactification.\footnote{Despite its simplicity, to our knowledge, this effective potential  is new in \cpn.  The reason is that  the natural variable  in terms of which this potential is expressed,  ${^L}\Omega$,  the line operator (\ref{hol}) associated with the holonomy of the $\sigma$-connection, is also new.}
\footnote{ The form of the  one-loop potentials, $V_{\pm}$,  can be deduced on physical grounds. 
First, it must be  order $N$ in the  large-$N$ limit. Hence,  at leading order,  it must be a sum over  single-trace operators $\tr \, {^L}\Omega^n$ with order one  coefficients.  Second, it must be invariant under center symmetry $\tr\, {^L}\Omega^n \rightarrow e^{i \frac{2 \pi n k}{N}} \tr \, {^L}\Omega^n $. 
Since center-symmetry is a symmetry of the compactified microscopic theory, it must also be a symmetry of the effective potential. One more piece of information is that for $N_f=1$, the theory is  supersymmetric ${\cal N}=(2,2)$ theory,  and the potential $V_{+} [{^L}\Omega]=0 $ to all orders in perturbation theory.   This dictates  $V_{+} [{^L}\Omega]  =   (N_f -1) \frac{1}{  L^2}  \sum_{n=1}^{\infty} a_n |\tr \,{^L}\Omega^n | $ in the spatial case  (\ref{spotential}). } In particular, in the thermal case, the minimum of one-loop potential   (\ref{tpotential}) is 
 \begin{equation}
 {^L}\Omega_0^{\rm thermal}  = e^{i \frac{2 \pi  k}{N}}  \left ( \begin{array}{ccccc}
1 & &&& \cr
  & 1 &&&  \cr
  && \ddots &&  \cr
      &&&&1
  \end{array} \right)  \; ,   \qquad (\rm thermal) 
  \label{minthermal}
\end{equation}
 where $k$ labels the center-position of the lump of eigenvalues. 
  This means that the eigenphases  of the 
 holonomy attract each other. See  Fig.~\ref{fig:holonomy}a.
 We identify this regime  as the deconfined  center-broken phase. 
 
The moral behind spatial circle compactification is the absence of thermal fluctuations, and only the presence of zero temperature quantum fluctuations. A  new phenomenon occurs, with far-reaching consequences.  The minimum of the one-loop potential (\ref{tpotential}) is located at 
 \begin{equation}
  {^L} \Omega_0^{\rm spatial}  =    \left ( \begin{array}{ccccc}
1 & &&& \cr
  & e^{i \frac{2\pi}{N} } &&&  \cr
  && \ddots &&  \cr
      &&&& e^{ i \frac{ 2\pi (N-1)}{N}}
  \end{array} \right) \; ,   \qquad (\rm spatial) 
  \label{minspatial}
\end{equation}
 This means that the eigenphases  of the 
 holonomy repel each other. See  Fig.~\ref{fig:holonomy}b.
   This simple result is   the analog  of  the ``adjoint Higgsing" at weak coupling  of gauge theories,  and is  ultimately responsible for the rest of this paper:     continuity between the small and large-$\mathbb S^1$ regimes in the sense of center-symmetry,  the fractionalization of 2d instantons to 1d kink-instantons, realization of  semi-classical renormalons, and identification of the resurgent structure in the \cpn model.

Note that the analog of this sharp  quantitative difference between the one-loop potentials (\ref{tpotential}) and (\ref{spotential}), based on either  thermal or spatial compactification,  has been obtained  previously  in  4d gauge theories in   \cite{Kovtun:2007py}, and employed in  \cite{Unsal:2007vu, Unsal:2007jx,  
Argyres:2012ka} to build a reliable  semi-classical expansion for non-perturbative effects in gauge theory.  We see that a parallel structure emerges in  \cpn.

Another reason why  the spatially twisted partition function $\widetilde Z (L)$ is more useful  than $ Z (\beta) $ for semi-classical analysis  is that the counter-part of the density matrix, which we may refer to as the ``twisted density matrix", is not positive definite  for $\widetilde Z (L)$. The usual  density matrix that enters the study of   $Z (\beta)$  is  positive definite, and is ultimately responsible  for  Hagedorn instability. If, for example, $\rho(E) =  \rho_{\cal B}(E)  +   \rho_{\cal F} (E)  \sim  e^{\beta^*  E}$, at large-$E$, then    $ Z (\beta) = \int^{\infty} dE \; e^{(\beta^* -\beta) E}$ diverges for $\beta < \beta^*$, indicating an instability  towards the deconfinement phase transitions.  On the other hand, with the  twisted partition function,   $\widetilde \rho(E) = \rho_{\cal B}(E)  -  \rho_{\cal F} (E)$, and even if the theory has exponential growth of states  in both bosonic and fermionic Hilbert spaces,   the twisted partition function $ \widetilde Z (L) = \int^{\infty} dE \; ( \rho_{\cal B}(E)  -  \rho_{\cal F} (E) ) e^{- L E}$ may be tame.  The fact that the center-symmetry is unbroken in the small-\sone\,  regime for periodic compactification can be traced to the non-positivity of the twisted density matrix.
  
\subsection{Thermal compactification: Center instability} 
\label{thermal1}
For the $N_f=0$ case, there is no distinction between the thermal and spatial compactification, and 
the minimum of the one-loop potential is at a center-broken configuration in the small $\mathbb S^1$ regime.  We can verify the correctness 
of the one-loop potential (\ref{tpotential})    by independent means, using basic statistical mechanics. 

The minimum of the one-loop potential (\ref{tpotential})  is at (\ref{minthermal}), $^L\Omega= {\bf 1} $. The value  of the potential at the minimum must give  the leading order  free energy density (or minus the pressure, $P$) of the hot \cpn model:
\begin{equation}
{\cal F} = V_{-} [^L\Omega_0^{\rm thermal}]   = \frac{2}{ \pi \beta^2} (N-1) \sum_{n=1}^{\infty} \frac{1}{n^2}   = -\frac{2}{ \pi \beta^2} (N-1) \zeta(2)  =
 -(2N-2)  \frac{\pi}{6}T^2 
\end{equation}
By comparison, a direct calculation using  a non-interacting gas of bosons gives the same leading order result:
\begin{equation}
{\cal F} =   (2N-2) T \int \frac{d{\bf p}}{2\pi} \;   \log (1- e^{-\beta |{\bf p}|}) =  -(2N-2)  \frac{\pi}{6}T^2  
\end{equation}
This is just the Stefan-Boltzmann result: the number of degrees of freedom $(2N-2)$ times the Stefan-Boltzmann factor  per bosonic quanta, which is  $\frac{\pi}{6}T^2$.

For the thermal theory with fermions, the analysis is similar.  The free energy density can be found either by evaluating the one-loop potential at its minimum $\Omega= {\bf 1} $, or by using statistical physics, 
\begin{equation}
{\cal F} =   (2N-2) T  \left(  \int \frac{d{\bf p}}{2\pi} \;   \log (1- e^{-\beta |{\bf p}|}) - N_f  \int \frac{d{\bf p}}{2\pi} \;   \log (1+  e^{-\beta |{\bf p}|})  \right)
\end{equation} 
Using  $\sum_{n=1}^{\infty}  \frac{(-1)^n}{n^2}  = -\frac{ \zeta(2)}{2}$, the free energy density can be written as 
\begin{equation}
{\cal F} = 
 -(2N-2)  \frac{\pi}{6} \left( 1+ \frac{N_f}{2}\right) T^2 
\end{equation}
The factor of $1/2$ in front of the fermion number arises due to Fermi-Dirac statistics. This is again the expected Stefan-Boltzmann free energy of the system. 

The $O(N^1)$-free energy   implies  that the  high-temperature [i.e., small-$\mathbb S^1_\beta$] regime of the  $\C\P^{\rm N-1}$ model is in the deconfined phase. At large-$\mathbb S^1_\beta$, the theory is expected to be in the confined phase with an  $O(N^0)$  free energy. The transition must take place at a critical (inverse) temperature $\beta_c= a  \Lambda^{-1}$ where $a$ is a pure $O(1)$ number.  The rapid-cross over between these two regimes at finite-$N$ becomes a sharp phase transition at $N=\infty$. 
The phase transition to the deconfined phase is interesting in its own right. However,  since the small and large $\mathbb S^1_\beta$  regimes are different phases, the information gained in the small circle regime does not help to understand the dynamics in the confined large $\mathbb S^1_\beta$ phase. In the large  $\mathbb S^1_\beta$ regime, where the theory is confined, and  a strong coupling center-symmetric $\s$-connection  holonomy (See  Fig.~\ref{fig:holonomy}c.)
 is operative,  one cannot use semi-classical methods due to strong coupling. What is needed is a weak-coupling  semi-classical regime which is continuously connect the the  strong-coupling regime.  This would indeed provide  a semi-classical  way to study the 2d non-perturbative dynamics. 

\subsection{Spatial compactification:  Center stability}
\label{section-oneloop-b}
We re-write the  holonomy dependent part of the   one-loop potential (\ref{spotential}) in the eigenvalue basis for $N_f >1$:
   \begin{eqnarray}
&&\frac{ V_{+} [\Omega]  \pi L^2}  { 2(N_f-1)}
 =    \sum_{n=1}^{\infty} \frac{1}{n^2}   
  \Big| \sum_{j=1}^N e^{i 2\pi n \mu_j } \Big| \, \qquad , \qquad 2 \pi \mu_j \equiv (\phi_j (0) -\phi_j(L)) 
 \label{spotential2}
\end{eqnarray} 
The right hand  side is non-negative definite. Because of the $1/n^2$ in the prefactor, the minimization is achieved by first minimizing  $|\tr\,( ^L\Omega)|$, then $|\tr\, ( ^L \Omega^2)|$, and 
all the way  $|\tr\, ^L\Omega^{\lfloor \frac{N}{2} \rfloor} | $, where  ${\lfloor x \rfloor} $ is the  floor function, the largest integer not greater than $x$.  Going all the way to ${\lfloor \frac{N}{2} \rfloor} $ is  both necessary and sufficient to lift all possible vacuum degeneracies, and one will not gain more by going to higher orders.  This procedure  determines  the global minimum of the potential given in (\ref{minspatial}).

The one-loop potential (\ref{spotential}) has a non-trivial dependence on the number of flavors, 
in close analogy with QCD(adj) with $N_f$ Majorana fermions in 4d \cite{Kovtun:2007py}. 
\begin{itemize} 
\item{ ${\bf N_f>1}$: the one-loop potential generates a repulsive interaction between the eigenvalues of the holonomy $^L\Omega$ as can be read-off easily from (\ref{spotential2})
 and the center symmetry is preserved. Namely, the minimum of the potential is at  (\ref{minspatial}) or 
 \begin{equation}
\tr\, ^L\Omega^n  = 0, \qquad {\rm for} \;  n \neq 0 \;  {\rm mod} (N)
\label{min}
 \end{equation}}
\item { ${\bf N_f=1}$: The one-flavor theory is the supersymmetric ${\cal N}=(2,2)$  $\C\P^{\rm N-1}$  model.  The perturbative potential vanishes at one loop order, and also  to all orders due to supersymmetry. Despite this, we later demonstrate that there is a non-perturbatively induced potential stabilizing the center-symmetry. Thus, in the $N_f=1$ case, (\ref{minspatial}) is also the center-symmetric vacuum of the theory.}
\item { ${\bf N_f=0}$: In the purely bosonic theory, since there is no difference between the spatial and thermal compactification, the center symmetry is always broken at small $\mathbb S^1$. 
Below, we discuss how to go around this obstacle,  either by using deformation or by integrating out heavy fermions.  }
\end{itemize}

\subsection{Deformed-\cpn\, and massive fermions}
\label{deformed}

Consider the theory with $N_f$ fermions with mass $m$. Then, the one-loop potential  
with periodic boundary conditions for fermions  takes the form 
    \begin{equation}
V_+[\Omega] = \frac{2}{\pi L^2} \sum_{n=1}^{\infty} \frac{1}{n^2}  \left[-1 + N_f (nLm)K_1(nLm) \right]
  (|\tr\,  ^L\Omega^n |-1) \; .
  \label{potmass}
\end{equation}
where $K_1(z)$ is the modified Bessel function. In the heavy and light fermion asymptotes, 
$  K_1 (z) \approx \sqrt \frac {\pi} {2z}  e^{-z}, z \rightarrow   \infty$, and  $  K_1 (z) \approx \frac {1}{z},  z \rightarrow   0 $, so that   (\ref{potmass}) reduces to (\ref{spotential}) with $N_f=0$   and 
 (\ref{spotential}) with $N_f\geq 1$ respectively. 

Provided $ mLN \lesssim 1$, the center symmetry will be stable.  This condition can be achieved at fixed-$N$ and fixed-$L$,  if $m$ is taken to be sufficiently small.  This can  be viewed as a small mass perturbation of the theory with massless fermions. 

A far more interesting (and at first glance, counter-intuitive)  case arises as follows. 
Take  $m \gg \Lambda$, so  large  that it is practically decoupled from the dynamics on $\R^2$.
This theory on $\R^2$ emulates the pure bosonic theory, and when compactified, indeed, has a center-breaking phase transition at $L_c= a \Lambda^{-1}$, as in pure bosonic  \cpn model. 
However, if one makes  $L$ sufficiently small, eventually  one will achieve $ mLN \lesssim 1$,
 so that the center  symmetry will stabilize. In other words, provided the hierarchy of scales, 
\begin{equation}
\Lambda \ll m \lesssim  \frac{1}{LN}
\end{equation}
$m$ is heavy with respect to $\Lambda$, but light with respect to $1/LN$.  The corresponding 
 small-$\mathbb S^1$ theory is a bosonic, center symmetric theory.   Therefore, we can also 
 consider the dynamics of bosonic center-symmetric theory at small-$\mathbb S^1$.
 
Inspired by the  massive-fermion induced stabilization, and deformed Yang-Mills theory \cite{Unsal:2008ch,Myers:2007vc} compactified on $\R^3 \times \mathbb S^1$,  we can construct  a {\it deformed} \cpn  model (or d\cpn).   The deformed model at small-$L$, should be viewed as the analytic continuation of the confined phase of large-$L$ undeformed $\C\P^{\rm N-1}$. The action of the deformed model is
\begin{eqnarray}
S^{\rm d\C\P} &&= S + \Delta S , \qquad 
\Delta S=  \frac{2}{ \pi L^2} \sum_{n=1}^{\lfloor N/2 \rfloor } \frac{a_n}{n^2}    (|\tr\, ^L\Omega^n |-1),  
\label{defCP}
\end{eqnarray} 
 where $a_n= O(N^0)$ are sufficiently large pure numbers, and $\lfloor x \rfloor$ is the integer part of $x$.  For example, one may take $a_n=2$, twice in modulus the value of the perturbative  one-loop result. The deformation respects all the symmetries of the original theory. In the small-$\mathbb S^1$ domain, it guarantees unbroken center-symmetry and semi-classical calculability. 
 
\subsection{Is  large-$N$ volume independence possible in a $\s$-model?}
\label{section-oneloop-c}
A class of non-abelian gauge theories, for example an $SU(N)$  theory, when studied on toroidal compactification of $\R^d$,  has properties independent of the compactification radius, provided  center-symmetry and translation symmetry are not spontaneously broken. This property is called large-$N$ volume independence or Eguchi-Kawai reduction \cite{Eguchi:1982nm, Kovtun:2007py}. When center-symmetry is unbroken,  then, in the reduced theory, the space-time  or momenta are encoded into the gauge structure of the non-abelian theory.  Vector models are,  to our knowledge,  not discussed in the volume independence context, because they do not possess a non-abelian gauge structure.  
 In the \cpn model, for example, the gauge structure is $U(1)$, and to an expert in volume independence, even talking about the possibility of having volume independence may sound absurd.  However, we suggest here that   the  \cpn models  should also exhibit volume independence, provided,  
 {\it (i)} center-symmetry is unbroken  for the toroidally  compactified theory, 
  {\it (ii)} translational invariance is not spontaneously broken for the theory on $\R^2$. 
 We expect the latter to hold in a Lorentz invariant theory.  In this work, we have showed that  \cpn model with multiple-fermions endowed with periodic boundary conditions do not break their  center-symmetries.  

In this subsection, we give an intuitive plausibility argument for volume independence in 
\cpn.   Since there is no non-abelian gauge structure in \cpn, there must be a new mechanism 
to generate space-time or momenta out of the reduced model. Indeed, there is, and this requires 
an interpretation of the vev that we obtained for the line operator (\ref{hol}). The key physical idea is that the vev of the order parameter (\ref{hol}), provided center-symmetry is unbroken,  generates  {\it momentum feeding}  into the system in units of   $\frac{2\pi  \mu_k}{L}=\frac{2\pi\, k}{N\, L}$, with $k\in {\mathbb Z}$,  which behave like  fractional momenta with respect to the "standard" Kaluza-Klein momenta  $\frac{2\pi k}{L}$, with $k\in {\mathbb Z}$.   
\begin{figure}[htbp]
\begin{center}
\scalebox{1.2}{
\rotatebox{-90}{
\includegraphics[width=5cm]{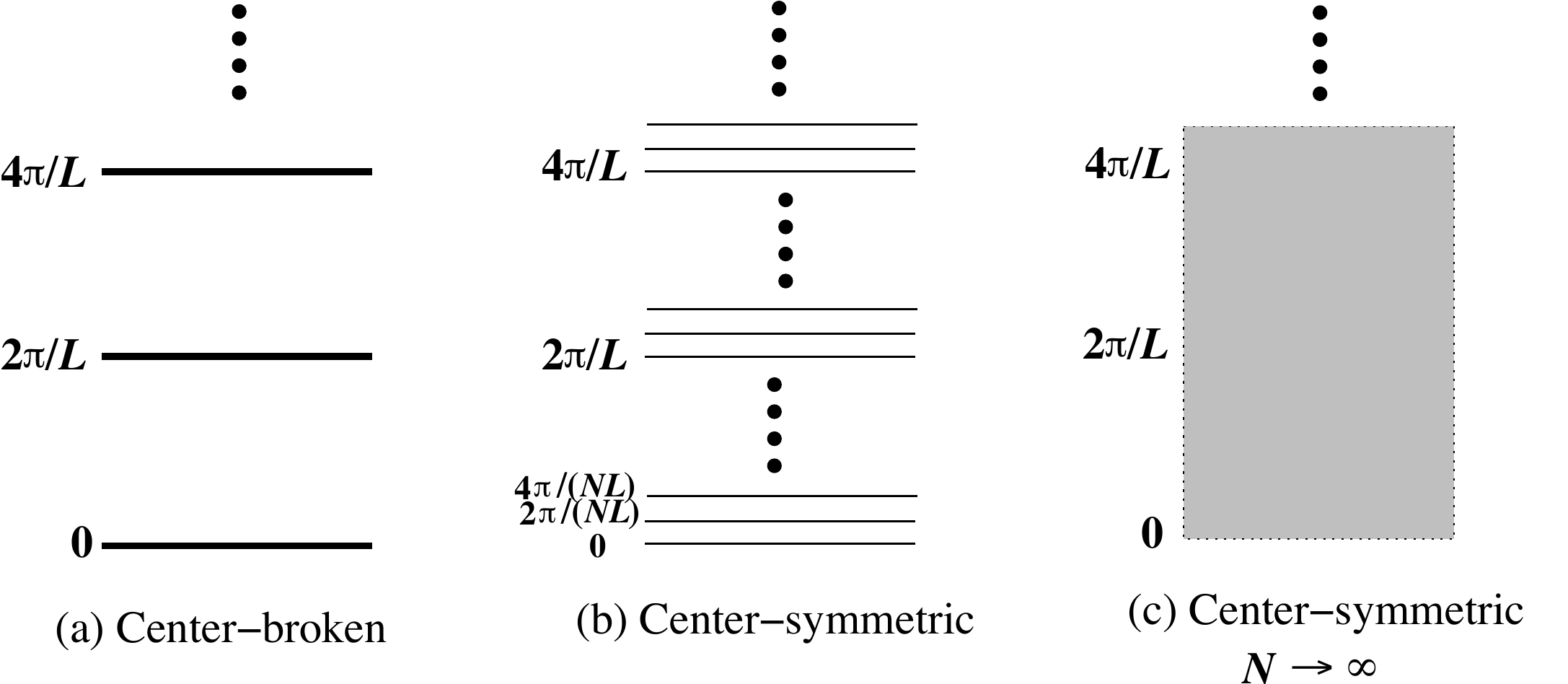}}}
\end{center}
\caption{
The perturbative   spectrum of the $\C\P^{\rm N-1}$ theory as a
function of  the background $\s$-connection holonomy.
(a) Weak-coupling trivial holonomy  (as well as classical theory)
gives the usual $2\pi/L$ level spacing.
 (b) Weak-coupling non-trivial holonomy   ($\Z_N$ symmetric
background)  at finite-$N$ produces a
  finer level spacing. (c)  $\Z_N$ symmetric background at $N=\infty$
leads to a continuous spectrum.   Classically, the  background for the
$\s$-connection holonomy is equivalent to
  twisted boundary conditions  on \cpn fields.
  For $N_f \geq 1$ theories, quantum mechanically, (b) and (c) are
stable upon spatial compactification and
  (a) is stable upon thermal compactification.  To achieve (b) and (c) in
the $N_f=0$ case, we deform the \cpn Lagrangian appropriately. (b)
admits a semi-classical analysis of the confined regime at finite $N$,
and (c) satisfies large-$N$ volume independence at $N=\infty$. (a) is
not suitable for the semi-classical study of the confined
regime/phase.
}\label{fig:states}
\end{figure}

As described in  Section \ref{section-model-b}, the spatial twist in the boundary condition can be removed in favor of a   background  $\s$-connection holonomy.  The form of the Kaluza-Klein   spectrum is crucially  dependent on the choice of twist matrix $^L\Omega$ or equivalently, the background field in the modified action  (\ref{action2}). In the center-broken  case (\ref{minthermal}),  $^L\Omega = 1$, Fig.~\ref{fig:states} a depicts the   KK tower with momenta  and spacing at integer multiples of $2\pi/L$. Each level has an  $O(N)$ degeneracy. This is the analog of the center-symmetry broken regime in gauge theories. In this case, the critical length scale that enters  the problem is $L$.  At length scales larger than $L$,  the non-zero frequency modes can be integrated out perturbatively.  
 
In contrast, when a $\Z_N$ symmetric  twist matrix $^L\Omega$, as in (\ref{minspatial}), is stable,  we find a much finer  KK spectrum with spacing of $2\pi/(NL)$ and $O(1)$ degeneracies. Expanding the periodic ${\tilde n}$-fields into their KK-modes along the compact direction, 
  \begin{equation}
{\tilde n}_j (x_1, x_2) =  \sum_{ k \in \Z} e^{\frac{i 2\pi k  x_2}{L}}  {\tilde n }_{j,k} (x_1) 
   \end{equation}
the quadratic terms in the  bosonic action take the form   
   \begin{equation}
S^{\rm quad} = \frac{2L}{g^2}  \sum_{k \in \Z} \int dx_1 \left|  \left(\partial_1  + i  \frac{2\pi}{L}   (\mu_j +k)  \delta_{\mu 2} \right) {\tilde n}_{j,k} (x_1)   \right|^2
\label{modified}
\end{equation} 
Therefore, a stable $\Z_N$-symmetric  background act likes momentum quantized in units of 
$\frac{2\pi}{LN}$. Alternatively, the stable twisted boundary conditions  shift the the phase acquired by a  excitation propagating around the spatial $S^1$ by the amount $\frac{2\pi j}{LN}$ for the mode $n_{j,k}$.  

The main observation is  that  the $SU(N)$ index of the \cpn field  and the  ordinary Kaluza-Klein momentum index intertwine, and the   $\C\P^{\rm N-1}$ field  $n_{j,k}$  breaks up to into  $N$ distinct pieces with shifted offsets in the frequency quantization.  The critical length scale that enters the problem is $NL$,  and  not $L$.   
The instanton and kink-instanton  effects discussed in the next section, like the perturbative effects, are sensitive to length scale $LN$, as opposed to $L$.

It is important to note that  in the  $N \to \infty$ limit with fixed $L$, the  perturbative spectrum  of the  $\C\P^{\rm N-1}$ model with $\Z_N$ symmetric   twist  approaches the continuous frequency spectrum of the decompactified theory theory on $\R^2$, as illustrated in Fig.~\ref{fig:states}c. This is also a property of gauge theories which satisfy volume independence in the   $N = \infty$ limit. In this limit, 
below the energy level $\Lambda$, perturbatively, there is a continuum band of states. It seems reasonable to expect that, in analogy with gauge theory, the  neutral sector observables  in the $\C\P^{\rm N-1}$  model should exhibit volume independence. In this limit, the effects due to compactification are $1/N$ suppressed, and the leading large-$N$ behavior of observables, including the non-perturbative mass spectrum of the theory, must be volume independent. 
In this regime,  the description in terms of microscopic degrees of freedom is strongly coupled, this is the absence of weak coupling description of long distance physics in \cpn. However, the macroscopic ``hadrons" of the theory should have a weakly coupled description, with couplings controlled by $\frac{1}{\sqrt N}$. 

Clearly, the relevant length scales in the problem and the approach to   the $N = \infty$ limit are critically dependent on the  twist-matrix, $\Omega$.  To recap, for $\C\P^{\rm N-1}$ on  $\R \times \mathbb S^1$, with a  $\Z_N$ symmetric   twist matrix $^L\Omega$, the physically relevant length scale appearing in finite volume effects is not $L$, but rather $NL$. The theory has two distinct characteristic regimes:
\begin{eqnarray}
    \frac{NL\Lambda}{2 \pi} \ll 1\,, & \qquad
    \mbox{semi-classical   $\Longrightarrow$ volume dependence;}
\\
      \frac{NL\Lambda}{2 \pi}  \gg  1\,, & \qquad
    \mbox{strongly coupled   $\Longrightarrow$ volume independence.}
\label{eq:ana}
\end{eqnarray}
The techniques of this paper allow us to study the semi-classical limit at arbitrary $N$, including the large-$N$ limit by analytical methods. 
For $N_f=0$, we will derive the mass gap in the semi-classical domain.  The result is an exact match to the well-known large-$N$ result on $\R^2$. 
Furthermore, we will demonstrate explicitly that the positions of the expected renormalon singularities on $\R^2$ for $N_f \geq 1$ are the same as the positions of the bion--anti-bion singularities 
in the semi-classical domain. 
These results, together with other features to be discussed in a future publication, 
suggest that all qualitative aspects of the volume independence domain is captured in the semi-classical domain. The results for the mass gap and renormalon pole positions exhibit also quantitative agreement between the large-$N$ results on $\R^2$ and  the semi-classical results in the compactified theory, furthermore providing the microscopic mechanisms underlying the large-$N$ results.


\section{Self-dual configurations}
\label{sec:selfdual}

The first part of this  section reviews the standard text-book  instanton construction for \cpn on $\R^2$ \cite{zak}, and highlights the non-existence of a dilute 2d instanton gas approximation. Next, on top of  our perturbative one-loop analysis on $\R \times \mathbb S^1$,   we perform a study of leading semi-classical configurations on $\R \times \mathbb S^1$. 

\subsection{2d instantons in \cpn}
\label{section-np-a}
The 2d instanton equations can be obtained by a standard Bogomolnyi factorization of the action density:
\begin{eqnarray}
\left(D_\mu n\right)^\dagger D_\mu n
=\left |\left(D_\mu\pm i \epsilon_{\mu\nu} D_\nu \right)n \right |^2\mp i \, \epsilon_{\mu\nu}\partial_\mu\left(n^\dagger \partial_\nu n\right)
\label{factor}
\end{eqnarray}
Thus, the self-dual instanton equations are
\begin{eqnarray}
D_\mu n=\mp i\, \epsilon_{\mu\nu}D_\nu n
\label{instanton}
\end{eqnarray}
For these instanton solutions, the action saturates the BPS bound:
\begin{equation}
S = \frac{2}{g^2}  \int \left(D_\mu n\right)^\dagger D_\mu n =  \frac{2}{g^2}  \left|  \mp i   \epsilon_{\mu \nu} 
\int \left(D_\mu n\right)^\dagger   D_\nu n \right|  
\geq \frac{4\pi}{g^2} |Q| 
\end{equation}
where $Q$ is the topological charge defined in (\ref{charge}). In describing instantons, it is convenient to define homogeneous coordinates for the \cpn fields,  
\begin{eqnarray}
n=\frac{v}{|v|}\quad \Rightarrow \quad A_\mu=\frac{i}{2}\left(\frac{v^\dagger\partial_\mu v-\partial_\mu v^\dagger v}{v^\dagger v}\right)
\label{homog}
\end{eqnarray}
where $v$ is an $N$-component column vector; then the instanton equations (\ref{instanton}) reduce to the Cauchy-Riemann equations $\partial_\mu v=\mp i\, \epsilon_{\mu\nu}\partial_\nu v$, which means that $v$ is holomorphic (instanton) or anti-holomorphic (anti-instanton):
\begin{eqnarray}
{\rm instanton:}\, v=v(z)\quad, \quad {\rm anti-instanton:}\, v=v(\bar z)
\label{cr}
\end{eqnarray}
Then for instanton and anti-instanton solutions we can write
\begin{eqnarray}
A_\mu=\pm \frac{1}{2}\epsilon_{\mu\nu}\partial_\nu \ln v^\dagger v
\label{holo}
\end{eqnarray}

On \rtwo,  the most general instanton with charge $k\in {\mathbb N}$ is expressed in terms of a holomorphic vector $v$ having entries that are polynomials in $z$, with maximal degree $k$, and no common roots. For example, in \cpone on \rtwo, the single instanton can be written 
\begin{eqnarray}
v=\begin{pmatrix}
1\cr (z-b)/a 
\end{pmatrix}
\quad\Rightarrow\quad
Q=\frac{1}{\pi}\int d^2 x\, \frac{|a|^2}{(|a|^2 +|z-b|^2)^2}=1
\label{r2-single}
\end{eqnarray}
There are two complex (four real)  moduli parameters entering  this solution.  
The parameter  $b$ characterizes the location of  instanton  in \rtwo,  $\rho = |a|$ encodes the size  modulus, and $\arg(a)$  is a $U(1)$ phase of the instanton.  More generally, in  the $\C\P^{\rm N-1}$  model on $\R^2$,  the 2d instanton  has $2N$ parameters, associated with $2N$ zero modes. These are  associated with the classical symmetries of the self-duality equation:  two  are the position of the instanton $({\bf a}_I \in \R^2)$ and arise due to translation invariance, one is the size modulus $(\rho \in \R^+)$ and is associated with invariance under dilatations, and the remaining $2N-3$ are internal orientational modes.
 \begin{eqnarray}
2N  &  \longrightarrow \ 
2+1+(2N -3)  =  ({\bf a}_I \in \R^2) + (\rho \in \R^+) 
+ ( {\rm orientation}) .  
\end{eqnarray}
 This is in  analogy with Yang-Mills  instantons on $\R^4$, where the count is $4N= 4+1+ (4N-5)$, with parallel physical interpretation.

For the \cpn theory on $\R^2$,   the existence of the size modulus $\rho$ implies that the instanton comes in all sizes  at no cost in action.  Therefore,  there is no precise sense in which 
the typical instanton separation  is  much larger  than the typical instanton size.  This prevents, on $\R^2$,  a meaningful  semi-classical  dilute instanton gas from first principles. On \rone $\times$ \sone, we will propose a way around this obstacle while staying continuously connected to $\R^2$.  
 
\subsection{Fundamental and affine kink-instantons in  \cpone\; } 
\label{kinkins}
${\bf CP^1}:$
First, we discuss the kink-instanton events in \cpone, and then give the generalization to \cpn. The
$\mathbb C \mathbb P^{\rm 1}$ model is locally equivalent to the $O(3)$  non-linear $\sigma$-model through the  identification of fields:   
\begin{equation}
\vec s(x) = n_i^\dagger \vec \sigma_{ij} n_j\qquad , 	\qquad   s^a(x) = n_i^\dagger  \sigma_{ij}^a n_j
\end{equation}
where   $\vec \sigma$ are the Pauli matrices. 
We would like to benefit from this equivalence while incorporating the  $\Z_2$-symmetric background (\ref{minspatial}).  It is crucial that this background must be  stable in the quantum theory, and this, as explained in Sections \ref{section-oneloop-b} and \ref{deformed}, can be achieved in the spatially compactified theory,  either by massless fermion-induced mechanism or by integrating out massive-fermions, i.e, the deformed bosonic theory. 

Let us, as before, trade the twisted boundary conditions in favor of a {\it stable} background field:  
Using  (\ref{aperiodic}) for $\C\P^{\rm 1}$,   we have  
\begin{eqnarray}
 \left ( \begin{array}{c}
n_1 \cr
 n_2
  \end{array} \right) =&&  \left ( \begin{array}{cc}
e^{i \frac{2\pi  \mu_1 x_2}{L}}   &  \cr
 &  e^{i \frac{2\pi  \mu_2 x_2}{L}} 
  \end{array} \right)  \left ( \begin{array}{c}
  \tn_1 \cr
 \tn_2
  \end{array} \right)   
 = \left ( \begin{array}{cc} e^{i \frac{2\pi  \mu_1 x_2}{L}}   &  \cr &  e^{i \frac{2\pi  \mu_2 x_2}{L}}   \end{array} \right)  
   \left ( \begin{array}{c} e^{-i \phi/2 } \cos \frac{\theta}{2} \cr 
 e^{+i \phi/2 } \sin \frac{\theta}{2}  \end{array} \right)    \cr
 =&& 
  \left ( \begin{array}{c}
 e^{i  \left( -\frac{\phi}{2} +  \frac{2\pi  \mu_1 x_2}{L} \right)}    \cos \frac{\theta}{2} \cr
 e^{i  \left( +\frac{\phi}{2} +  \frac{2\pi  \mu_2 x_2}{L} \right)}    \sin \frac{\theta}{2} 
  \end{array} \right),  \qquad  \theta\ \in  [0, \pi], \;\; \phi \in [0, 2\pi] 
  \label{par1}
\end{eqnarray}
where $\theta(x_1, x_2)$  and $\phi (x_1, x_2)$ are periodic fields of $x_2$. Now, the $s_a$ fields  take the form
\begin{eqnarray}
 \left ( \begin{array}{l}
s_1 \cr
 s_2 \cr
 s_3
  \end{array} \right)  = \left ( \begin{array}{l}
 \sin \theta \cos \left(\phi +  \xi x_2 \right)   \cr
 \sin \theta \sin \left(\phi +  \xi x_2 \right)
    \cr \cos \theta  
  \end{array} \right)  \, , 
\quad   \qquad \xi \equiv  \frac{2\pi}{L}   (\mu_2 -\mu_1) 
      \label{par2}
\end{eqnarray}
The Lagrangian obtained  in this manner, on ${\R \times \mathbb S^1}$, is given by  
\begin{equation}
 S= \frac{2}{g^2} \int_{\R \times \mathbb S^1_L}  |D_\mu n_i|^2  =
 \frac{1}{2 g^2} \int_{\R \times \mathbb S^1_L}  |\partial_\mu \vec s|^2  = 
\frac{1}{2 g^2} \int_{\R \times  \mathbb S^1_L}  (\partial_\mu \theta)^2 + 
\sin^2 \theta   (\partial_\mu \phi + \xi \delta_{\mu2})^2 
\label{action3}
\end{equation} 
which is identical to (\ref{action2}). As before,  the   twisted boundary conditions are undone in favor of a  twisted-background  $ \partial_2 \phi_{\rm background} = \xi$.  

Since the fields $\theta, \phi$ entering (\ref{action2}) are  manifestly periodic, we can reduce this Lagrangian to simple quantum mechanics by truncating it to its zero Kaluza-Klein mode. (We  drop the higher Kaluza-Klein modes momentarily.) The action of the associated zero mode quantum mechanics is (this does not capture all the interesting effects, which we will restore momentarily by keeping the relevant KK-modes)  
\begin{equation}
S^{\rm  zero}=   \frac{L}{2 g^2} \int_{\R}  (\partial_t \theta)^2 + 
\sin^2 \theta   (\partial_t \phi)^2  + \xi^2   \sin^2 \theta   
\label{zeroth}
\end{equation} 
The equations of motions associated with this action are\footnote{Setting $\xi=0$, this theory  reduces   to the quantum mechanics of a  particle on a sphere $S^2$, and described the 
  high-temperature deconfined regime  of the thermally compactified theory at scales larger than $T^{-1}$.  
   The zero mode of  the  \cpone model, 
   \begin{align}
(\phi, \theta)(\tau) :  \R/\Z \rightarrow X \qquad  {\rm where} \;\;    \R/ \Z =S^1_\beta, \;\; X= S^2 \;.
\end{align}
    is just the quantum mechanics of a particle on $X= S^2$, and more generally,  $X= $\cpn. 
Resurgence in this  type of  quantum mechanical system is  recently examined by Kontsevich  from path integral point of view \cite{Kontsevich}. However, the  study of resurgence in this quantum mechanics  is not the relevant one for the purpose of understanding the 2d QFT in its semi-classical regime. The 
quantum mechanical theory relevant  to \cpn on $\R^2$ is the one in which $\xi$ is nonzero.  
\label{fn:k}}:
\begin{eqnarray}
&&\ddot \theta - \frac{1}{2}\sin 2 \theta [  (\dot \phi)^2 + \xi^2 ] =0 \cr
&&
\ddot \phi  + 2 \dot \phi  \cot \theta =0 
\label{Eeom}
\end{eqnarray} 
Setting  $\phi$=constant upon which the second equation is satisfied,  the first one reduces 
to the usual equation for a kink in a  one-dimensional problem:
\begin{eqnarray}
&&\ddot \theta - \frac{\xi^2}{2} \sin 2 \theta  =0 
\label{kinksecond}
\end{eqnarray} 
We can find the action of this configuration by using Bogomolny's method: Let $V(\theta)=({\cal W}')^2$ where ${\cal W} = \xi \cos \theta$. Then
\begin{equation}
S^{\rm  zero}=   \frac{L}{2 g^2} \int_{\R}  (\dot \theta)^2 + ({\cal W}')^2   
= \frac{L}{2 g^2} \int_{\R}  \left[ (\dot \theta \pm {\cal W}')^2  \mp 2  \dot \theta {\cal W}' \right]
  \geq \left|  \frac{L}{ g^2}  \int d {\cal W} \right|
\end{equation}
The  kink-instanton, which we refer to as  $\K_1$,  ($\K_j$ in the general case),   satisfies the first order differential equations $\dot \theta \pm {\cal W}' =0$,  or  $\dot \theta \pm \xi \sin \theta=0$.  
Differentiating this equation once with respect to Euclidean time, we recover (\ref{kinksecond}).  Using  $\int d {\cal W} = 2  \xi $ for the kink interpolating from $\theta=0$ to $\theta=\pi$, 
we find the  action of the $\K_1$ kink-instantanton as 
\begin{equation}
\K_1 : \qquad S_1=    \frac{L}{g^2}  (2  \xi)=   
 \frac{4\pi}{g^2} \times (\mu_2-\mu_1)  \equiv S_{\rm I}  \times  (\mu_2-\mu_1)  = \frac{S_I}{2}
 \label{kinkac}
\end{equation} 
In the last step, we used the actual value of the background $\mu_i$. The action of the kink is {\it half} that of the 2d instanton.   The fact that the action of the kink is  determined by the separation between the eigenvalues of the $^L\Omega$  matrix  (\ref{minspatial})  holds more generally.  The kink configuration is an interpolation between Euclidean times
$x_1= \mp  \infty$
\begin{equation}
\K_1: \qquad  \left ( \begin{array}{c}
\tn_1 \cr
 \tn_2
  \end{array} \right) (-\infty)   =  \left ( \begin{array}{c}
 1\cr
0
  \end{array} \right) \qquad  \left ( \begin{array}{c}
\tn_1 \cr
 \tn_2
  \end{array} \right) (+\infty)   =  \left ( \begin{array}{c}
 0\cr
1
  \end{array} \right) 
\end{equation}
and we denote its anti-kink by $\bar \K_1$.

The kink-instanton has two-zero modes associated with the two-global symmetries of the quantum mechanics:
\begin{itemize}
\item {The position modulus $a \in \R$  
associated with translational invariance along $x_1$ direction, }
\item{Angular modulus $\phi \in U(1)$ associated with the  shift symmetry 
$\phi \rightarrow  \phi +c$ of the action (\ref{zeroth}).  Note that  the one-loop potential (\ref{potmass}) in terms of the holonomy  $\Omega$, given in (\ref{hol}), also respects the shift symmetry, as it depends on $\phi_i$  as $\phi_i (x_1, 0) - \phi_i (x_1, L) $, always as in a difference equation, despite the fact that it is not strictly derivatively coupled.} 
\end{itemize}
 The angular modulus  is reflected in the classical Euclidean equations of motions   (\ref{Eeom})  as the choice $\phi=$constant.

{\bf Affine kink-instanton:} Ordinarily, when one performs  dimensional reduction of  a QFT, the Kaluza-Klein modes which carry momenta in compact direction by an amount $\frac{2\pi}{L}$ decouples, as it takes a divergent amount of energy to excite the modes associated with it. 
However, and although not appreciated broadly, when certain conditions are satisfied, such as unbroken center-symmetry as in gauge theory or \cpn\,, this argument is invalid.    This is 
discussed in Section \ref{section-oneloop-c}. Fig.\ref{fig:states}b and Fig.\ref{fig:states}c are manifestations of this fact.  In particular, the lightest mode which may be instrumental in the low-energy dynamics may be hidden in the first KK-mode (not the zeroth KK-mode).

In finding the kink-instanton solution in (\ref{action3}), we reduced  the periodic fields 
 $\tn_1$ and $\tn_2$   to their  zero momentum mode sector, and this leads to (\ref{zeroth}) and the ensuing kink-instanton solution with charge $Q= \mu_2 - \mu_1$.  Instead, now, use 
\begin{eqnarray}
 \left ( \begin{array}{c}
n_1 \cr
 n_2
  \end{array} \right) = 
    \left ( \begin{array}{l}
 e^{i  \left( -\frac{\phi}{2} +  \frac{2\pi  \mu_1 x_2}{L} \right)}    \cos \frac{\theta}{2} \cr
 e^{i  \left( +\frac{\phi}{2} +  \frac{2\pi  (-1 +\mu_2)  x_2}{L} \right)}    \sin \frac{\theta}{2} 
  \end{array} \right), 
    \label{par3}
\end{eqnarray}
which carries a single extra  unit of KK-momentum in the $x_2$ direction. Now, substitute this into  
the \cpone\, action in (\ref{action3}). Then, dimensionally reduce the corresponding Lagrangian down to QM,  by declaring the fields  $\theta, \phi$ entering  (\ref{par3}) to be  independent of the compact spatial $x_2$ coordinate. The resulting action is, 
\begin{equation}
S^{\rm  first}=   \frac{L}{2 g^2} \int_{\R}  (\partial_t \theta)^2 + 
\sin^2 \theta   (\partial_t \phi)^2  + ( \xi')^2   \sin^2 \theta,     
\label{first}  \qquad  \xi' =  \frac{ 2 \pi [-1+ (\mu_2 - \mu_1)]} {L}
\end{equation} 
The kink (not anti-kink) solution of   (\ref{first}) interpolates from $\pi$ to $0$, 
\begin{equation}
\K_2: \qquad  \left ( \begin{array}{c}
\tn_1 \cr
 \tn_2
  \end{array} \right) (-\infty)   =  \left ( \begin{array}{c}
 0\cr
1
  \end{array} \right) \qquad ,\qquad  \left ( \begin{array}{c}
\tn_1 \cr
 \tn_2
  \end{array} \right) (+\infty)   =  \left ( \begin{array}{c}
 1\cr
0
  \end{array} \right) 
\end{equation}
 and has topological charge  and action 
\begin{equation}
Q= 1-(\mu_2-\mu_1) = \frac{1}{2}, \qquad 
S_2=    \frac{L}{g^2}  |2  \xi'|=   
 \frac{4\pi}{g^2} \times (1- (\mu_2-\mu_1))  =  \frac{S_I}{2}
  \label{kinkac2}
\end{equation} 
It is important to note that  $\K_2$ is not the same as $\bar \K_1$. In particular,  $\K_2$ is not an anti-kink, as it has positive topological charge. 

The simplest way to see these differences is to artificially  deviate  $\mu_1 - \mu_2 $ from $\frac{1}{2}$. Then, the action of $\K_1$ and $\bar \K_1$ remain the same, and  the actions of $\K_2$ and $\bar \K_2$ are the same, but these are not equal to each other. At $\mu_1 - \mu_2 = \frac{1}{2}$
the two types of kink configurations is distinguished by their topological charges, or equivalently, by  the  $\Theta$-dependence of the associated amplitudes. These will be discussed in more detail when we discuss the role of the  $\K_i$-events   in dynamics. 

To summarize, the leading  self-dual kink configurations  in \cpone are:
\begin{eqnarray}
\K_1:  \qquad [\theta: 0 \rightarrow \pi]\quad , \;\;  Q= \textstyle{\frac{1}{2}},  \qquad \qquad  \qquad 
\bar \K_1:  \qquad [\theta: \pi \rightarrow 0]\quad , \;\;  Q= \textstyle{-\frac{1}{2}} \cr 
\K_2:  \qquad [\theta: \pi \rightarrow 0]\quad , \;\;  Q= \textstyle{\frac{1}{2}},  \qquad \qquad  \qquad 
\bar \K_2:  \qquad [\theta: 0 \rightarrow \pi]\quad , \;\;  Q= \textstyle{-\frac{1}{2}} 
\end{eqnarray}
  There actually exists an infinite  tower  of both types of kinks, with higher topological charge, and identical asymptotes.  This tower may be useful to establish  a duality between the particle/soliton  topological defects on  $\R^{1,1}$ and  kink-instanton topological defects on 
  $\R \times \mathbb S^1_L$ along the lines of  \cite{Poppitz:2011wy}. This direction will be pursued separately.

\subsection{Embedding   \cpone\;  kink-instantons into  \cpn}
\label{subsec:ki-cpn}

In the presence of a quantum mechanically stable twisted background or equivalently, twisted boundary conditions in the compact spatial   $x_2$ direction, we find that a sufficiently large 2d instanton with $Q=1$   in \cpn decomposes into up to $N $ constituent kink-instantons, each of which carry $Q=\frac{1}{N}$.   More precisely, 
\begin{eqnarray}
&& \rho  <  LN  \ll \Lambda^{-1}  \qquad \text {small-instanton, no fractionalization}  \cr
&& LN  \lesssim \rho  \ll \Lambda^{-1}   \qquad \text {large-instanton, fractionalization to kinks}
\label{factorization}
\end{eqnarray}
The observation regarding  fractionalization of 2d instantons in \cpn   recently appeared in interesting 
works \cite{Bruckmann:2007zh, Brendel:2009mp, Harland:2009mf} for the thermally compactified theory, and  part of our current analysis has been inspired  by these results. However, as noted earlier, 
our work differs from  \cite{Bruckmann:2007zh, Brendel:2009mp, Harland:2009mf} in the sense that we consider a zero temperature  spatial compactification, and  consequently, the  $\Z_N$-symmetric background is stable against quantum fluctuations, 
as opposed to the thermal case  in which the   $\Z_N$-symmetric background is unstable against thermal fluctuations. 
 
Kink-instantons in \cpn can be constructed by embedding \cpone\, kink-instantons into \cpn. 
The fundamental  and affine kinks are characterized by the  simple and affine roots of $SU(N)$ algebra. This mimics the construction of $SU(N)$ BPS monopole-instantons on  $\R^3 \times \mathbb S^1$ from fundamental monopoles by the embedding of $SU(2)$ monopoles, again characterized by roots of $SU(N)$, including the affine root. 

For \cpn, in order to describe the kink-instantons, it is  convenient  to  use complexified 
 hyper-spherical coordinates (\ref{chs}), with $ \phi_i \in [0, 2\pi)$ and  $\theta_i \in [0, \pi]$:
\begin{eqnarray} 
\tn=&& \begin{pmatrix}
e^{i \phi_1} &0& 0&\dots &0 \cr
0& e^{i \phi_2 }  & 0&\dots &0 \cr
0& 0& e^{i \phi_3 }   \dots &0 \cr
\vdots\cr
0&0& 0&\dots &e^{i \phi_N}
\end{pmatrix}  
\left( \begin{array}{l}
  \cos \frac{\theta_1}{2} \\
 \sin\frac{\theta_1}{2}\, \cos \frac{\theta_2}{2} \\
\sin \frac{\theta_1}{2} \, \sin\frac{\theta_2}{2}\, \cos\frac{\theta_3}{2} \\ 
\vdots \cr
\sin \frac{\theta_1}{2} \, \sin \frac{\theta_2}{2} \, \sin \frac{\theta_3}{2} \,\dots \sin \frac{\theta_{N-1}}{2}
\end{array} \right) 
\cr
n=&&  \begin{pmatrix}
e^{i \frac{2 \pi \mu_1 x_2}{L}} &0&\dots &0 \cr
0& e^{i \frac{2 \pi \mu_2 x_2}{L}} & \dots &0 \cr
\vdots\cr
0&0&\dots &e^{i \frac{2 \pi \mu_N x_2}{L}} 
\end{pmatrix}    \tn
\end{eqnarray}
This representation makes it easy to see that there is 
a kink-configuration for each simple root of the $SU(N)$ algebra. 

{\bf  Embedding ansatz:}
Perform the following truncation  which reveals the existence of kink-instantons:
\begin{equation}  
\theta_1=  \ldots =\theta_{k-1}=\pi,   \;\;\;    \theta_k = \theta_k(x_1), \;\;\;
 \theta_{k+1}= 0,  \;\; \; \{\theta_{k+2},  \ldots,  \theta_{ N-1} \} \rightarrow  {\rm arbitrary} 
 \label{embed1} 
 \end{equation} 
Substituting this into the action (\ref{action2}),   
\begin{equation}
\label{kth}
S =  \textstyle \frac{L}{2 g^2} \int_{\R}  (\partial_t \theta_k)^2 + 
\sin^2 \theta_k   [\partial_t (\phi_{k+1} - \phi_{k}) ]^2  +  \left(  \frac{ 2 \pi (\mu_{k+1} -\mu_{k}) }{L}  \right)^2  
\sin^2  \theta_k     
\end{equation} 
which can be identified with (\ref{zeroth}) with simple  matching, $\theta_k \equiv  \theta, \; (\phi_{k+1} - \phi_{k}) \equiv \phi,  \;  \frac{2 \pi (\mu_{k+1} -\mu_{k})}{L} \equiv  \xi_k$.
 This is the embedding of the  \cpone\, kink into   \cpn, in analogy with the embedding of the $SU(2)$ monopole-instanton into $SU(N)$. 

The asymptotic values of the $\tn$ configurations are
\begin{eqnarray}
\label{embed3}
\K_k  : \qquad 
\tn (-\infty) =   \left ( \begin{array}{c}
 0\cr  
\vdots \cr
1 \cr
0 \cr
\vdots \cr
0
  \end{array} \right) =
  e_{k} \; , \qquad  \tn (\infty) =  \left ( \begin{array}{c}
 0\cr  
\vdots \cr
0 \cr
1 \cr
\vdots \cr
0
  \end{array} \right) = e_{k+1} \; ,  \qquad  e_{N+1} \equiv e_1,  \;\;\;  k=1, \ldots, N \qquad 
  \end{eqnarray}
The differences of the asymptotes  are  associated with the simple  and the affine co-root of the $SU(N)$ algebra  and can be used to uniquely label a kink event:
\begin{eqnarray}
\K_k  : \qquad  \Delta \tn  =\tn (\infty)  -\tn (-\infty) =   \left ( \begin{array}{c}
 0\cr  
\vdots \cr
-1 \cr
+1 \cr
\vdots \cr
0
  \end{array} \right)
=
 e_{k+1} - e_{k}  \equiv \alpha_k\qquad ,  \qquad k=1, \ldots, N
\end{eqnarray}
The action of the  kink-instanton $\K_k$  can easily be obtained by using (\ref{kinkac}) and (\ref{kth})
\begin{equation}
\K_k  : \qquad  S_{ k}=    \frac{L}{g^2}  (2  \xi)=   
 \frac{4\pi}{g^2} \times (\mu_{k+1}-\mu_k)   = \frac{S_I}{N}\qquad, \qquad   k=1, \ldots, N
 \label{kinkac22}
\end{equation} 
where in the last-step, we used the center-symmetric  background (\ref{minspatial}).   The crucial point here is that the action of the kink-instanton is $1/N$ of the action of the 2d-instanton. This will play a major role in the determination of the mass gap and the renormalon structure in the semi-classical domain.

\subsection{Small thermal circle:  non-fractionalization of 2d instantons}
\label{section-np-a2}

\begin{figure}[htb]
\includegraphics[scale=0.36]{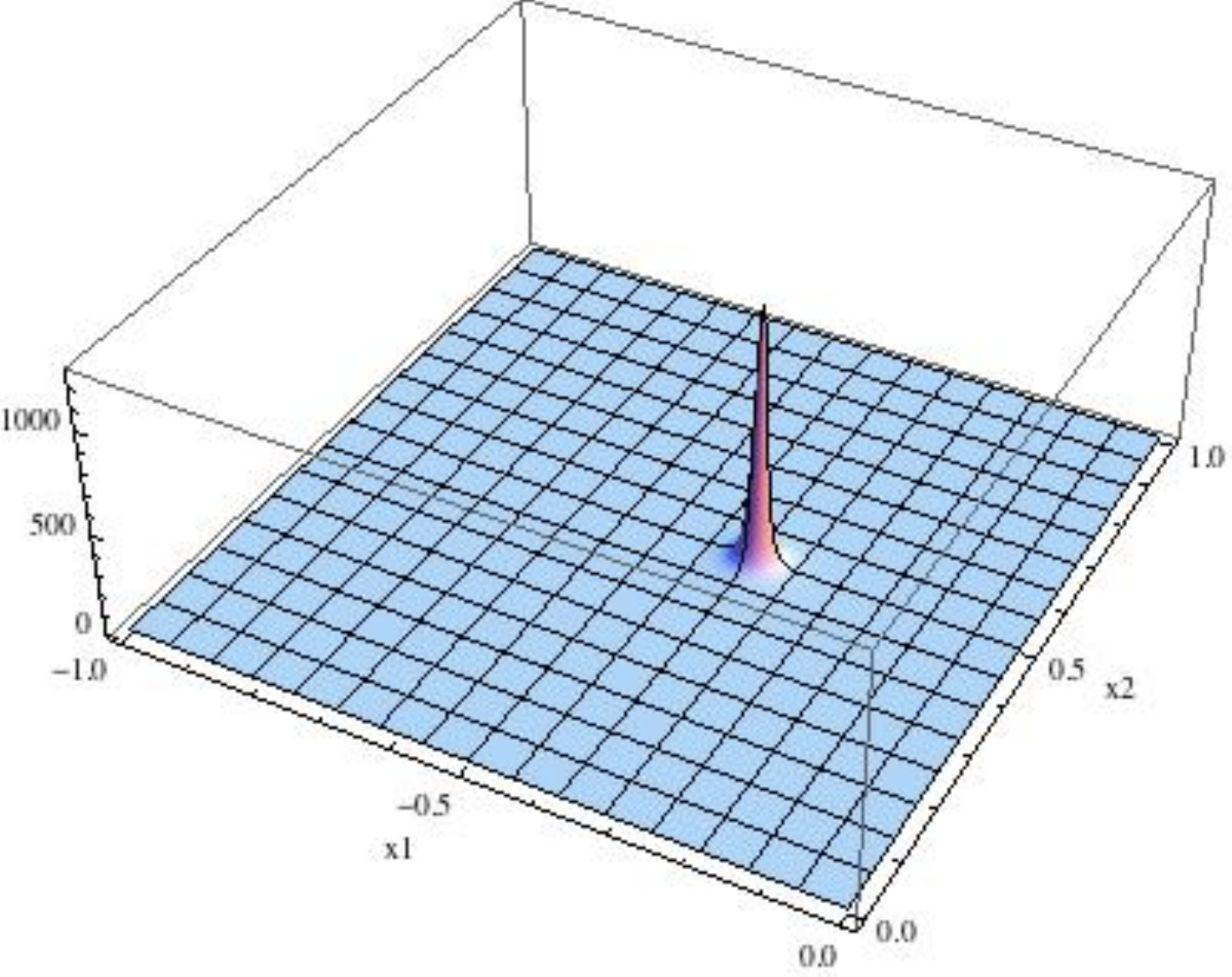}
\includegraphics[scale=0.36]{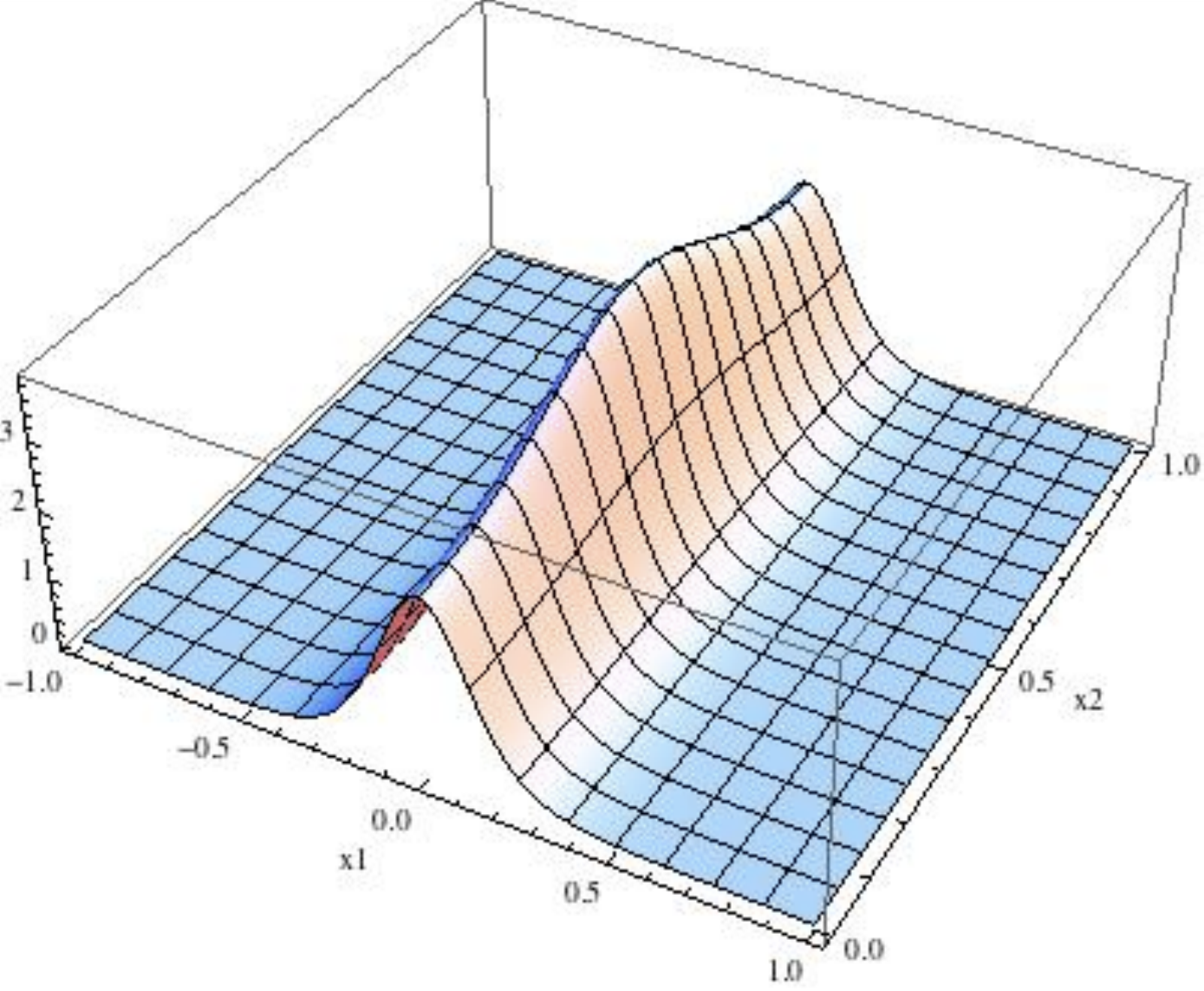}
\includegraphics[scale=0.36]{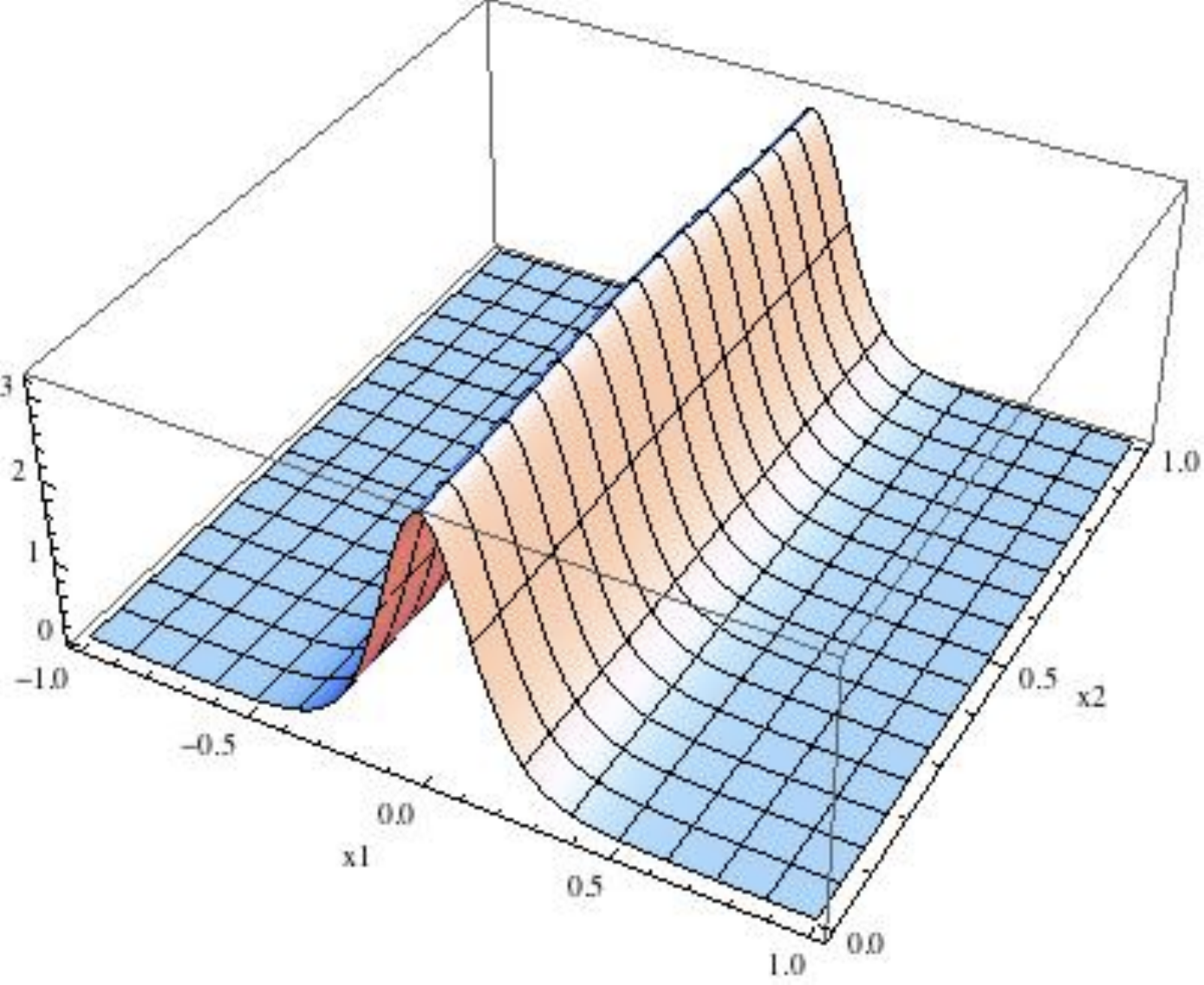}
\caption{The topological charge density of the single  untwisted  \cpone\,  instanton 
(\ref{untwisted1})  on \rone$\times$\sone at fixed $\lambda_2=1$ and $L=1$. The plots are for 
 $\lambda_1=10$, $\lambda_1=1/10$, and $\lambda_1=1/100$, respectively.  The fractionalization of the instanton does not occur for an untwisted background.}
\label{fig1}
\end{figure}

As discussed in the Introduction in Section \ref{subsec:goodbad}, thermal compactification does not permit one to smoothly connect weak-coupling semi-classical results to the strong-coupling phase.
To further clarify this point, directly in the context of the \cpn model, consider for example the simplest, untwisted, $Q=1$ instanton for \cpone, where the second component of  the homogeneous \cpn coordinate $v$  (\ref{homog}) is a first order polynomial in $e^{-\frac{2\pi}{L} z}$ (the formulas here are for spatial compactification, which we apply later, but the formulas are  not important for the point we wish to make here):
\begin{eqnarray}
&&v=\begin{pmatrix}
1\cr
\lambda_1+\lambda_2\, e^{-\frac{2\pi}{L}z} \cr
\end{pmatrix} 
\cr
&&
 v^\dagger v=1+|\lambda_1|^2+|\lambda_2|^2 e^{-\frac{4\pi}{L} x_1}+2|\lambda_1 \lambda_2| e^{-\frac{2\pi}{L} x_1}\cos\left(\frac{2\pi}{L}x_2-{\rm arg} \lambda_1+{\rm arg}\,\lambda_2\right)
\label{untwisted1}
\end{eqnarray}
The corresponding topological charge density $q(x_1, x_2)=\frac{1}{2\pi}\epsilon_{\mu\nu}\partial_\mu A_\nu$ is plotted in Figure \ref{fig1} for various values of  the $\lambda$ parameters. For large $|\lambda_1|$ the instanton looks like a single instanton in ${\mathbb R}^2$, while for small $|\lambda_1|$ it looks like the topological charge for a kink in the non-compact $x_1$ direction.
In fact, when $\lambda_1=0$, $v^\dagger v$ is independent of the compact coordinate $x_2$, so $A_1=0$, and  $A_2$ has a characteristic kink profile:
\begin{eqnarray}
A_2&=&\frac{1}{2}\partial_1 \ln v^\dagger v =\frac{\pi}{L}\left[\tanh\left(\frac{2\pi}{L}\left(x_1-\frac{L}{2\pi}\ln |\lambda_2|\right)\right)+1\right]\\
q&=&\frac{\pi}{L^2} \, {\rm sech}^2\left(\frac{2\pi}{L}\left(x_1-\frac{L}{2\pi}\ln |\lambda_2|\right)\right) 
\end{eqnarray}
This change in the form of the single instanton occurs because the compact  direction sets a maximal size for an instanton, and so a small  instanton looks like an instanton on ${\mathbb R}^2$, while a large instanton   looks like a 1d kink. However, as noted above, in quantum theory, this effect should only  be trusted when  $\beta < \Lambda^{-1}$, as emphasized in (\ref{scales-th1}).  In the low-temperature confined phase, we must use a center-symmetric holonomy, as opposed to  (\ref{minthermal}). However, in strong coupling, the eigenvalues of (\ref{hol}) are strongly fluctuating and a weak coupling analysis is not applicable.   

On the other hand,  we have already shown that  a  ${\mathbb Z}_N$-symmetric twist is stable in  a  compact spatial direction, at small--$\mathbb S^1_L$.    The difference between the weak-coupling center-symmetric regime and the strong coupling center-symmetric regime is analogous to the weak coupling  adjoint Higgs regime  vs. the strongly coupled ``unbroken" regime of gauge theories on $\R^3 \times \mathbb S^1_L$ \cite{Unsal:2008ch}. In other words, the role of the global part of the $SU(N)$ gauge symmetry in  gauge theory on $\R^3 \times \mathbb S^1$   is played by the  $SU(N)$-global  symmetry in the \cpn model. In particular,  in the quantum theory at large-$\beta$ [low temperature] one cannot just use the weak coupling center-symmetric configuration (\ref{minspatial}) to reveal the kink-constituents of a 2d  instanton. Doing so naively,  would result in a fractionalization of instantons at a scale $\beta > \Lambda^{-1}$, however, there is no clear interpretation of what an instanton with size modulus $ \rho \sim  \beta > \Lambda^{-1}$ actually means. No such semi-classical configuration actually exists. 

\subsection{Small spatial circle: fractionalization of 2d instantons}
\label{section-np-b}

We have already shown in Section \ref{kinkins} that the theory in a  $\Z_N$ symmetric background at weak coupling has $N$-types of elementary kink configuration with action $S_0= \frac{S_I}{N}$. 
In this section, we re-derive  this result in an alternative way. We show that  a  2d  instanton decomposes into $N$ kink-instantons  in the presence of a stable $\Z_N$ symmetric spatial
twist.  

Returning to the  \cpone\, example, we now incorporate a {\it spatial}  twist by multiplying  the second component of the homogeneous coordinate (\ref{homog}) $v_2$ by a factor $e^{\frac{2\pi}{L}\,\mu_2\,z}$, where $\mu_2=1/2$, 
\begin{eqnarray}
v_{\rm twisted}=\begin{pmatrix}
1\cr
\left(\lambda_1+\lambda_2\, e^{-\frac{2\pi}{L}z}\right)\, e^{\frac{2\pi}{L}\,\mu_2\,z} \cr
\end{pmatrix}
\label{tw1}
\end{eqnarray}
To satisfy the twisted boundary condition (\ref{twist}) it would be enough to take a factor $e^{\frac{2\pi i}{L}\, \mu_2\, x_2}$, but for an instanton $v$ must be holomorphic, so we need to take $e^{\frac{2\pi}{L}\, \mu_2\, z}$, which therefore prescribes also a certain dependence on the non-compact direction $x_1$. This is the essence of how the twisted spatial boundary conditions affect the structure of instantons on \rone$\times$\sone.

\begin{figure}[htb]
\includegraphics[scale=0.36]{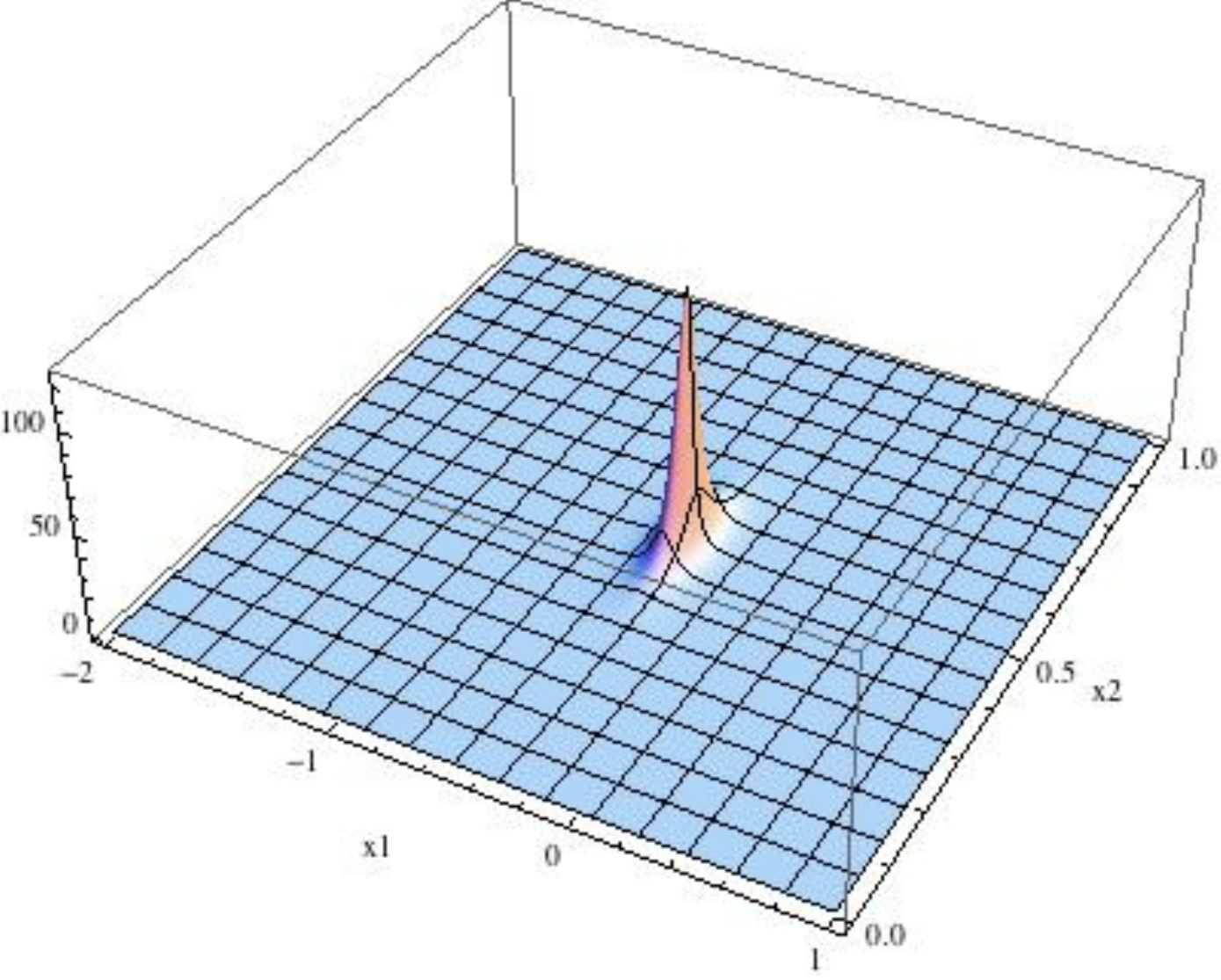}
\includegraphics[scale=0.36]{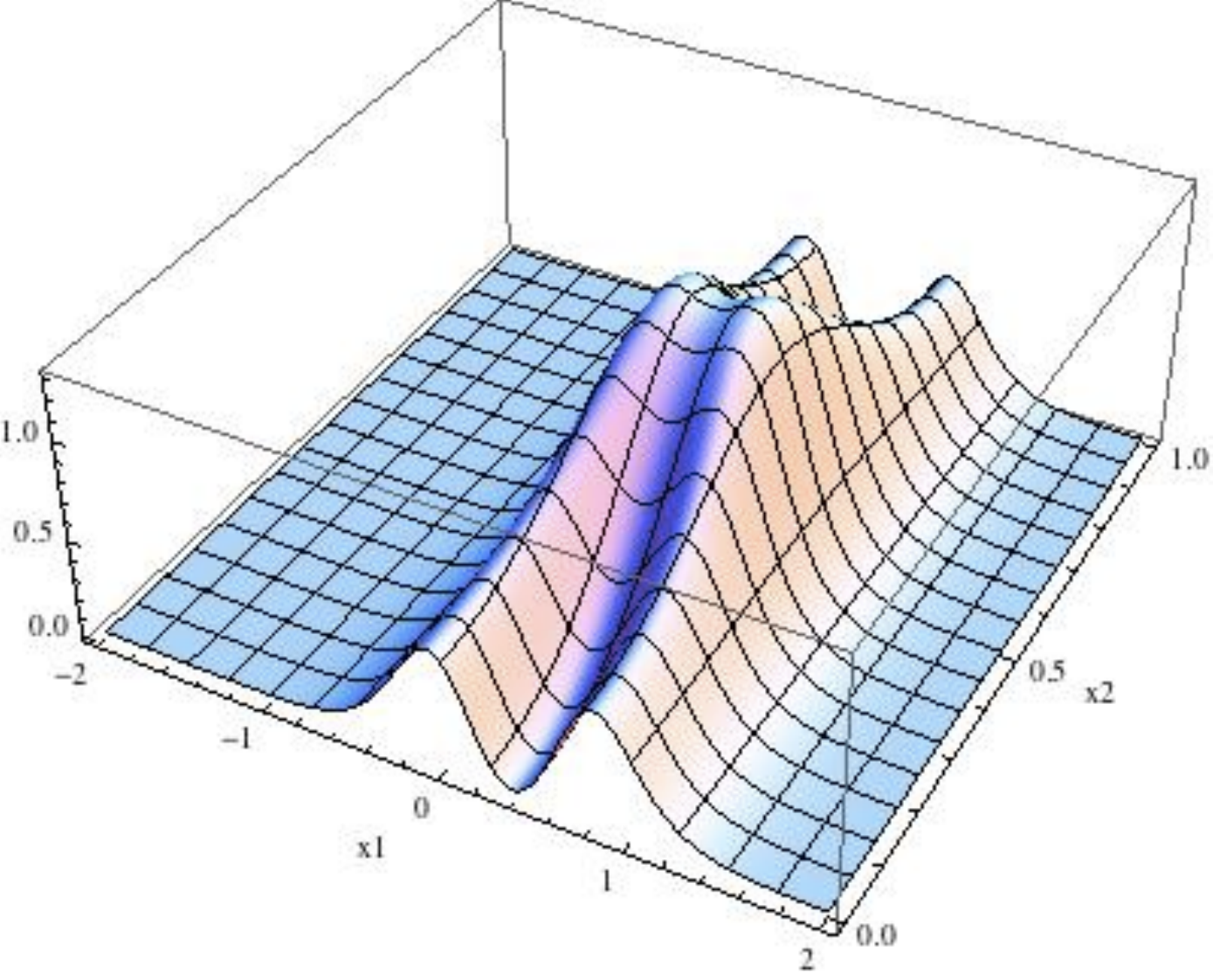}
\includegraphics[scale=0.36]{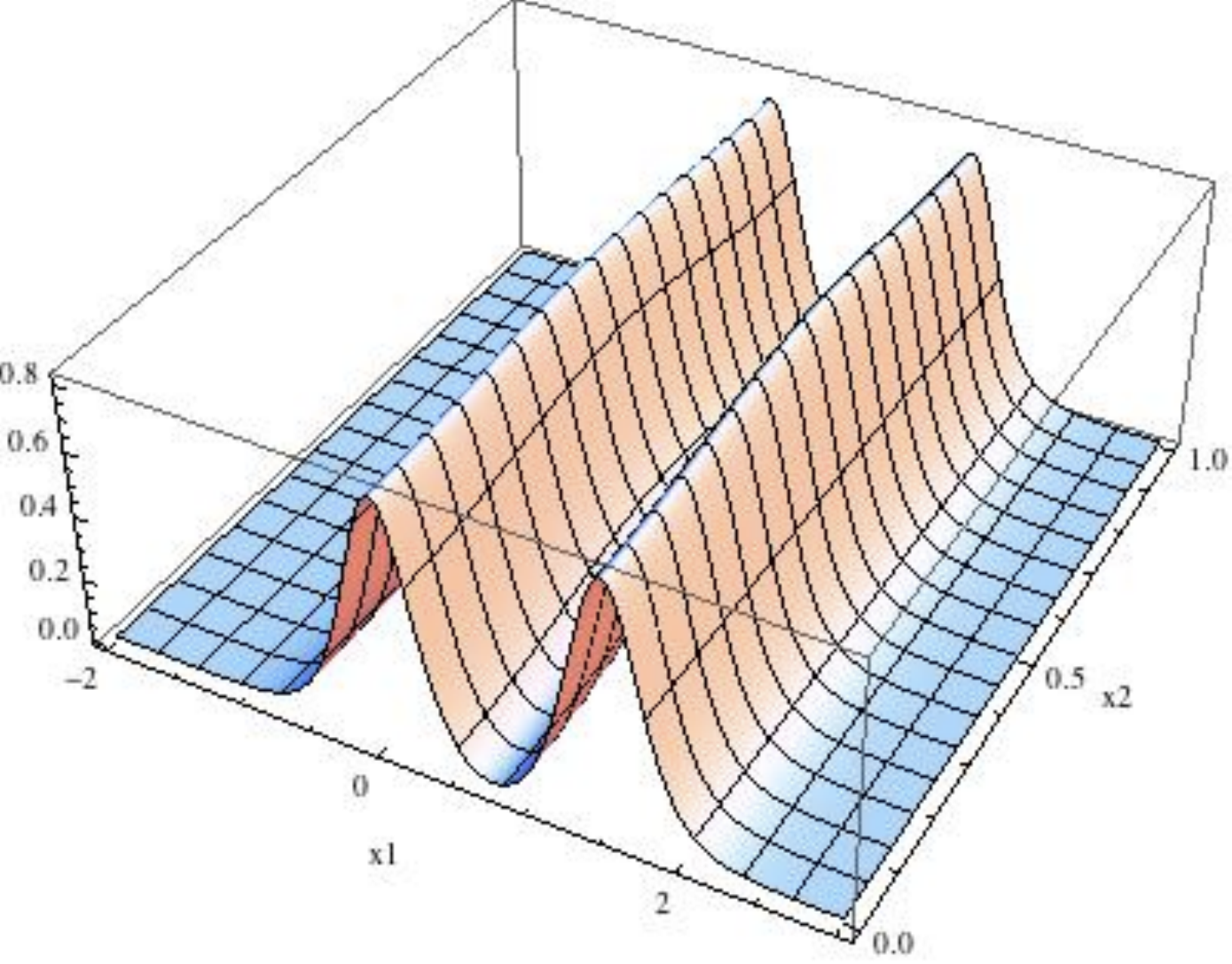}
\caption{ Small and large $Q=1$ instantons in  \cpone in a weak coupling center-symmetric background.   Large instantons split into two $Q=\frac{1}{2}$ instantons. 
 }
\label{fig4}
\includegraphics[scale=0.60]{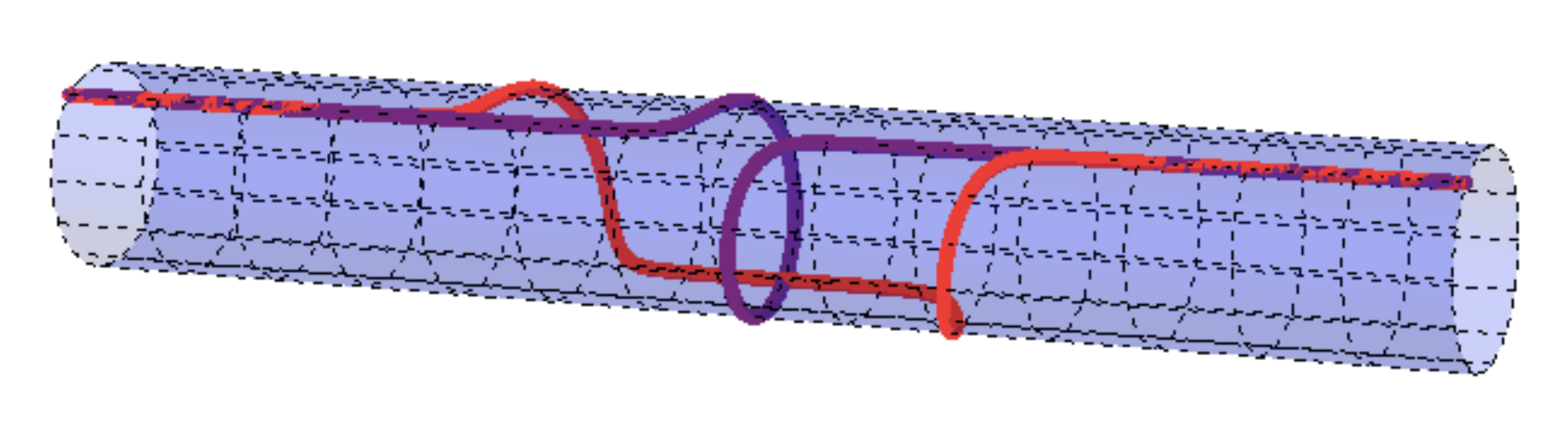}
\caption{
The Wilson loop for a small $Q=1$ instanton  is shown in purple. The large instanton splits into two separate kink-instantons. Each wraps half-way around the cylinder.
}
\label{fig5}
\end{figure}

The twisted instanton (\ref{tw1}) has charge $Q=1$, but at long distances it splits into two distinct kink-instantons, each of charge $1/2$. In general, for \cpn\,  a charge $Q=1$ decomposes into $N$ distinct kink-instantons, each of topological charge $1/N$. To see how this works for the twisted \cpone\, instanton in (\ref{tw1}), note that 
\begin{eqnarray}
v_{\rm twisted}^\dagger v_{\rm twisted}=1+|\lambda_1|^2 e^{\frac{2\pi}{L} x_1}+|\lambda_2|^2 e^{-\frac{2\pi}{L} x_1}+2|\lambda_1 \lambda_2| \cos\left(\frac{2\pi}{L}x_2-{\rm arg} \lambda_1+{\rm arg}\,\lambda_2\right) \qquad 
\end{eqnarray} 
$A_1$ is manifestly periodic in $x_2$, so it does not contribute to the topological charge. On the other hand,
\begin{eqnarray}
A_2=\frac{1}{2}\partial_1 \ln v^\dagger v=\frac{\pi}{L}\frac{|\lambda_1|^2 e^{\frac{2\pi}{L} x_1}-|\lambda_2|^2 e^{-\frac{2\pi}{L} x_1}}{1+|\lambda_1|^2 e^{\frac{2\pi}{L} x_1}+|\lambda_2|^2 e^{-\frac{2\pi}{L} x_1}+2|\lambda_1 \lambda_2| \cos\left(\frac{2\pi}{L}x_2-{\rm arg} \lambda_1+{\rm arg}\,\lambda_2\right)} \qquad 
\end{eqnarray}
Thus, $A_2\to \pm \frac{\pi}{L}$ as $x_1\to \pm \infty$, and so $Q=1$. However, inspection of the form of $A_2$ shows that $A_2$ behaves like two separate kinks, each of charge $1/2$, one located at $x_1\approx -\frac{L}{\pi}\ln \lambda_1$, and the other at $x_1\approx \frac{L}{\pi}\ln \lambda_2$.
The corresponding topological charge densities are plotted in Figure \ref{fig4}.

\begin{figure}[htb]
\includegraphics[scale=0.38]{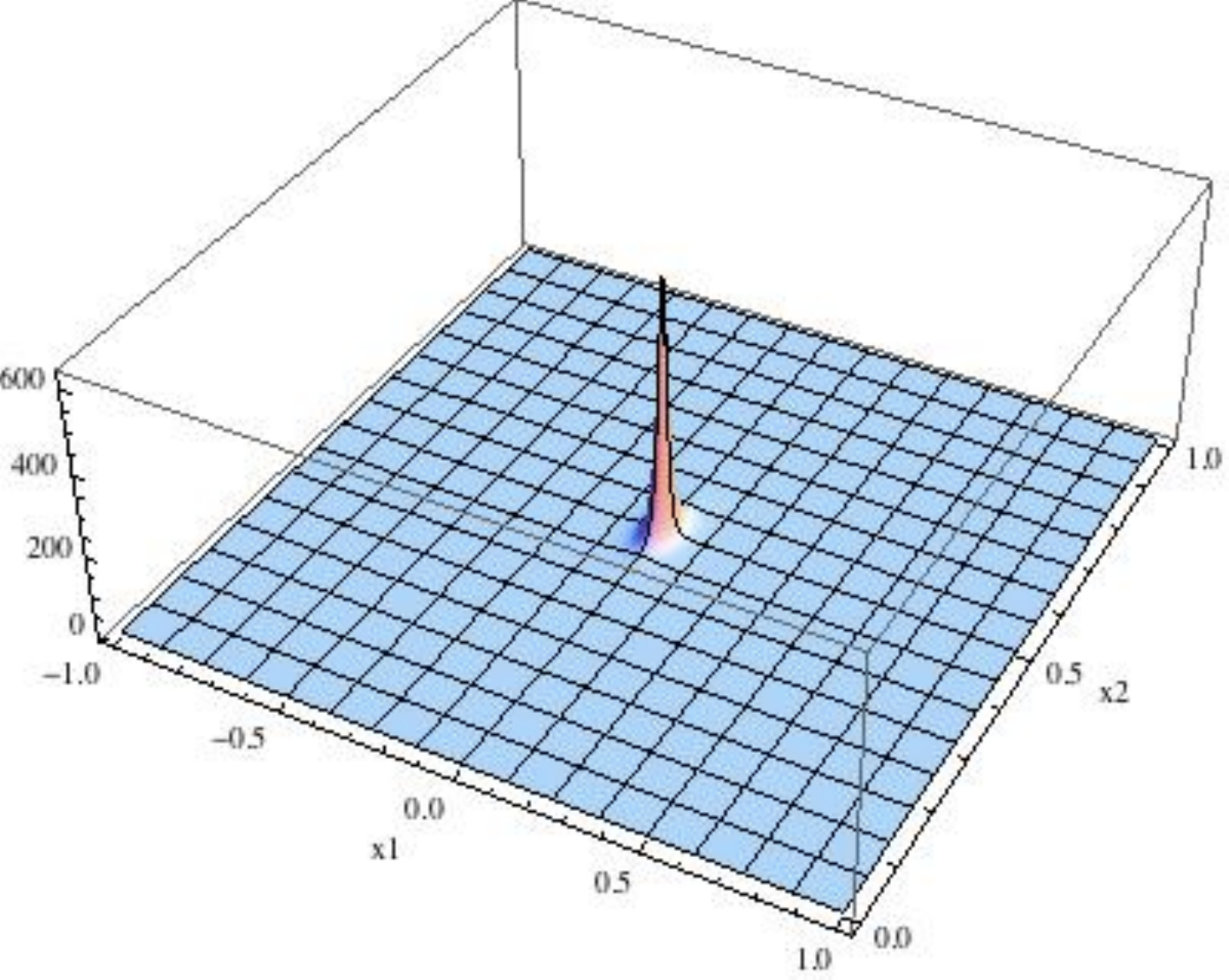}
\includegraphics[scale=0.38]{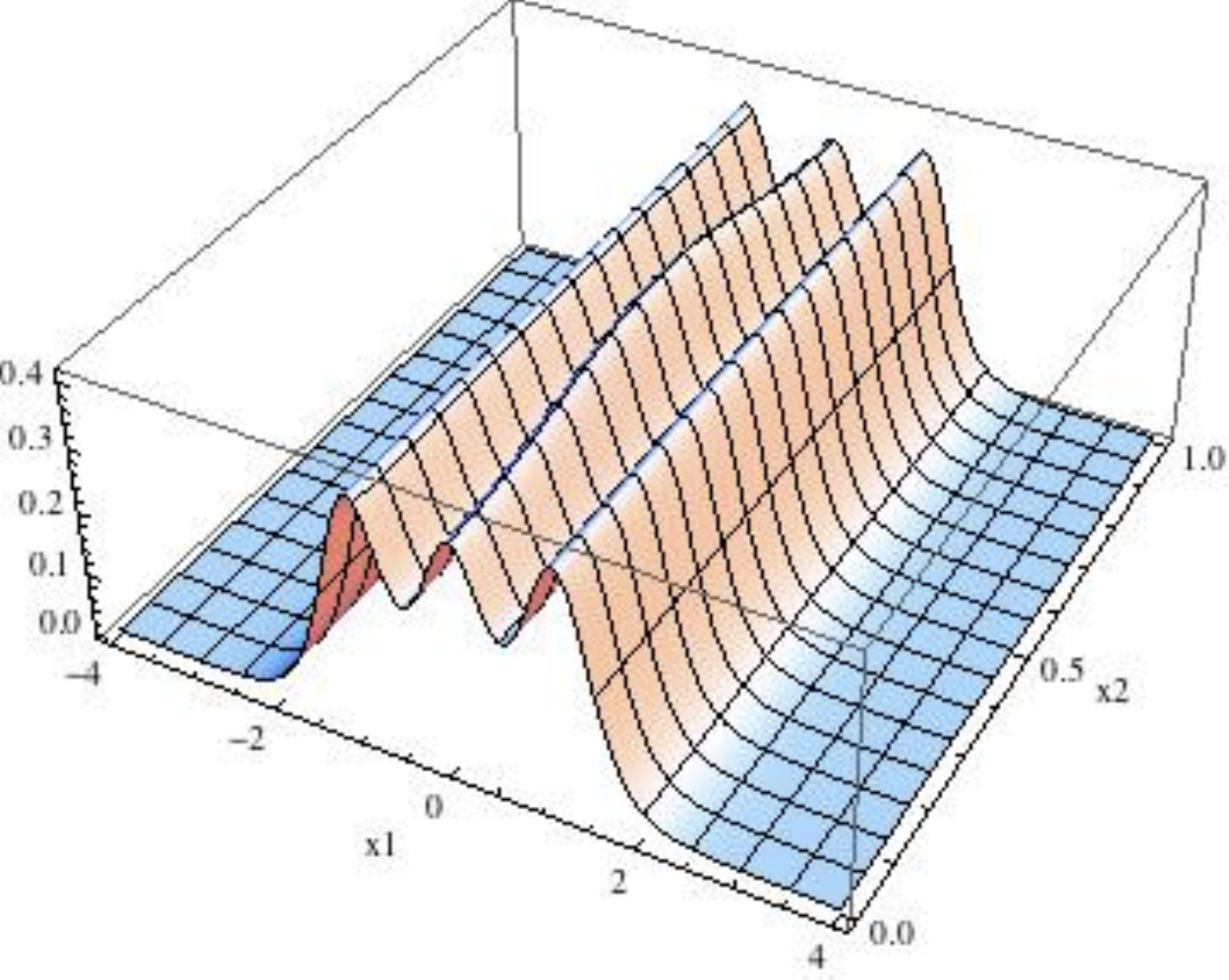}
\includegraphics[scale=0.38]{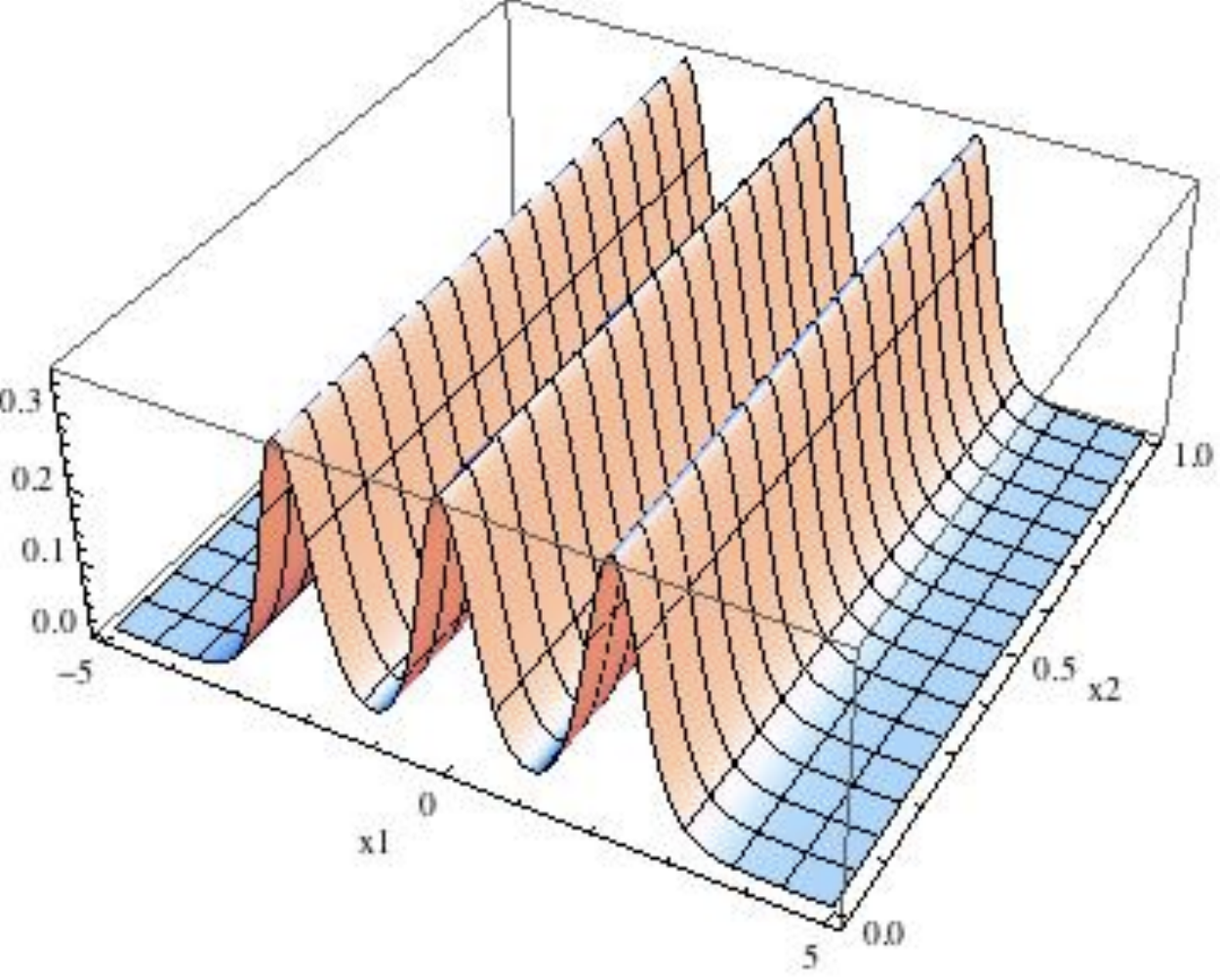}
\caption{Same as Fig.\ref{fig4}, but now   for ${\mathbb C}{\rm P}^{2}$.  Large instantons split into three  $Q=\frac{1}{3}$ kink-instantons, as the scale changes.
}
\label{fig6}
\begin{center}
\scalebox{3}{
\rotatebox{0}{
\includegraphics[width=4.5cm]{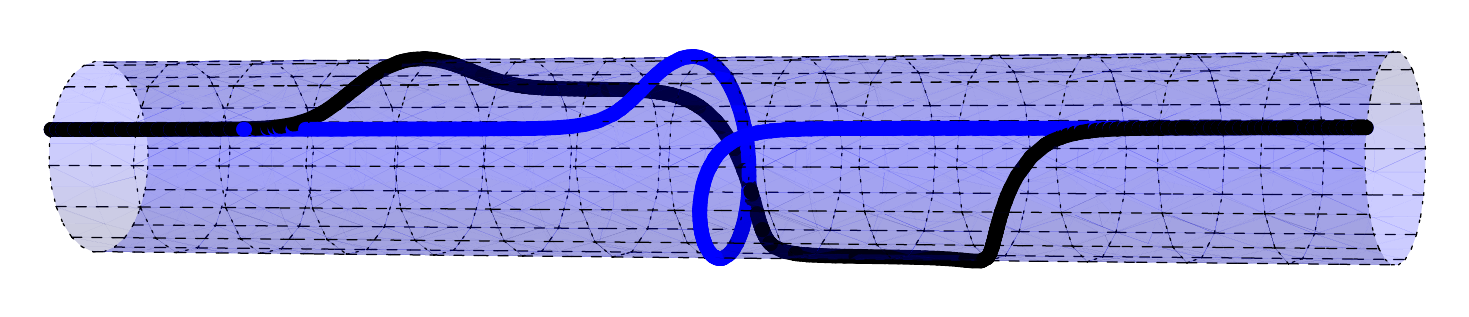}}}
\end{center}
\caption{Same as Fig.\ref{fig5}, but now  for ${\mathbb C}{\rm P}^{2}$. The large instanton splits into three separate kink-instantons, as the scale changes.
}\label{fig7}
\end{figure}

Another useful way to visualize these kink-instanton constituents is using the Wilson loop (\ref{wilson1}). As the Euclidean time coordinate $x_1$ goes from $x_1=-\infty$ to $x_1=+\infty$, the Wilson loop winds around on a unit circle. For the twisted $Q=1$ \cpone\, instanton in (\ref{tw1}), the Wilson line $W(x_1)$ is plotted in Figure \ref{fig5}.  A small instanton on ${\mathbb R}\times {\mathbb S}^1_L$ behaves like a  single instanton on ${\mathbb R}^2$, and winds fully around the circle. A sufficiently large instanton, on the other hand,  decomposes into its two constituents, each of which winds half-way around the circle, but displaced from one another in the non-compact  $x_1$ direction. The ${\mathbb C}{\mathbb  P}^{2}$ case ($N=3$) is illustrated in Figures. \ref{fig6} and  \ref{fig7}. Again, the small instanton is essentially the same as a small instanton on  $\R^2$, however, the large-instanton splits into three constituent kink-instantons, each of which has topological charge $Q= \frac{1}{3} $.

\subsection{Matching and reinterpreting  the bosonic  zero modes} 

 As reviewed in Section \ref{section-np-a}, in  the $\C\P^{\rm N-1}$  model on $\R^2$,  the 2d instanton zero modes  are associated with the classical symmetries of the self-duality equations:  2 are the position of the instanton $({\bf a}_I \in \R^2)$ and arise due to translation invariance, one is the size modulus $(\rho \in \R^+)$ and is associated with invariance under dilatations, and the remaining $2N-3$ are internal orientational modes. This is equally true for the small instantons (\ref{factorization}) on $\R \times \mathbb S^1_L$: 
\begin{eqnarray}
2N  &  \xrightarrow{\text{short-distance}}\ 
2+1+(2N -3)  =  ({\bf a}_I \in \R^2) + (\rho \in \R^+) 
+ ( {\rm orientation}) .  
\end{eqnarray}

As emphasized earlier, for the theory on $\R^2$,   the existence of the size modulus $\rho$ implies that the instanton comes in all sizes  at no cost in action, and prevents a meaningful long-wavelength description of a dilute instanton gas from first principles. However, in the small $\R^1 \times \mathbb S^1$ regime of  $\C\P^{\rm N-1}$, the instanton has a maximal size set by the eigenvlaue separation of the holonomy matrix 
$^L\Omega$. In this regime, and at long distances, the 2-d instanton is described  as  a composite of 
$N$  separate 1d kinks. The $ 2N $ bosonic zero modes of  the 2-d instanton on $\R^2$   matches the counting of the zero modes of the $N$ kinks of the theory on $\R \times \mathbb S^1$.  Each kink has two zero modes: One is the position of the kink, and arises due to translational invariance, and the other is an angular zero mode, associated with an internal symmetry. Therefore, 
the $2N$ collective coordinates split as 
\begin{eqnarray}  
2N& \ \xrightarrow{\text{long-distance}}\ 
N [1 +1] = N [(a \in \R) + (\phi \in U(1))]  .    
\end{eqnarray}
In particular the  size modulus  of the 2-d instanton is no longer present in the long distance description   of the $\C\P^{\rm N-1}$    on small $\R \times \mathbb S^1$.  This permits a meaningful dilute gas expansion.   

{\bf Collective coordinates of kink-instantons:}
The one-loop measure for integrating over configurations of a  type-$j$ kink-instanton is 
\begin{align}
 d\m_{\rm B}^j  d\m_{\rm F}^j=
e^{-S_j} \cdot
\frac{d a\, d\phi}{(2 \pi)} \prod_{f=1}^{N_f} d^2 \xi_f \cdot
\m^{2-N_f} \cdot 
J_a J_\phi (J_\xi)^{-N_f} \cdot 
\left[{\det}'(-D^2_{j})\right]^{N_f-1} .
\nonumber
\end{align}
\begin{list}{$\bullet$}{\itemsep=0pt \parsep=0pt \topsep=2pt}
\item $a\in\R$ is the kink-instanton position, $\phi\in U(1)$ is an  angle,  $\xi_f$ are the Grassmann-valued fermionic zero modes.  

\item $\m$ is the (Pauli-Villars) renormalization scale.  Each bosonic zero mode gives a contribution proportional to $\m$  and each Grassmann zero modes gives a contribution proportional to $\m^{-1/2}$, yielding  $\m^{2-N_f}$.

\item 
The  collective coordinate Jacobians for  bosonic and fermionic  coordinates are:
 $J_a =S_j^{1/2}$,  $J_{\phi} = L S_j^{1/2} [2\pi\ba_j(\f)]^{-1} = \frac{LN S_j^{1/2}}{2\pi}$, 
  $J_\xi = 2S_j$.
  
\item The primed determinant  is the result of integrating over the Gaussian non-zero modes. 
When $N_f=1$, the bosonic and fermionic primed  determinants cancel precisely due to supersymmetry and the absence of non-compact scalars.  This  also helps us to deduce the result for general $N_f$. 
\end{list}

The regularized (primed) determinant  in the background of a  type-$j$ kink-instanton
depends  linearly on the renormalization scale, 
${\det}'(-D^2) \sim \m$.  Since the the fields of a type-$j$ kink-instanton reside entirely within a \cpone\; 
sub-manifold of \cpn, the   only scale which appears in the classical equations is $2\pi \ba_j(\f)/L= \frac{2\pi}{LN}$.
Since the determinant is dimensionless, it must therefore have the form
\begin{align}\label{fluctdet}
[{\det}'(-D^2)]^{N_f-1}= C^{N_f-1} \left( \frac{\mu LN}{2 \pi}\right)^{N_f-1}
 \,,
\end{align}
where $C$ is a pure number of order one.  

\begin{list}{$\bullet$}{\itemsep=0pt \parsep=0pt \topsep=2pt}
\item The exponent  of the renormalization scale $\m$ 
  coming from the collective coordinates  $\mu^{2-N_f}$ and from integrating out non-zero modes (the primed determinant)   $\mu^{N_f-1}$ combine to give, (also keeping the exponential of the kink-action)
  \begin{align} 
  \underbrace{\mu^{2-N_f}}_{\rm zero -modes}  \times  \underbrace{\mu^{N_f-1}}_{\rm UV-det'} e^{-S_0} \sim  
  \mu e^{-S_I/N}  
  \end{align} 
  
  \item 
   Putting this all together,  and realizing that at the center-symmetric vacuum, $S_j= S_0=\frac{S_I}{N}$, 
the one-loop type-$j$ kink-instanton measure can be written as
\begin{align}
d\m_{\rm B} d\m_{\rm F} =
\frac{C^{N_f-1}}{\pi}    \mu e^{-S_0} 
\left( \frac{ LN} {2 \pi} \right)^{N_f} S_0^{1-N_f} 
da  d \phi \, \prod_{f=1}^{N_f} d^2  \xi_f ,
\label{Eq:measure}
\end{align}

\end{list}

Consequently, the kink-instanton amplitude takes the form  (neglecting interactions among kinks) 
$\K_j \sim  \mu \; e^{-S_0}$
 in the bosonic theory, and ${\cal K}_j \sim  \mu  e^{-S_0}  \psi^{2N_f}$ in the theory with fermions. 
 The crucial point is the appearance of the renormalization group invariant scale  
 $\mu \; e^{-S_0} = \Lambda$ through the kink amplitudes, and $\Lambda^2$ through the charged and neutral bion amplitudes. 
 The product of the  amplitudes of $N$ types of kink-instantons produces $I_{2d}  \sim \prod_{j=1}^{N} \K_j \sim \mu^N   e^{-N S_0} =  \mu^{\beta_0}  e^{-S_I} = \Lambda^{\beta_0}$, as expected, the  2d instanton factor,  $\Lambda^N$. The kink and bion amplitudes will be crucial 
  crucial in understanding the microscopic origins of various 
 observables in \cpn models, such as the non-perturbatively induced mass gap,  chiral symmetry realization and 
 the renormalon singularity structure.

\subsection{1-defects: Kink-instantons  } 
\label{kink-1}
 We have seen in Section \ref{subsec:ki-cpn} that in  the $\C\P^{\rm N-1}$  model on $\R \times \mathbb S^1_L$, there are $N$ types of elementary kink-instanton events associated with the simple and affine co-roots of the $SU(N)$  algebra. The vacuum of the theory, surprisingly, is the  co-root lattice  of $SU(N)$, that is the kink-instanton events  are valued in the co-root lattice $\G_r^\v$.
 \begin{equation}
\tn \longrightarrow \tn + \alpha_i,    \qquad  \alpha_i \in \G_r^\v
 \end{equation}
 The amplitudes associated  with these kink-events in a $\Z_N$-symmetric background (\ref{minspatial}) are given by (ignoring interactions among the  kink-instantons), 
\begin{eqnarray}
{\cal K}_j=&& \exp\left[- \frac{S_{I}}{N} \right]  \qquad  {\rm where}  \qquad  S_{I} = \frac{4\pi}{g^2} - i \Theta,   
\qquad   j=1, \ldots, N
\label{kink1}
\end{eqnarray}
Since $\Theta$ is an angular variable with period $2 \pi$,   $\frac{\Theta }{N} $ does assume $N$ different values. Therefore, the kink-instanton amplitude associated with each $j$ is already a  multi-branched 
quantity. 
\begin{eqnarray}
{\rm fixed-} j : \qquad 
 {\cal K}_j=&& \exp \left[- \frac{4\pi}{g^2N} + i \frac{\Theta + 2\pi\, k}{N} \right]\quad , \qquad  k=1, \ldots, N
\label{kink12}
\end{eqnarray}
 Indeed, 
when we discuss the $\Theta$ dependence of  observables, such as the vacuum energy density, mass gap or topological susceptibility,  it will be crucial to note the fact that $\Theta$ dependence enters as $\frac{1}{N}$ of the 2d instanton effect and the kink-instanton contributions are multi-branched. 

As noted, (\ref{kink12})   does not take into account the interaction between kinks. This can be restored  
 by writing  the bosonic kink amplitudes as 
\begin{eqnarray}
{\cal K}_j&=& e^{-\alpha_j \cdot Y }, \qquad  j=1, \ldots N-1, \qquad    \cr 
   {\cal K}_N&=& \eta e^{-\alpha_N \cdot Y }, \qquad \eta=  e^{- \frac{4\pi}{g^2} + i \Theta },
\label{kink2}
\end{eqnarray}
where 
\begin{equation}
 \langle e^{-\alpha_j \cdot Y } \rangle =  e^{-\frac{4\pi}{g^2}  (\mu_{j+1}-\mu_{j}) }
\end{equation}
$Y$ is an  $N$-component complex field, with $2N$ real variables. It is related to the original variables as follows:
\begin{equation}
Y (x_1)= \Re Y(x_1) - i \,\Im Y(x_1) 
\end{equation} 
where $  \Re Y(x_1)   =\{  {\cal A}_{2,1}, {\cal A}_{2,2}, \ldots,  {\cal A}_{2,N} \}    $ is the $N$-component sigma-model  connection defined in (\ref{gt2}), and $\Im Y$ is an $N$-component field  which accounts for the $\{\theta_1, \ldots  \theta_{N-1}\}$  induced interactions, and is non-locally dual to the $\theta$-field.  Although both real and imaginary parts of $Y$ are $N$-component objects, due to the constraint $\sum_{j=1}^{N} \alpha_j \cdot Y  =0$, there are only  $2(N-1)$ independent degrees of freedom entering 
to the $N$-types of kink amplitudes, which agrees with the  number of degrees of freedom 
entering  to the original $n$ or  $\tn$ fields.

\subsection{Index theorem for Fredholm-type  Dirac operator on  $\R \times \mathbb S^1_L$}
\label{sec:index}
 In the theory with  $N_f$ fermions,  the 2d instanton  has $2N N_f$ fermionic  zero modes, as dictated by the Atiyah-Singer index  theorem. The number  of fermionic zero modes of kink-instantons  is determined by an appropriate modification of the Nye-Singer index theorem, along the same lines as  Refs. \cite{Nye:2000eg, Poppitz:2008hr}.   
 
Here, we give the the index formula  for topological excitations on $\R^1\times S^1$ without derivation: 
\begin{align}
{\rm ind} (\Dslash)  =&  \dim \ker \Dslash- \dim \ker \Dslash^{\dagger}  \cr 
=&
 \int {\rm ch}_1(F) - \frac{1}{2} \eta[D_{S^1, ^L\Omega}]
\label{index}
\end{align}
where ${\rm ch}_1(F)$ denotes the first Chern character, $\eta[D_{S^1, ^L\Omega}]$ is the eta-invariant  (spectral asymmetry)  of the one-dimensional  Dirac operator    at the end of the cylinder  (at ${ |x_1|} =  \infty$) coupled to the non-trivial $\sigma$-connection holonomy.  The index is well-defined if   and only if the operator  $D_{S^1, ^L\Omega}$ is Fredholm, i.e.,  $[^L\Omega ({\bf \infty}) -1]$ must be  invertible, or non-degenerate.  Typically, neither contribution to the index is actually an integer, and yet, the sum always is, as the index counts the number of fermionic zero modes associated with  a kink. 

The index theorem can be derived by using  axial current non-conservation---an exact operator identity valid on any two-manifold---and  relies on the existence of a    $\sigma$-connection holonomy that satisfies the  Fredholm condition.  
The main result is that the index associated with $k_i$ many kinks of type $\a_i$ in \cpn theory with $N_f$ flavors of fermions is given by
\begin{align}
{\rm ind} [k_1, k_2, \ldots, k_N] = N_f  \sum_{i=1} 2 k_i
\label{index2}
\end{align}
i.e., each elementary kink-instanton carries $2N_f$ fermionic zero modes. This result is identical to the index for the monopole-instantons in QCD(adj) on $\R^3 \times S^1$,  see Appendix B of 
\cite{Nye:2000eg, Poppitz:2008hr}.   

 In the $N_f=1$  case  corresponding to compactified  ${\cal N}=(2,2)$ theory,  each kink-instanton  ${\cal K}_j$ has two zero modes. The importance of just two zero modes 
 this will be discussed in the next section.  The kink-instanton amplitude in the $N_f$ flavor theory is  given by
\begin{eqnarray}
{\cal K}_j&=& e^{-\alpha_j \cdot Y } \det_{f,f'}   [(\alpha_j \cdot   \psi_{-}^f )   (\alpha_j  \cdot  \psi_{+ }^{f'}) ]
 \qquad  j=1, \ldots N-1, \qquad    \cr 
   {\cal K}_N&=& \eta e^{-\alpha_N \cdot Y }  \det_{f,f'}  [ (\alpha_N \cdot   \psi_{-}^f )   
   (\alpha_N  \cdot  \psi_{+ }^{f'}) ]  \qquad \eta=  e^{- \frac{4\pi}{g^2} + i \Theta },
\label{kink3}
\end{eqnarray}
Note that the product of the $N$-types of kink-instanton amplitudes gives the  action and the fermion zero mode structure of the the 2d instanton amplitude, namely
\begin{eqnarray}
\prod_{j=1}^{N} {\cal K}_j =  I_{2d}
\end{eqnarray}

The fermion-bilinear in these  amplitudes transforms non-trivially under  $\Z_{2N N_f}$, and is singlet under $\Z_{2N_f}$, and continuous global symmetries  (\ref{symm})
 \begin{eqnarray}
\Z_{2N N_f} &&: \det_{f,f'}   [(\alpha_j \cdot   \psi_{-}^f )   (\alpha_j  \cdot  \psi_{+ }^{f'}) ]     
 \longrightarrow e^{i \frac{2 \pi}{N} } \det_{f,f'}   [(\alpha_j \cdot   \psi_{-}^f )   (\alpha_j  \cdot  \psi_{+ }^{f'}) ]
\label{transform}
\end{eqnarray}
 (Recall the discussion around (\ref{symm}), the genuine discrete chiral symmetry, which may be spontaneously broken is just $[\Z_N]_A$ and it is broken down to $\Z_1$ on $\R^2$.)   Since the kink-amplitude only respects the   $\Z_1$ subgroup of $U(1)_A$, one may wonder if this 
  implies   that the anomaly free sub-group is  $\Z_1$ and not $\Z_N$. However, this is not so 
in a way which  differs crucially  from QFT on $\R^3 \times   \mathbb S^1_L$.\footnote{In ${\cal N}=1$ SYM and QCD(adj)  on $\R^3 \times \mathbb S^1_L$, the transformation of the fermion bilinear is as in  (\ref{transform}) and $\Z_N$ symmetry is still exact  because the ``flux" part of the monopole-amplitude transforms in the opposite direction, 
\begin{equation} 
\Z_{2N} :\quad   e^{-\alpha_j \cdot Y }   \quad \longrightarrow\quad    e^{- i \frac{2 \pi}{N} }  e^{-\alpha_j \cdot Y } 
 \end{equation}
 rendering   $ e^{-\alpha_j \cdot Y }   (\alpha_j \cdot   \psi_{-} )   (\alpha_j  \cdot  \psi_{+ })$ a singlet under $\Z_{2N}$ symmetry.   This is possible because in QFT, $\sigma\equiv \Im Y(x)$ is the dual photon and in the absence of the monopoles, has a shift symmetry, $\sigma \rightarrow \sigma + \epsilon$. This shift symmetry intertwines and reduces to $[\Z_N]_A$ in the presence of monopoles.  
 This mechanism is no longer  valid on small $\R^1 \times \mathbb S^1$, because $\Im Y$ is a massive field at tree level, and there is no analogous  shift symmetry to  undo  the chiral rotation (\ref{transform}). 
 In quantum mechanics, a new mechanism is operative. This requires a separate discussion of its own.}
  $[\Z_{N}]_A$ is an exact symmetry of the quantum theory, which is broken spontaneously on $\R^2$. 
  On small-$S^1$ where the low energy theory is described by a  quantum mechanics, the theory develops 
  $N$ superselection sectors, and  in fact, the existence of $N$-chirally broken vacua holds at any (finite) $\mathbb S^1_L$.

 \section{Application: ${\cal N} = (2,2)$ \cpn\,  theory on $\R \times \mathbb S^1_L$} 
  \label{sec:susy}
  
 The power of the semi-classical transseries expansion, whenever applicable, transcends supersymmetry, and applies to all QFTs which admit a weak coupling limit.  In this section, we  study the dynamics of  the supersymmetric theory by using  kink-instantons (\ref{kink-1}),  the index theorem (\ref{index}) and basic supersymmetric techniques. This will help us to check the semi-classical methods applied to topological molecules with the results of the supersymmetric approach. 
   
As discussed earlier, the one-loop potential (\ref{spotential}) for the $\s$-connection holonomy  (\ref{hol}) is zero for the supersymmetric  ${\cal N} = (2,2)$ \cpn\, theory with supersymmetry preserving periodic boundary conditions for fermions:
   \begin{eqnarray}
&&V_{+} [^L\Omega]  =0  \qquad  
 \label{spotsusy}
\end{eqnarray} 
 This is  true  to all orders in perturbation theory because of supersymmetry. 

 As noted in (\ref{kink3}), for the $N_f=1$ case,  each kink-instanton  ${\cal K}_j$    event carries  two fermion zero modes, including the affine-one as a result of the index theorem (\ref{index}). The $N$ types of kink-instanton amplitudes are given by  
\begin{eqnarray}
{\cal K}_j&=& e^{-\alpha_j \cdot Y }   [(\alpha_j \cdot   \psi_{-})   (\alpha_j  \cdot  \psi_{+ } ) ]
 \qquad ,\quad  j=1, \ldots N-1, \qquad    \cr 
   {\cal K}_N&=& \eta e^{-\alpha_N \cdot Y }    [ (\alpha_N \cdot   \psi_{-} )   
   (\alpha_N  \cdot  \psi_{+ } ) ] 
\label{kinkSYM}
\end{eqnarray}
where we dropped the prefactors for convenience. The existence of two zero modes implies that such  kink-instanton amplitudes can induce a  superpotential.  The  associated  superpotential is  
 \begin{equation}
{\cal W}= \sum_{j=1}^{N-1}   e^{-\alpha_j \cdot Y } + \eta  e^{-\alpha_N \cdot Y }  
\label{superpotential}
 \end{equation}
 which is  called the  affine Toda superpotential.  
  
 The non-perturbatively induced bosonic potential  $V (Y) = \sum_{j=1}^{N} \left| \partial_{Y_j}   {\cal W} \right|^2 $   leads to a $\Z_N$-symmetric ground state, i.e., the same $\Z_N$-symmetric background (\ref{minspatial}) as in the $N_f>1$ theories. In order to see this,   let  $V_i = \a_i \cdot Y$, and re-write the superpotential as 
  \begin{equation}
{\cal W}= \sum_{j=1}^{N-1}   e^{-V_j } + \eta  \prod_{j=1}^{N-1} e^{V_j } 
\label{superpot2}
 \end{equation}
 The minimum can be found by extremizing the superpotential,    which yields   
 \begin{equation}
  e^{-V_j } =  \eta  \prod_{j=1}^{N-1} e^{V_j }   \Longrightarrow   e^{-V_j } = \eta^{1/N} 
\label{supermin}
 \end{equation}
  leading to the center-symmetric background 
  \begin{equation}
 \label{minsusy}
  \langle  \Re (Y_{j+1} - Y_{j}) \rangle = \frac{4 \pi}{g^2N} ,  \qquad  \mu_{j+1} - \mu_{j}  = \frac{2 \pi}{N}   
  \end{equation}
  given in (\ref{minspatial}).  This provides a self-consistency condition to the use of semi-classics 
  in supersymmetric  \cpn.
  
  The superpotential expanded around the minimum  (\ref{minsusy}) can be re-written as
   \begin{equation}
{\cal W} (  \langle Y_{j} \rangle + Y_j )
= \eta^{1/N}  \sum_{j=1}^{N}   e^{-\alpha_j \cdot Y } =  \eta^{1/N}  \sum_{j=1}^{N}   e^{-( Y_{j+1}- Y_j )} 
\label{superpotential2}
 \end{equation}
 where we parametrized the fluctuations around the minimum  $\langle Y_{j} \rangle$ by $Y_j$. 
Therefore, the bosonic potential is
   \begin{align}
V (Y) &= \sum_{j=1}^{N} \left| \partial_{Y_j}   {\cal W} \right|^2  = 
 \eta^{2/N} \sum_{j=1}^{N}  \left| 
e^{- \a_j  Y} - e^{-\a_{j-1}Y} \right|^2    \cr 
&=  \eta^{2/N} \sum_{j=1}^{N}  \left[ 2 e^{- \a_j  (Y+ Y^*) } -   e^{-\a_j Y - \a_{j-1}Y^*}    -   e^{-\a_j Y^* - \a_{j-1}Y}    
\right]
\label{bosonic}
 \end{align}
 A few interpretational comments are in order:   
 
 \begin{list}{$\bullet$}{\itemsep=0pt \parsep=0pt \topsep=2pt}
\item   The topological configurations (\ref{kink3}) with $N_f=1$, leading to a superpotential, are 1-defects, or kink-instantons.

\item The bosonic potential is induced  by  non-selfdual 2-defects, correlated kink--anti-kink events, called   bions $  \cB_{ii}= [\K_i \bar \K_{i}]$ and  $ \cB_{i,i-1}= [\K_i \bar \K_{i-1}]$.   These 2-defects do not carry any fermonic zero modes, hence they are capable of generating a bosonic potential, as opposed to a superpotential.

\item The existence of the semi-classical  kink-instantons  and  bions relies on 
the $\Z_N$ symmetric (or more generally, non-degenerate)  background such as  (\ref{minspatial}), but not on supersymmetry.  These defects generalize to arbitrary $N_f$. In Section \ref{sec:Bions}  we give a general argument for the construction of the two-types of bions. 
\end{list}

\subsection{Chiral ring and  condensate,  and mirror symmetry} 
\label{sec:mirror}
In order to calculate the chiral condensate,  it is useful to introduce  a 
Veneziano-Yankielowicz (VY)   Lagrange multiplier superfield $S$ \cite{Veneziano:1982ah}.  
Then,  the superpotential  can be re-written as 
  \begin{equation}
{\cal W} (V_i, S) = \sum_{j=1}^{N}   e^{-V_j } + S ( \tau - \sum_{j=1}^{N}  V_j)     
\label{superpot3}
 \end{equation}
Integrating out $S$, we land on (\ref{superpot2}). Instead,  integrating out the $V_i$ super-fields,   one obtains the VY-type superpotential, given by 
  \begin{equation}
{\cal W} ( S) =  -S N \log S - S \tau  +NS 
\label{superpot4}
 \end{equation}
 
   It is now straightforward to obtain the chiral condensate by using  the VY-superpotential. Integrate out the $S$ -field, which amounts to   $S^N= e^{- \tau}$, a quantum modified chiral ring relation, similar to its 4d counterpart \cite{Cachazo:2002ry}.
  In dimensionful  units,
  \begin{equation}
  S^N= \Lambda^N, \qquad S= \Lambda e^{i \frac{2 \pi k}{N}}\quad , \quad k=1, \ldots, N 
  \end{equation}
   where $\Lambda$ is the strong scale.  The lowest component of $S$ is the chiral condensate. 
   
 The chiral condensate can also be recovered as follows. Substituting  $S^N= e^{- \tau}$ into (\ref{superpot4}), we obtain the superpotential in terms of the holomorphic parameter $\tau$:
 \begin{equation}
{\cal W} ( \tau) = N e^{- \tau/N}
\label{superpot5}
 \end{equation}
 and the chiral condensate reads   $\langle  \psi_+ \psi_- \rangle = - \partial_\tau {\cal W} =  e^{- \tau/N}$. The supersymmetric theory, (as we will see the non-supersymmetric theories with $N_f > 1$) 
possess $N$-vacua $|\Psi_k^0 \rangle$, associated with spontaneous chiral symmetry breaking:
\begin{align}
 \langle \Psi_k^0  | \psi_+ \psi_- |\Psi_k^0   \rangle =  \Lambda e^{i \frac{2 \pi k}{N}},  \qquad k=1, \ldots, N 
 \end{align}

The affine Toda superpotential (\ref{superpot3}) has been obtained  
earlier, on $\R^2$, by using   mirror symmetry in string theory, see  \cite{Hori:2000kt}.  In our framework, this follows from a simple duality in quantum mechanics on $\R \times \mathbb S^1_L$.  This duality, in our formulation, amounts to re-writing of a quantum mechanical system with multiple degenerate minima, in terms of a dilute gas of kink-instantons and the  two-types of bions. In string theory, the mirror of \cpn is obtained by using the standard
$R \to 1/R$  duality and the dynamical generation of a superpotential by vortices. 
 It would be interesting to understand the connection between these two derivations in more detail.

\section{Infrared renormalons and topological molecules }
 \label{sec:Kinks}

\subsection{Prelude: Large orders in perturbation theory, Borel summation and the Stokes phenomenon}
\label{sec:large}

Perturbation theory in almost all interesting quantum field theories,  quantum mechanics and even in 
ordinary integrals with multiple saddles, leads to divergent  asymptotic expansions \cite{LeGuillou:1990nq}.   The divergence encodes physical information about the  saddles of ordinary integrals, or path integrals of quantum mechanics and quantum field theory, as a consequence of Darboux's theorem \cite{Dingle:1973,BerryHowls}.  We recall a few relevant definitions and motivate  (known)  generalizations of those  definitions by using simple quantum mechanics. 

Let $P(g^2)$  denote a perturbative asymptotic series that satisfies the ``Gevrey-1" condition:
 \be
 \label{Gevrey-1}
  P(g^2)  = \sum_{q=0}^\infty a_{q} g^{2q}, \qquad {\rm Gevrey-1:} \;\;\; |a_{q}| \le C R^q q!
  \ee 
for some positive constants $C$ and $R$ \cite{Sternin:1996, Costin:2009}. Known examples of perturbative series that arise in quantum mechanics and QFT  satisfy the ``Gevrey-1" condition  \cite{LeGuillou:1990nq}. We denote the Borel transform of $P(\l)$ by $BP(t)$ and define it as 
\begin{align}\label{}
BP(t) := \sum_{q=0}^\infty \frac{a_{q}}{q!} t^q .
\end{align}
The formal Borel transform determines ``a germ of a holomorphic function" at $t=0 $, with a finite radius of convergence.  
Next, one analytically continues the obtained germ  $ BP(t)$ to the  whole  complex  $t$-plane, called the Borel plane. We also assume that the analytic continuation of the Borel transform  $ BP(t)$ is ``endlessly  continuable". That roughly means that the function is represented by an analytic function with a discrete 
set of singularities (poles or cuts) on its Riemann surface. 
The  Borel resummation of  $P(g^2)$, when it exists,  is defined as  the Laplace transform  of the analytic continuation of the germ:  
\begin{equation} 
\B(g^2) = \frac{1}{g^2} \int_0^\infty BP(t) e^{-t/g^2} dt \; .
\end{equation}

In quantum theories with multiple-degenerate vacua, (but no instability of any kind),  perturbation theory is typically a non-alternating Gevrey-1 series, hence non Borel resummable \cite{Brezin:1977gk,Stone:1977au,zinn-book,h2plus,silverstone,LeGuillou:1990nq}.  Non-Borel summability means that there is no unique answer in perturbation theory;  i.e.,  resummed perturbation theory  does not produce a unique answer for a physical observable which ought  to be unique, for example, the ground state energy. Of course, this is senseless.  This means that perturbation theory (re-summed or otherwise) is insufficient to define the theory.

 In certain cases, a perturbative sum which is not Borel summable becomes Borel summable upon continuation $g^2 \rightarrow -g^2$, see Fig.~\ref{fig:continuation}. 
  In simple quantum mechanics,  let us mention an example that is directly  relevant for our purpose \cite{Stone:1977au}. Perturbation theory for the periodic potential
 $V(x) = \frac{1}{g^2}\sin^2(gx)$  is non-Borel summable, whereas  perturbation theory  for $V(x) = \frac{1}{g^2}\sinh^2(gx)$ is Borel summable.  [Recall and compare with the 0-dimensional partition functions discussed in Section \ref{subsec:examples}]. Both series are, of course, asymptotic and divergent. The difference between the two is that the asymptotic series which arises in the first case is  {\it non-alternating}, whereas the series in the latter is just the {\it  alternating}  version of the former. Let us refer to the  Borel resummed series for the latter, Borel resummable series, as  $\B_0(g^2)$.  Then, we can define the perturbative sum for the non-alternating series as the analytic continuation of $\B_0(g^2)$ in the $g^2$ complex plane from negative coupling, $g^2<0$, to  the positive real axis, $g^2>0$. This can be done in one of the two ways as shown in  Fig.~\ref{fig:continuation}.  
  Approaching the positive real axis clock-wise (from above)  and counter-clock-wise (from below). 
\begin{equation}
\label{Borel}
\B_0(|g^2| \pm i \e)  = \Re\,\B_0(|g^2|) \pm i\, \Im\,\B_0(|g^2|)   \qquad {\rm where\,}  \; \;   \Im\,\B_0(|g^2|) \sim e^{-2 S_I} \sim e^{-2 A/g^2} 
\end{equation}
is the ambiguous part, and is a manifestation of non-Borel-summability [compare with (\ref{contin})].

A definition of the Borel sum equivalent to what we described above  through analytic continuation in the complex $g^2$-plane  is the {\it directional (sectorial) Borel sum}. Define 
\begin{equation}\label{directionalBorel}
{\cal S}_{\th} P (g^2)   \equiv \B_{ \th}(g^2) = \frac{1}{g^2} \int_0^{\infty \cdot e^{i\th}} BP(t)\, e^{-t/g^2} dt ,
\end{equation}
\begin{figure}[htb]
\centering{\includegraphics[scale=0.5]{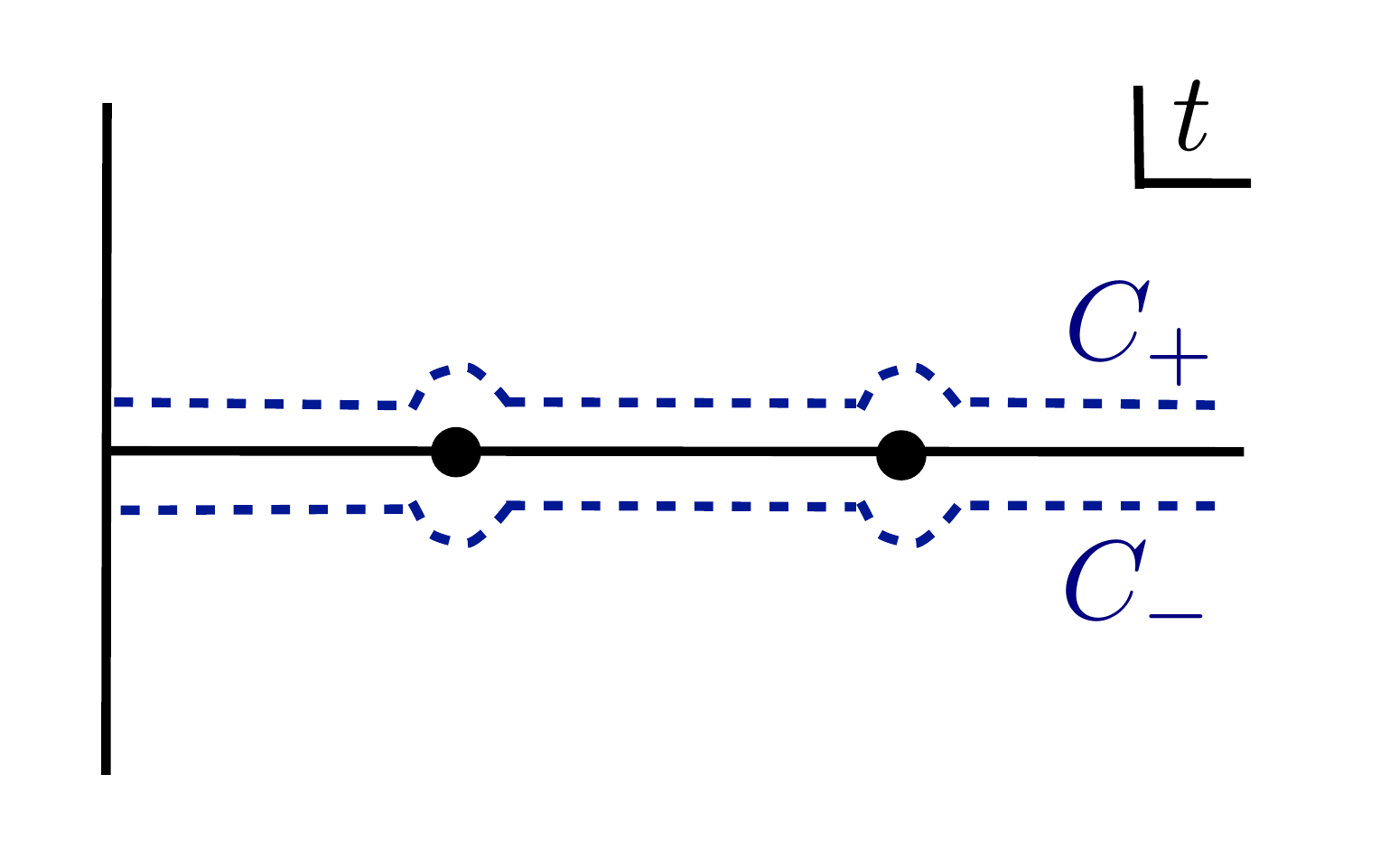}}
\caption{ Lateral, or right and left,  Borel sums. Dark circles are singularities (poles or branch points). 
Whenever a singularity exists between the right and left Borel sums,  the theory is non-Borel summable. 
The singular direction in the t-plane corresponds to  a Stokes line in the complex $g^2$-plane, see Fig.\ref{fig:continuation}. The difference of the  sectorial sums in passing from $\theta=0^-$ to  $\theta=0^+$ is the Stokes ``jump" across a Stokes ray. 
}
\label{fig:lateral}
\end{figure}

A special case of this is the {\it lateral Borel sum}. The function $\B_{\theta \pm}(g^2)$ 
is  associated with contours just above and just below the ray at angle $\theta$, and is called right (left) 
Borel resummation. If there are no singular points in the $\theta$ direction, then the left and right Borel sums are equal, and the theory is sectorial Borel summable in the $\theta$-direction.  A theory for which there are no singularities on $\theta=0$ is called Borel summable in physics. 
In many cases, there is a 
 ray of singular points of the Borel transform $BP(t)$, as shown in Figure \ref{fig:lateral}. Then, the theory is non-Borel summable, but left and right Borel summable. 
   The ambiguity described above, associated with whether we approach the real positive axis from above or below in the complex $g^2$-plane, in the  latter language, maps to the  choice of the integration contour in the Laplace-transform. The integral is, of course, dependent on the choice of the contour, yielding (\ref{Borel}).  The overall procedure can be summarized through  the diagram: 
\be
\xymatrix{
  P (\l)   \ar[rd]_{\rm Borel- sum} \ar[rr]^{\rm B - transform}  & &  BP(t)   \ar[ld]^{{\cal L}^{\theta}-{\rm transform}} \\
  &{\cal S}_{\th} P(g^2)   & }
  \ee
where $B$- is the Borel transform operator, and  ${\cal L}_{\theta}$ is Laplace transform along the ray at some angle $\theta$. 

In cases where there are singularities in the $\theta=0$ direction, the $\B_{0 \pm}(g^2)$ are different  sums 
differing in their   exponentially small imaginary parts, and still, both sectorial sums are associated with the 
asymptotic expansion (\ref{Gevrey-1}).  The divergence of the original series, in the original $g^2$ plane, 
just means that the perturbative expansion is taking place on a 
 singular Stokes ray in the complexified coupling constant plane.  Singular directions in the Borel t-plane correspond to Stokes lines in the $g^2$-plane, where the  sum exhibits the Stokes phenomenon crossing the ray.  The Stokes phenomenon in the $g^2$-plane, is mathematically, the origin of the ambiguity of the Borel sum.  The understanding of the connection of the distinct sectorial sums entails the understanding of the jumps across this direction. This is achieved via the {\it Stokes automorphism}, or {\it passage automorphism}
labelled by  $\underline{\frak{S}}_\theta$. 
 This corresponds to composing the  Laplace transform in a given direction, say $\theta^{+}$, with the inverse Laplace transform in another direction, say $\theta^{-}$,  \cite{cnp93}. 
\be
{\cal S}_{\theta^+} = {\cal S}_{\theta^-} \circ \underline{\frak{S}}_\theta \equiv {\cal S}_{\theta^-} \circ \left( \mathbf{1} - {\rm Disc}_{\theta^-} \right),
\label{stokesauto}
\ee
where $ {\rm Disc}_{\theta^-} $ is the full discontinuity of the Borel sum across $\theta$.
 The Stokes jump is purely non-perturbative, and it is  roughly 
 \be 
 {\rm Disc}_{\theta^-}  \,\B \sim e^{-t_1/g^2} +e^{-t_2/g^2} +  \ldots   
 \ee
where  $t_i \in e^{i \theta} \R^{+}$ is an ordered sequence of singularities along the singular direction.

In physical applications,  $\B_0(|g^2| \pm i \e)  \equiv {\cal S}_{ \pm0} P(g^2)   $ may be a physical observable in quantum mechanics, such as the energy of an eigenstate.    If we take Borel-resummation literally, and if there are indeed singularities at $t_i \in \R^{+}$, then the Borel sum 
 yields an answer which {\it i)} has imaginary component,  {\it ii)} the imaginary part is two-fold ambiguous.  On the other hand, we are dealing with a quantum mechanical system which does not have any instability, and the energy eigenvalues {\it must be} real.

In simple quantum mechanical systems,  it is understood how the non-perturbative ambiguity in perturbation theory is cancelled with the ambiguity in instanton-anti-instanton events\footnote{It is important to note that 
the cancellation is with a {\it topologically neutral} instanton-anti-instanton event. Instantons, as they carry topological charge, cannot mix with the perturbative vacuum. It  is important to realize and appreciate this difference. In particular, we will also discuss theories in which there are instantons, but the theory is Borel summable. This is due to the absence of a correlated instanton-anti-instanton event. For example, 
\cpn with extended supersymmetry, such as ${\cal N}= (4,4)$  is of this type.}
: see the important collection of works
 \cite{Bogomolny:1977ty,Bogomolny:1980ur,Brezin:1977gk,Stone:1977au,Balian:1978et,ZinnJustin:1981dx,zinn-book,Balitsky:1985in,h2plus,silverstone,Achuthan:1988wh}.  In these theories, the non-perturbative molecular instanton-anti-instanton events are also  ambiguous, and obtained through the analytic continuation and 
 back depicted in Fig.~\ref{fig:continuation}. 
  (The instanton and anti-instanton events are not ambiguous, only their topologically neutral composites are.) This ambiguity arises from the quasi-zero mode integrals and will be explained more generally in Section \ref{sec:qzm}.  The instanton-anti-instanton molecule amplitude is 
 \be 
   [\cI  \bar \cI ]_{\th=0^{\pm}} =  \Re\, [\cI  \bar \cI ] + i\,  \Im\, [\cI  \bar \cI ]_{\th=0^{\pm}} 
  \ee 
 As shown explicitly  in quantum mechanics, these two classes of ambiguities, between the  
 the re-summed perturbation theory  around the perturbative vacuum and the non-perturbative amplitudes associated with neutral topological molecules, cancel precisely at the $O( e^{-2S_I})$ level, yielding our simplest ``{\it confluence equation}'', (see  (\ref{rBZJ}) and (\ref{rBZJ2}) for generalizations):
  \begin{align}
\label{borelmon}
\Im\, \B_{0, \th=0^{\pm} }  +   \Im \, [\cI  \bar \cI ]_{\th=0^{\pm}}=0 \;  , \qquad  {\rm up \; to} \;  O( e^{-4S_I})
\end{align}
leading to an unambiguous physical observable: 
\begin{equation}
\label{obser}
O(g^2) =  \Re\, \B_0(|g^2|)  +       \Re\, [\cI  \bar \cI ]   + \ldots  \;  , \qquad  {\rm up \; to} \;  O( e^{-4S_I})
\end{equation}
Resurgence is the statement that this structure repeats itself at all non-perturbative orders, and a consistent semi-classical trans-series expansion for observables in quantum mechanics can be obtained.

\subsection{How to tame a theory with IR renormalons? } 
\label{sec:taming}

The ambiguity that arises in QM is related to the $q!$  factorial growth  of the number of Feynman diagrams combined with the non-alternating character of the series \cite{Bender:1976ni}. This ambiguity is canceled by  instanton-anti-instanton events. In   ``realistic QFTs'' including   asymptotically free theories such as QCD and   non-linear sigma models which appear as low-energy description of  certain spin systems [e.g., quantum anti-ferromagnets],  however, the situation is quite different, as described by 't Hooft \cite{'tHooft:1977am}. 
  In particular, apart from the   $q!$  factorial growth of the number of Feynman diagrams, in this class of theories,  perturbation theory  is so wild that a certain subset of diagrams also yields $q!$  growth  due to integration over momenta. This situation arises in renormalizable QFTs, and these new divergences are called ``renormalons'' \cite{Beneke:1998ui}.  There are both UV and IR renormalon singularities, associated with high and low momentum integration with respect to a given scale. In asymptotically free theories, the ambiguities that are due to IR renormalons are on the positive real axis, $t\in \R^{+}$, of the Borel plane, and furthermore are located much closer to the origin than the $[\cI \bar \cI]$ Borel pole. Until very recently \cite{Argyres:2012vv}, it was  not known whether there exists a  first principles non-perturbative approach  to cancel these perturbative ambiguities, despite the existence of substantial literature on renormalons, see  \cite{Beneke:1998ui} and references therein.   We now address this problem in the context of  \cpn  model.

According to 't Hooft's analysis \cite{'tHooft:1977am} and its generalizations  to non-linear sigma models    \cpn and  $O(N)$,
\cite{Beneke:1998ui, David:1980gi},  perturbation theory generically develops ambiguities of the form 
\begin{equation}
\label{renormalon1}
\Im\, \B_0(|g^2|) \sim \, e^{-2n S_I  /\beta_0}  \sim  \pi \, e^{-2 n S_I  /N}, \qquad  n= 2, 3, \ldots,  
\qquad 
\end{equation}
which are exponentially more important than the $[\cal I \bar {\cal I}]$ singularity.
We make two simple observations: 
\begin{itemize}
\item  This renormalon ambiguity is exponentially more important than the instanton--anti-instanton $[\cal I \bar {\cal I}]$ ambiguity, 
  \be
e^{-2  S_I  /N} \gg e^{-2  S_I  } 
\ee   
The BZJ approach that  suffices in quantum mechanics  would only cure  the $[\cal I \bar {\cal I}]$ ambiguity in QFT, not the much larger and more important renormalon  ambiguity.
\item  On $\R^2$, there are no-known semi-classical (or even non-semi-classical) configurations  
with action  ${2nS_I  /N}, n=2, 3, \ldots $  that can fix the IR renormalon ambiguity.
\end{itemize} 
The formalism developed in  this work helps us to  solve this problem on $\R \times  \mathbb S^1_L$ in a regime  of QFT continuously connected to $\R^2$.     Many authors previously considered some form of compact space to fix the IR-problem of the 2d instantons  \cite{Jevicki:1979db, Affleck:1980mb, Munster:1982sd,Aguado:2010ex}, however, the above mentioned problems did not find a solution there. What is new in our approach is {\it continuity}, an idea developed for  gauge theory in  \cite{Unsal:2007vu,Unsal:2007jx, Argyres:2012vv,Poppitz:2011wy}. 
   
 {\bf Main underlying idea of continuity: } 
Compactify the theory  on  $\R \times  \mathbb S^1_L$,  and find the conditions under which there are 
  no phase transitions  
or rapid crossovers as the  $\mathbb S^1_L$ size is dialed to small values. 
  Since we are reducing the theory to simple quantum mechanics and to finite volume, a sharp phase transition would only occur in the infinite-$N$ limit. At finite-$N$, this phase transition 
 would turn into a rapid cross-over, which is an equally drastic change of the long-distance theory. 
 Our continuity argument avoids both types of drastic change in the dynamics and smoothly connects the small-$L$ regime to the large-$L$ regime. 
 Since the theory is asymptotically free,  at sufficiently small $L$, the theory is rendered weakly coupled, and since there is no phase transition or rapid cross-over,  the non-perturbative phenomena must also be continuous. For example, for the mass gap, there cannot be any drastic changes, and indeed, we will demonstrate this non-trivial claim explicitly.  Then, the non-semi-classical notion of an IR renormalon on $\R^2$  {\it must}  find a semi-classical realization on $\R \times  \mathbb S^1_L$. In this way, we may indeed find a semi-classical configuration continuously connected to the 
 renormalon singularity.  This is the main and simple physical idea in our formalism.

We have already shown that in the semi-classical regime, there are kink-instanton configurations 
$\K_i$  with action $\frac{S_I}{N}$. However, there is no ambiguity associated with these configurations, and their topological charge  or  topological $\Theta$ angle dependence  suffice to distinguish them from the perturbative vacuum. Hence, the kink-instantons are {\it not} the realization of the IR -renormalons. Rather, the IR renormalons  must be  associated with certain  topological configurations (or  {\it molecules}), 
indistinguishable from perturbation theory. 
\begin{figure}[htbp]
\begin{center}
\scalebox{1.2}{
\rotatebox{-90}{
\includegraphics[width=6cm]{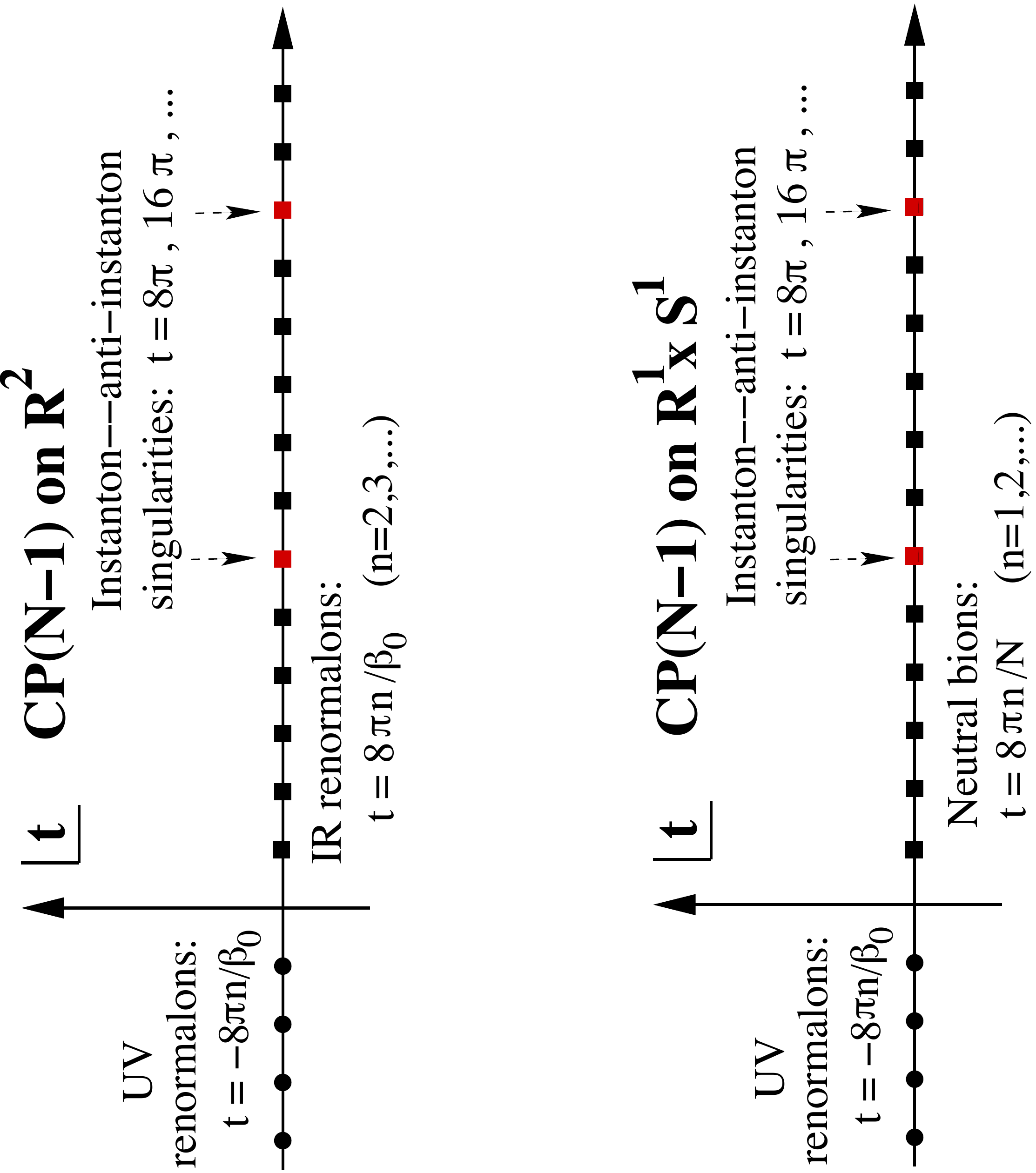}}}
\end{center}
\caption{
Upper figure: The conjectured structure of the Borel plane for
$\C\P^{\rm N-1}$
on $\R^2$. Lower figure:  The semi-classical singularities  associated
with the neutral bion molecules in
$\C\P^{\rm N-1}$   on small  $\R \times S^1$.   For $N_f=0$, the
weak-coupling regime has an extra singularity closer to origin than
the leading renormalon pole on $\R^2$.   For $N_f \geq 1$,  the
location of the semi-classical  and non-semi-classical renormalon
singularities  coincide.
Although the theory moves from a weakly coupled description to a
strongly coupled one, the structure of the Borel plane singularities
either do not change at all or change extremely mildly.  We take this
as  evidence that  the neutral bion molecules are the semi-classical
realization of renormalons. This also gives us hope that even the
theory on
$\R^2$ may potentially be solvable at arbitrary $N$.
}\label{fig:Borel}
\end{figure}

In the next section, we give a classification of $n$-defects (1-defects are kinks.). We show that 
there is an ambiguity in the semi-classical amplitude of certain  $n$-kinks,  which is  identifiable with the infrared renormalon singularities. It is not true that all $n$-kink configurations  are ambiguous. For example, in the list of topological configurations for the bosonic $N_f=0$ theory,  only the configurations with subscript $\pm$ have ambiguous imaginary parts,  and  the ambiguities that arise in the Borel summation of the perturbation theory cancel with the ambiguities of these molecular events. On the other hand, in a theory with fermions, the appearance of the first non-perturbative ambiguity is delayed by one order. A few examples of the topological configurations and  the  (non)existence  of their ambiguities are given in the following lists: 
\begin{align}
N_f =0: &\qquad \{\K_i,  [\cB_{ij}], \;   [\cB_{ii}]_{\theta=0^{\pm}},  \; \; [\cB_{ij} \cB_{ji}]_{\theta=0^{\pm}}, 
\;    [\cB_{ij} \cB_{jk}\cB_{ki} ]_{\theta=0^{\pm}}, \ldots, [ {\cal I}  \overline{\cal I}]_{\theta=0^{\pm}} , \ldots  \} \cr
N_f  \geq 1: &  \qquad \{\K_i,  [\cB_{ij}], \;   [\cB_{ii}],  \; \;  \qquad [\cB_{ij} \cB_{ji}]_{\theta=0^{\pm}}, 
\;    [\cB_{ij} \cB_{jk}\cB_{ki} ]_{\theta=0^{\pm}}, \ldots, [ {\cal I}  \overline{\cal I}] , \ldots  \} \
\end{align}
In other words, when $N_f=0$ we first see the non-perturbative ambiguities in the neutral bion amplitude $[\cB_{ii}]$, while for $N_f\geq 1$ the non-perturbative ambiguities first arise in the neutral correlators of two bions. The location of the ambiguities in the semi-classical molecules matches the location of the  renormalon singularities on   $\R^2$ for $N_f \geq 1$ theories, and for $N_f=0$, the semi-classics has an extra singularity closer to the origin than the leading renormalon pole on $\R^2$. See  Figure \ref{fig:Borel}.
  
The elegance of this analysis  is that a very difficult problem in QFT,  tied with the renormalon singularities, reduces to a relatively simpler  problem in quantum mechanics without much change in the structure of the Borel plane singularities.   The crucial physical elements permitting this analysis in QFT are continuity and compactification with spatially twisted boundary conditions.  

\subsection{Classification of bions and Cartan matrix} 
\label{sec:Bions}

The nodes of the extended Dynkin diagram of the $SU(N)$ algebra ${\hat A}_{N-1}$ provides a unique labeling of the kink-instanton events on $\R \times \mathbb S^1$.    As described earlier, 
the difference of the asymptotes of the kink event $\K_j$ is given by 
\be
\Delta \tn =  \tn (\infty) - \tn (-\infty) = \alpha_j \in \G_r^\v 
\label{charge2}
\ee
Since for a given kink-instanton  $\K_j$,  $\Delta \tn= \alpha_j $ cannot be  deformed   to zero, or 
to another $\alpha_{j'}$ where  $j' \neq j$, by small perturbations, the co-root  $\alpha_j \in \G_r^\v $ must be considered as  a topological charge.\footnote{This is the counterpart of the magnetic charge of monopole-instanton. 
Recall that  a monopole-instanton on $\R^3 \times S^1$ is  labelled by two topological charges:  the second Chern number, which is   $\frac{1}{N}$, and a magnetic charge $\alpha_j \in \G_r^\v$. Reducing QFT on $\R^3 \times S^1$ down to  quantum mechanics  on $\R \times T^2 \times S^1$, we may define the magnetic flux change associated with 
the monopole events as  $\Delta \Phi = \int_{T^2} B \in \G_r^\v$ in quantum mechanics as well. 
Identifying $\Delta \Phi$ with   $\Delta \tn$, this  compactification can be used to find an exact mapping among topological configurations between  \cpn and gauge theory. This is actually the primary reason why the classification of the topological configurations in compactified gauge theory and \cpn model are identical.}

The study of the topological molecules in the weakly coupled domain of the   $\C\P^{\rm N-1}$  model 
on $\R \times \mathbb S^1$  is parallel  to the study of the same class of molecules in center-symmetric QCD(adj) on  $\R^3 \times \mathbb S^1$. Here, we follow  Ref.\cite{Argyres:2012ka}.
 As in gauge theory, there exists a one-to-one mapping between the molecules at second order in the semi-classical expansion and non-vanishing entries of the extended Cartan matrix, $\hat A_{ij} := (\alpha_i,\alpha_j)$, $i, j =1, \ldots, N$.   The extended Cartan matrix determines the interaction between kinks of different types, and plays a crucial role in classifying the molecular kink-instanton/kink-anti-instanton  events. For brevity and due to their similarity to bions in gauge theory, we also refer to these {\it universal}  correlated events that appear in the second order in the semi-classical expansion as ``bions''.  As in gauge theory, there are two types of bions:
\\

{\bf {Definition 4: Charged bions and neutral bions}}

\begin{itemize}
\item  {\it Charged bions:} For each pair $(i,j)$ such that the entry of the Cartan matrix is negative 
$\hat A_{ij}<0$ (as a result, the bosonic interaction $V_{ij} \sim - \hat A_{ij}<0$   is repulsive at short separations),  there exists a bion $[ {\K}_i \bar\K_j]$,  associated with the 
 correlated tunneling-anti-tunneling 
 event
\be 
\tn \longrightarrow \tn + \alpha_i - \alpha_j    \qquad  \alpha_i \in \G_r^\v
\ee
The amplitude associated with such an event is 
\begin{eqnarray}
\label{bion1}
{\cal B}_{ij} = [ {\K}_i \bar\K_j] \sim  
e^{-S_i(\varphi)-S_j(\varphi)} e^{i \sigma(\alpha_i -\alpha_j)}\quad .
\end{eqnarray}  
There is no ambiguity for a charged bion, as  shown in  Section \ref{sec:2defects}. This is the counter-part of the magnetic bion in compactified gauge theory. 

\item {\it  Neutral bions : }  For each $i$,  such that the entry of the Cartan matrix is positive 
$\hat A_{ii}>0$  (as a result, the bosonic interaction $V_{ii} \sim - \hat A_{ii} < 0$   is attractive  at short separations),        there exists a bion $ [\K_i\bar\K_i]$ with  vanishing topological charge and  associated with the  correlated tunneling-anti-tunneling  event of the same type
\be 
\tn \longrightarrow \tn + \alpha_i - \alpha_i   \qquad  \alpha_i \in \G_r^\v
\ee
The real part of the amplitude for the neutral bion is unambiguous 
\begin{eqnarray}
\label{bion2}
\Re {\cal B}_{ii} = \Re [{\K}_i \bar\K_i] \sim e^{-2 S_i(\varphi)}.
\end{eqnarray}  
However, the  neutral bion amplitude develops an imaginary   ambiguous part, as will be discussed in Section \ref{sec:neutral}, along with its physical implications.

We define the  right (and left) neutral  bion amplitudes  as the  bion amplitude evaluated at $g^2 \pm i 0$ according to the continuation shown in Fig. \ref{fig:continuation}.
These   are unambiguous, but  the two differ by a non-perturbative jump $e^{-2 S_i(\varphi)}$ for the bosonic theory . This will be crucial in the non-perturturbative cancellation of ambiguities.  

\end{itemize}

The derivations of these assertions will be given in  Sections \ref{sec:2defects} and \ref{sec:neutral}.  For the \cpn theory, both tunneling events actually carry zero topological charge (\ref{charge})
\be 
Q_T=1/N + (-1/N)=0  \qquad, \qquad   \text {for both types of  bions}  
\ee
However,  the {\it charged} bion   is still associated with a co-root charge. Correspondingly, 
the charged bions  can be assigned a  more refined topological quantum number, physically associated with where it starts and ends in the $\tn$ landscape of classical vacua, which are different points. In contrast, the neutral bion starts at some configuration, tunnels in the   $\alpha_i$ directions and returns back to the original point.   
For the theory with fermions, this first ambiguity disappears for $N_f \in \Z^{+}$ and the first ambiguity appears for a 4-defect. 

\subsection{Zero, quasi-zero and non-zero modes of $n$-defects}
\label{sec:qzm}

In the semi-classical regime, the path integral can be formally rewritten as a sum over a dilute gas of 
$n$-defects. In Euclidean space where  1-defects can be seen as Euclidean particles,  $n$-defects  should be viewed as Euclidean molecules, hence the terminology  {\it topological molecules} that we use from time to time.  In Section \ref{kink-1}  we discussed 1-defects and their contribution  to the low-energy theory.  These 1-defects are solutions to a self-duality equation. They are exact solutions to the classical equations. The $n$-defects are only approximate quasi-solutions. 

In the background of a general $n$-defect, within  the path-integral formulation, one needs to perform 
a sum over all fluctuations, meaning that one needs to obtain information about the eigen-spectrum of fluctuations. In general semi-classical analysis, the eigen-spectrum has three types of modes:
\begin{itemize}
\item {Zero-modes: zero eigenvalues of the eigenspectrum,  associated with the moduli } 
\item{Quasi-zero modes:  parametrically small compared to typical eigenvalue }
\item{Non-zero modes:  Modes for which the eigenspectrum is of order one in natural units.} 
\end{itemize}
For 1-defects,  kinks, as discussed in Section \ref{kinkins}, there are only two moduli, the position moduli  $a \in \R$  and the angular moduli $\phi \in U(1)$; the latter is integrated trivially. Of course, there are also non-zero modes associated with the small fluctuations operator, in the background of the 1-defect. These modes and associated determinants can be dealt with in the Gaussian approximation. The quasi-zero mode does not exist for 1-defects. 

For $n$-defects with $n\geq 2$, all three types of modes exist.  The quasi-zero-modes need to be treated exactly within semi-classics.  The reason is as follows: there is a characteristic length scale entering the quasi-zero mode analysis. Denote this scale by  $\ell_{\rm qzm} $. As we will see,  this scale is much larger than the characteristic kink size $r_{\rm k}$ , but much smaller than the typical inter-kink separation $d_{\rm k-k}$: 
\begin{eqnarray}\label{hierarchy2}
r_{\rm k} \ll   \ell_{\rm qzm}   \ll d_{\rm k-k} 
\end{eqnarray}
The integral over the quasi-zero mode is mainly supported at the scale  $\ell_{\rm qzm}$.   Performing the quasi-zero-mode integral, one can treat the 2-defect as a point operator  when considering  physics at length scales larger than $\ell_{\rm qzm}$.  This is the way that we construct the bion operators ${\cal B}_{ij}$ and other $n$-defect operators. The reason that semi-classical analysis in compactified \cpn is reliable at small-$\mathbb S^1_L$ is the  hierarchy of length scales 
  \begin{eqnarray}\label{hierarchy}
\begin{matrix}
r_{\rm k}&\ll&r_{\rm b} \sim   \ell_{\rm qzm} &\ll&d_{\rm k-k}&\ll&d_{\rm b-b},\\ 
\downarrow   &&\downarrow&&\downarrow && \downarrow  \\
L&\ll& L\, \log\left(\frac{1}{g^2} \right)&\ll&L\, e^{S_0}&\ll& L\, e^{2S_0}.
\end{matrix}
\end{eqnarray}
 that arises from a careful treatment of the quasi-zero modes. In this formula,  $r_\text{b}$ is the size of a bion, and  $d_\text{b-b}$ is  the typical  inter-bion separation.
 This physical  hierarchy of length scales is also naturally built into the mathematical trans-series expansion.

\subsection{2-defects: Charged bions}
\label{sec:2defects}
The interaction (correlation)  between $\K_i$ and $\bar \K_j$ has two components. One is due to the exchange of the bosonic fields and the other is due to the exchange of fermion zero modes. 
Following the explanation in Section 5.1 of \cite{Argyres:2012ka}, and generalizing the Appendix A43.2 of Zinn-Justin's book \cite{zinn-book}, we find that the  interaction  induced by bosonic exchange  between $\K_i$ and $\bar \K_j$ kinks separated at a distance $\tau$ is given by 
\be 
S_{\rm int}(\tau) =  - 8  \xi  \frac{   \alpha_i. \alpha_j}{ g^2    } e^{-\xi \tau} , \qquad \xi \equiv  \frac{2 \pi (\mu_{i+1} - \mu_i) }{L}  = \frac{2 \pi}{LN}
\ee 
The   fermion zero-mode exchange induces an interaction between the two kinks as well, which can be read-off from the connected correlator:
\begin{align}
 \Big\langle
\prod_{f=1}^{N_f} [\a_i(\psi_f)]^2(t- \tau/2) \,
\prod_{f=1}^{N_f} [\a_j(\bar\psi_f)]^2(t+ \tau/2) \Big\rangle 
=  \left(\frac{\alpha_i. \alpha_j}{2}\right)^{2N_f}  \left( \frac{g^2}{2L}\right)^{2N_f}  e^{-2N_f \xi \tau}
\nonumber
\end{align}
 Consequently, the bion  amplitude may be written as in (\ref{bion1}) with coefficient involving an integral over the quasi-zero mode (separation between the  two events.)
\begin{align}\label{bioncoeff}
\cA_{ij} =
\cA_i\cA_j\,   \left(\frac{\alpha_i. \alpha_j}{2}\right)^{2N_f}  \left( \frac{g^2}{2L}\right)^{2N_f}
 2 \int_{0}^{\infty}  d\tau  \, e^{-V_\text{eff}^{ij} (\tau)},
\end{align}
where 
\begin{align}\label{mmbar}
V_\text{eff}^{ij}(\tau) =  - 8  \xi  \frac{   \alpha_i. \alpha_j}{ g^2    } e^{-\xi  \tau}  + 2N_f \xi \tau 
\end{align}
The factor of $2$ in (\ref{bioncoeff}) comes from the integration over the solid angle in 1d, 
$\int d\Omega =2$, as the interaction of the constituents of the bions only depends on separation. 
For $N\geq 3, \hat A_{ij} = -\frac{1}{2}$ for non-vanishing off-diagonal entries of the extended Cartan matrix  (for $N=2$, $\hat A_{ij} = -1$),   the quasi-zero mode integral given in (\ref{bioncoeff}) is 
 \begin{align} 
 I(g^2) &=  \int_{0}^{ \infty }  d\tau   \exp \left[ - \left( \frac{  4  \xi  }{ g^2    } e^{-\xi \tau} +  2N_f \xi \tau  \right)  \right]  = \left(\frac{g^2}{4 \xi} \right)^{2N_f} \int_0^{\frac{4 \xi}{g^2} } du  \; e^{-u} \; u^{2N_f-1} \cr
 &  \underbrace{\longrightarrow}_{g^2 \ll1 } 
  \left(  \frac{g^2}{4 \xi} \right)^{2N_f} \Gamma(2N_f) =   \left(  \frac{g^2N}{8 \pi } \right)^{2N_f} \Gamma(2N_f) 
  \label{quasi}
  \end{align}  
  The charged-bion amplitude is, therefore, 
  \begin{align}\label{bion12}
\cB_{ij} = -\cA_{ij} e^{-S_i(\f)-S_j(\f)} 
e^{2\pi i \s(\a_i^\v-\a_j^\v)} 
\end{align}
Various comments are in order regarding (\ref{quasi}): 
  For \cpn and $N_f \geq 1$,  for  non-vanishing negative 
entries of the extended Cartan matrix,  $ \alpha_i. \alpha_j <0$, the bosonic interaction between the constituents is repulsive, whereas  the fermion zero mode induced exchange is attractive. Therefore, there is a characteristic scale dominating the integral. 
\begin{align}
V_\text{eff}^{'} (\tau)  = 0 \quad \Longrightarrow \quad   \tau^* =  \frac{1}{\xi} \log \left(\frac{4\pi }{g^2N N_f} \right)  \qquad, \qquad r_{\rm b} = r_{\rm k} \log \left(\frac{4\pi }{g^2N N_f} \right)   \qquad N_f \geq 1
\label{sizeb}
\end{align}
We interpret this as the size of the charged bion.  

The integral  (\ref{quasi})  appeared in the study of molecular instantons in supersymmetric and non-supersymmetric quantum mechanics in \cite{Balitsky:1985in}.  For $N_f=0$, where the integral at
large-$\tau$ is not cut-off, this appeared already in the work of Bogomolny and Zinn-Justin \cite{Bogomolny:1980ur, ZinnJustin:1981dx} in bosonic quantum mechanics.
In this case,  Bogomolny \cite{Bogomolny:1980ur} realized that the  divergence arises from the double-counting of uncorrelated  instanton-anti-instanton events at large separations  and upon a careful treatment of the partition function, this divergence subtracts off as 
$I(g^2) =  \int_{0}^{ \infty }  d\tau  \left[ \exp \left[ - \left( \frac{  4  \xi  }{ g^2    } e^{-\xi \tau}   \right)  \right] -1 \right]$. Using this integral, let us evaluate the  the size of the charged bion in the case $N_f=0$. 
Using integration by parts,  
$I(g^2) =    \frac{  4  \xi^2  }{ g^2    }   
\int_{0}^{ \infty }  d\tau  \tau   \exp \left[ - \left( \frac{  4  \xi  }{ g^2    } e^{-\xi \tau}   + \xi  \tau \right)  \right]  $.
Following (\ref{sizeb}), the characteristic size of the charged bion in the  bosonic theory 
  is given by 
\begin{align}
r_{\rm b} = r_{\rm k} \log \left(\frac{8\pi }{g^2N} \right) \; ,   \qquad N_f=0
\label{sizeb2}
\end{align}

The result of the quasi-zero mode integral can be found either by working out this integral exactly as done in 
bosonic quantum mechanics \cite{Bogomolny:1980ur}, or 
via an  equivalent prescription:   take the $N_f=\epsilon \rightarrow 0$ limit  in (\ref{quasi}) and subtract the pole term: 
 \begin{align} 
 I(g^2 ) =    \left(  \frac{g^2N}{8 \pi } \right)^{2\epsilon} \Gamma(2\epsilon) = \frac{1}{2 \epsilon} + \left(  \log   \left(  \frac{g^2N}{8 \pi } \right)  - \gamma \right) + O(\epsilon)  \longrightarrow  
 \left(  \log   \left(  \frac{g^2N}{8 \pi } \right)  - \gamma \right) 
  \label{quasi0}
  \end{align}  
  The two result indeed agree exactly.

The mechanism of pairing in \cpn with $N_f\geq 1$ is  the same as in QCD(adj)  with $N_f\geq 1$ \cite{Unsal:2007jx}, as well as quantum mechanics with fermions  \cite{Balitsky:1985in};  and for  
 $N_f=0$,   the deformed bosonic \cpn, the pairing takes place in the same way as in deformed YM 
\cite{Unsal:2012zj}, as well as bosonic quantum mechanics \cite{Bogomolny:1980ur}. This mechanism 
is a  universal feature  of semi-classical analysis.  

\subsection{2-defects: Neutral bions and first non-perturbative ambiguity}
\label{sec:neutral}
For each diagonal entry  of the Cartan matrix, there exists a {\it neutral} bion  $\cB_{ii} = [\K_i \bar \K_i]$.   In our Lie algebra convention, $\hat A_{ii}=1$,  and the quasi-zero mode integral takes the form 
 \begin{align} 
 \widetilde I(g^2) &=  \int_{0}^{\infty}  d\tau   \exp \left[ - \left(  -  \frac{  8 \xi  }{ g^2    } e^{-\xi \tau} +  2N_f \xi \tau  \right)  \right]    
 \end{align}
 However, as in gauge theory, both  the bosonic interaction  as well as the fermion zero mode induced interaction between constituents are actually attractive, and  naively, the integral is dominated  at  small separations with respect to $r_{\rm b}$.  Consequently,  a semi-classical  $[\K_i \bar \K_i]$ configuration seems meaningless. A related issue is that  all topological quantum numbers that we may associate with $\cB_{ii} $ are actually  zero. It is indistinguishable from the perturbative vacuum, and potentially may mix with the perturbative contribution to a given observable. Understanding this precise connection leads to a quantitative  and rigorous theory of semi-classics.

 In order to make sense out of the $\cB_{ii} $ molecule, we apply the generalized BZJ-prescription: 
 deform the contour of integration so that the kink anti-kink has a repulsive component, or equivalently, rotate $g^2 \rightarrow g^2 e^{i \theta}$, and take, for example, $\theta=\pi$. Then, again, the interaction has a  repulsive component.  Perform the integration and then, continue back to the original $g^2$. The result of the BZJ-prescription is, 
\begin{align}\label{sign}
\til I(g^2,N_f) \to I(-g^2,N_f) =    \left( - \frac{g^2N}{8 \pi } \right)^{2N_f} \Gamma(2N_f) 
\end{align}  
For positive integer number of flavors  $N_f \geq 1$, this result is unambiguous. 
 However, for $N_f=0$, subtracting the pole due to the uncorrelated kink-anti-kink events, we obtain 
   \begin{align} 
 \til I(g^2, N_f=0 )  =     \left(  \log   \left(  - \frac{g^2N}{8 \pi } \right)  - \gamma \right)  = I(g^2) \pm i \pi
  \label{quasi02}
  \end{align}  
  The same ambiguity is obtained by Bogomolny in bosonic quantum mechanics 
  \cite{Bogomolny:1980ur}.  
  Thus, we learn that the kink-anti-kink amplitude in the bosonic theory is two-fold ambiguous. The left and right bion amplitude is therefore 
     \begin{align} 
  [\K_{i}  \bar \K_{i} ]_{\th=0^{\pm}} =& \Re\,[\K_{i}  \bar \K_{i} ] + i \,
   \Im\, [\K_{i}  \bar \K_{i} ]_{\th=0^{\pm}}  \cr
= &  \left(  \log   \left(   \frac{g^2N}{8 \pi } \right)  - \gamma \right) 
2 {\cal A}_i^2 
  e^{-2S_0}  \pm i \pi  2 {\cal A}_i^2  e^{-2S_0} 
    \qquad 
    \label{firstambiguity}
     \end{align}
This is the first of many non-perturbative ambiguities that we will see in the semi-classical analysis.  On its own, such ambiguities are disastrous, as they would render the semi-classical expansion meaningless. 
However, this ambiguity, and the many other ambiguities that we will find in semi-classical configurations  are actually  the resolution of a long-standing puzzle, the IR-renormalons in perturbation theory around the perturbative vacuum.

Consider a typical observable in the  bosonic  \cpn theory.  This  observable will receive contributions to all orders in perturbation theory, and also  non-perturbative contributions. Let us denote the  lateral  (left and right)  Borel summation  for perturbation theory by  $\B_{0, \th=0^\pm}$.  Let us write $g^2 = |g^2| e^{i \th}$, where $\th$ is the phase of the complexified coupling. In order for QFT to make sense, these two types of ambiguities must cancel: 
\begin{align}\label{borelbion}
\Im\,\B_{0, \th=0^\pm}  + \Im\, [\cB_{ii}]_{\th=0^\pm}
=0\; ,\quad {\rm up \; to}  \; e^{-4S_0}
\end{align}
In words, this equation means: The sum of the left  (right) Borel resummation and left (right) neutral  bion amplitude is unambiguous at order  $e^{-2S_0}= e^{-2S_I/N}$.  The limit $\theta \rightarrow 0^{\pm}$  
is accompanied by a Stokes jump for the Borel resummation, which is mirrored with a jump in the neutral 
bion amplitude, such that the sum of the two  gives a unique result, with a smooth limit up to ambiguities at order $e^{-4S_0}$.   We will indeed confirm this important confluence relation by explicit computation in two different ways.   As a physical effect,  the neutral bion amplitude leads to a repulsion
between the eigenvalues of the $\s$-connection holonomy, as manifest from the supersymmetric example (\ref{bosonic}). This is the same as the role that the neutral bion plays in non-abelian gauge theory \cite{Poppitz:2012sw,Argyres:2012ka}.

  \subsection{4-defects: Bion-anti-bion molecules and more ambiguities}
  \label{sec:4defects}
 According to the general classification stated in \S 5.5 of Ref. \cite{Argyres:2012ka}, in theories with massless fermions, both   charged    and neutral bion events are unambiguous. 
 In theories with $N_f \geq 1$,   the first ambiguity in semi-classical expansion  arise at $4^{\rm th}$ order,  as opposed to $2^{\rm nd}$ order as it was the case in bosonic theory. Below, we show that the quasi-zero mode integral does not yield an imaginary part for  $[\cB\cB]$, but does yield an imaginary part for $[\cB\bar\cB]$. The quasi-zero mode integrals are of the form 
\begin{align}
&\label{bb}   
I (g^2) =  \int_0^{\infty}  d\tau\, \exp\left(- V(\tau)  \right)    
\qquad \text{for $[\cB\cB]$, and}\\
\label{bbbar}  
& \til I(g^2) = \int_0^{\infty}  d\tau\, \exp\left(+ V(\tau)  \right)    
\qquad \text{for $[\cB\bar\cB]$},
\end{align}
where 
\begin{align}\label{bbpot}
V(\tau) = (\m_{\cB} , \m_{\cB})  \frac{  8 \xi  }{ g^2    } e^{-\xi \tau} 
\end{align}
and $\m_\cB = \a_i - \a_j \in  \G_r^\v$ is the charge of the  bion $\cB_{ij}$.  

This type of  integral, as noted earlier, is addressed in bosonic quantum mechanics by Bogomolny \cite{Bogomolny:1980ur}.  Both integrals are divergent at large separation, and the latter is dominated by $\tau \rightarrow 0$ where molecular configurations are meaningless.  The first of these problems is due to double-counting of the uncorrelated $[\cB]$-$[\cB]$ or $[\cB]$-$[\bar\cB]$ events, and is  subtracted off. 
 
  \begin{figure}[htb] 
  \centering{\includegraphics[scale=0.3]{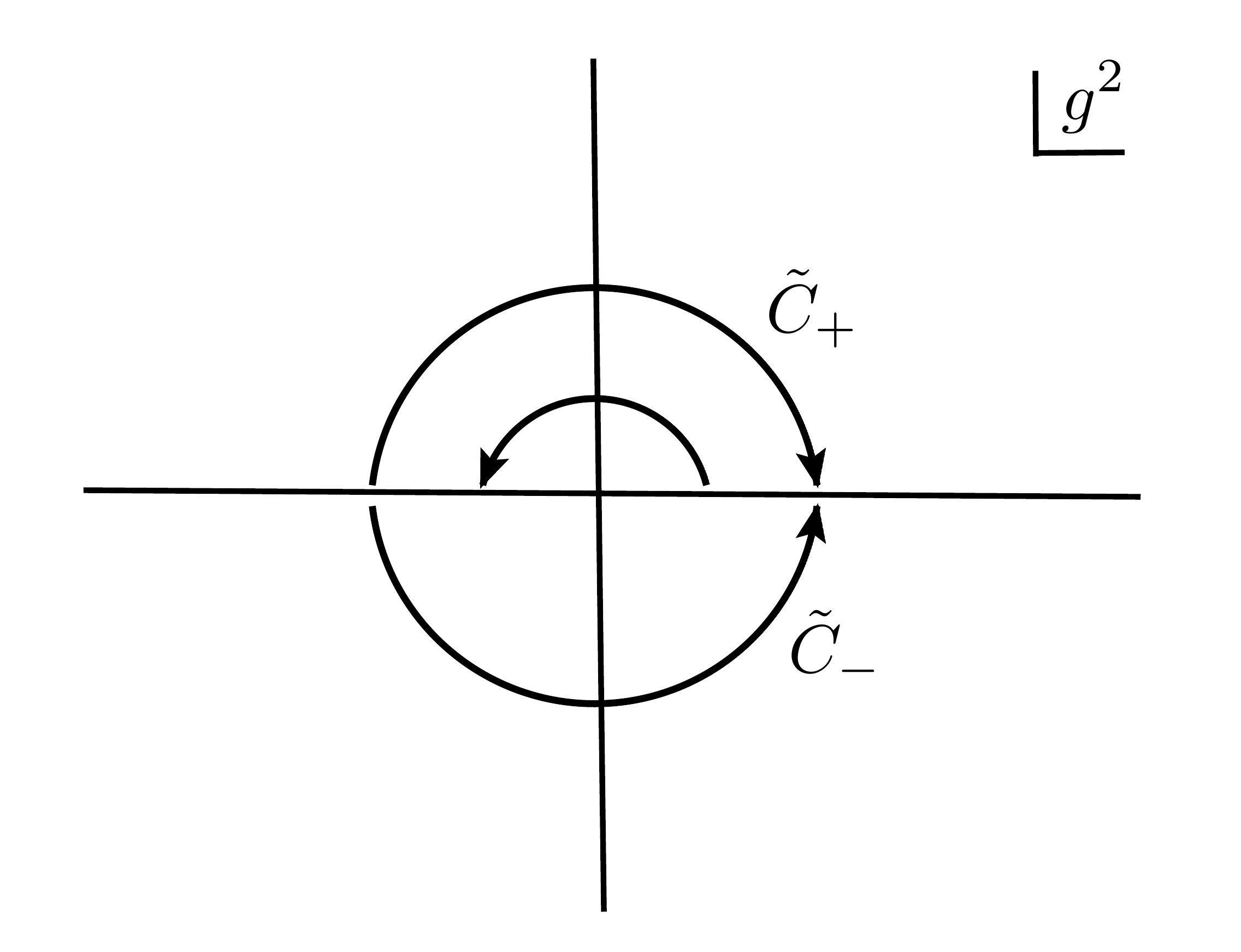}}
  \caption{Defining left (right) bion-anti-bion amplitude   $ [\cB_{ij}  \bar \cB_{ij} ]_{\th=0^{\pm}}$, we proceed as in the construction of left (right)  Borel resummation  $\B_{0, \th=0^\pm}$.} 
  \label{fig:continuation2}
   \end{figure}

 The short-distance domination of $\til I(g^2)$ can be taken care of by modifying the integration contour, or by rotating $g^2 \rightarrow -g^2$, where the bion-anti-bion interaction becomes repulsive, and continuing the integral back to positive $|g^2| + i 0^{\pm}  $.  The result, as was the case with (\ref{firstambiguity}), is two-fold ambiguous:
    \begin{align} 
  [\cB_{ij}  \bar \cB_{ij} ]_{\th=0^{\pm}} = \Re \,    [\cB_{ij}  \bar \cB_{ij} ]  
  + i \,
   \Im \,[\cB_{ij}  \bar \cB_{ij} ]_{\th=0^{\pm}}  \sim  e^{-4S_0}  \pm i\, \pi \, e^{-4S_0} 
    \qquad 
    \label{secondambiguity}
     \end{align}
     Consider a typical observable in  \cpn theory with $N_f \geq 1$ fermions. We expect that this observable will receive contributions to all orders in perturbation theory, as well as  non-perturbative contributions. Denote the  lateral  Borel summation  for perturbation theory by  $\B_{0, \th=0^\pm}$.  Then write $g^2 = |g^2| e^{i \th}$, where $\th$ is the phase of the complexified coupling. For QFT to make sense, these two ambiguities must cancel:
\begin{align}\label{borelbion2}
\Im\, \B_{0, \th=0^\pm}  
+ \Im\,  [\cB\bar \cB]_{\th=0^\pm}
=0 \; , \quad {\rm up \; to}  \,\, e^{-6S_0}
\end{align}
This confluence relation is the counter-part of the leading ambiguity cancellation (\ref{borelbion}) in the $N_f=0$ theory to $N_f\geq 1$.  In the next sections, we explicitly derive (\ref{borelbion}). 

\section{Resurgence in \cpn QFT}
\label{sec:resurgence-cpn}
\subsection{Borel-\'Ecalle summability at leading order}
 \label{sec:atwork}
 We derived the actions describing the low energy dynamics in Eqs. (\ref{zeroth}) and (\ref{first}) and described 
 the embedding of the \cpone\;  kink   into \cpn theory in Section \ref{subsec:ki-cpn}. The Euclidean action describing this embedding is 
 given in (\ref{kth}). Passing to a Minkowskian formulation, we  write down the Hamiltonian 
 associated with the action  (\ref{kth}). It is given by 
  \begin{equation}
H^{\rm  zero}_{\a_k}=    \textstyle  \frac{g^2}{2} P_\theta^2+ 
 \frac{\xi^2}{2g^2}  \sin^2  \theta    +   \frac{g^2}{2 \sin^2 \theta} P_\phi^2, \qquad  \xi= \frac{2 \pi}{N},   \qquad  ({\rm set} \; L=1)
\label{zero3}
\end{equation} 
We are interested in  the ground state properties of this Hamiltonian.  The field $\phi$ is a cyclic coordinate and 
  $P_\phi$ is the associated angular momentum with eigenstates $m_\phi =0, \pm 1, \pm 2, \ldots$. The ground state in the $\phi$-sector is $m_\phi=0$.  As we will justify {\it a posteriori}, the energy gap  of low lying modes is non-perturbative in $g$: it is $e^{-\frac{4 \pi}{g^2N}}$. Therefore,  
  within the   Born-Oppenheimer approximation, we drop  
  the  high $\phi$-sector modes. 
    Then, the relevant  low energy Hamiltonian reduces to 
   \begin{equation}
H^{\rm  zero}_{\a_k}  =   - \textstyle  \frac{1}{2}   \frac{d^2 } {d\theta^2}+ 
 \frac{\xi^2}{4g^2}  [1- \cos (2 g  \theta) ]   
\label{zero3b}
\end{equation} 
and the Schr\"odinger equation takes the form 
\be
\psi'' + \left( p + \frac{\xi^2}{2g^2}   \cos (2 g  \theta) \right) \psi=0,  \qquad p = 2E - \frac{\xi^2}{2g^2}
\ee
The asymptotic perturbative expansion for the ground state energy in units of the natural frequency $\xi$ is  evaluated in Ref.\cite{Stone:1977au}  by using the methods developed by  Bender and Wu \cite{Bender:1969si}
 \begin{equation}
 {\cal E}(g^2)  \equiv 
E_0 \xi^{-1}   = \sum_{q=0}^{\infty} a_{q} {(g^2)}^{q}, \qquad a_q \sim - \frac{2}{\pi} \left( \frac{1} {4\xi}\right)^{q} q! \left( 1 - \frac{5}{2q}  + O(q^{-2}) \right)  
\label{bw}
 \end{equation}
The series is Gevrey-1, non-alternating, and hence non-Borel summable.  This is a manifestation of the fact that we are expanding the ground state energy  along a Stokes ray in the complex-$g^2$ plane. 
\begin{figure}[htb]
\centering{
\includegraphics[scale=0.4]{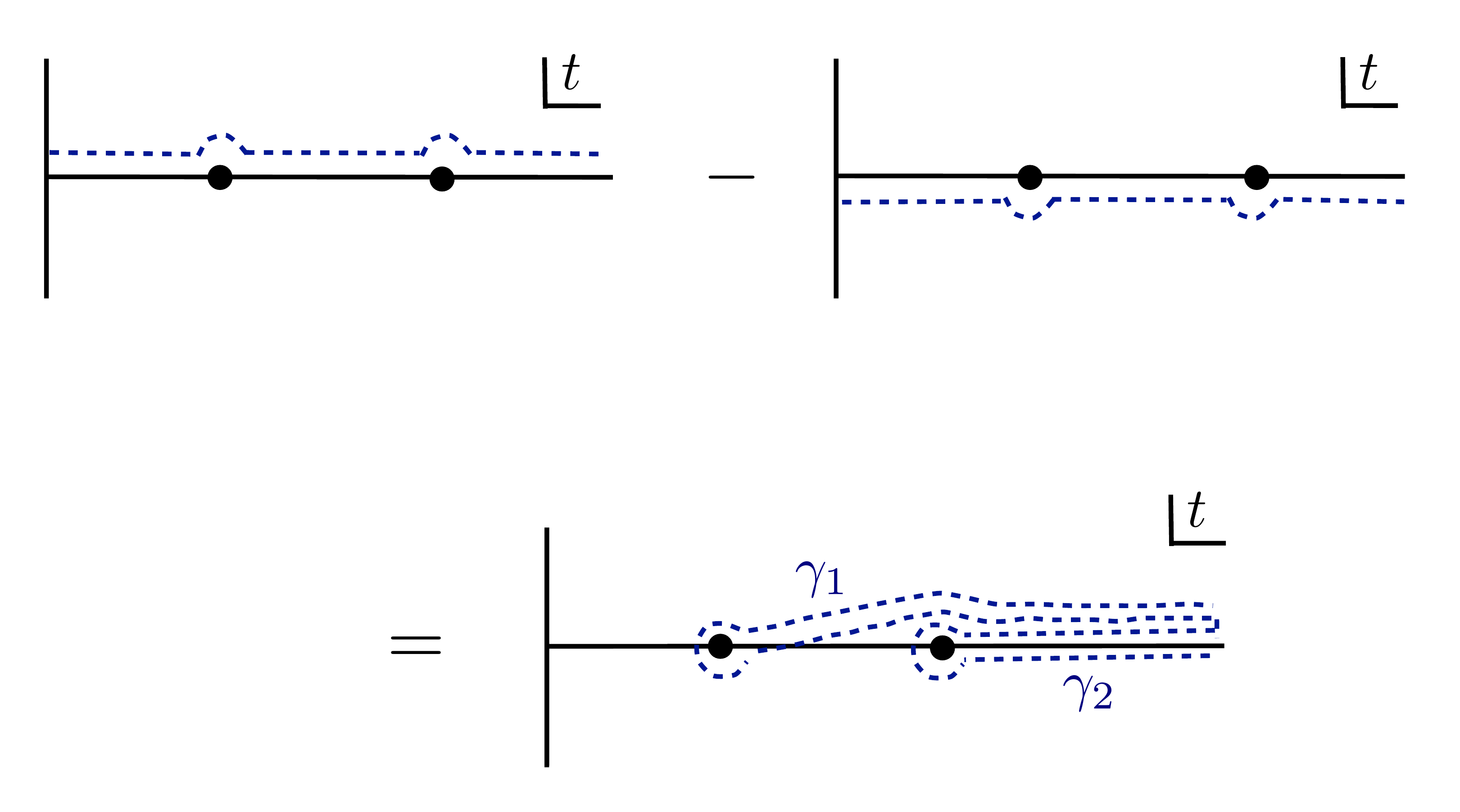}}
\caption{The difference between the left and right Borel sum defines the Stokes automorphism, as a sum over Hankel contours.  This is the discontinuity in perturbation theory. It  is  cancelled by the discontinuity of the neutral bion and bion-anti-bion amplitudes. } 
\label{fig:Hankel}
\end{figure}
The Borel transform (for the leading $q!$ divergence)  is given by 
\be
B   {\cal E} (t) = 
- \frac{2}{\pi}  \sum_{q=0}^\infty \left(  \frac{t}{ 4 \xi} \right)^q   
 = - \frac{2}{\pi} \frac{1}{1- \frac{t}{4 \xi}}  
\ee
and has a pole  singularity   on the positive real axis $\R^{+}$.  The transform of the subleading term $(q-1)!$ generates a $\log(1- \frac{t}{4 \xi})$ branch point at the same position. Hence, the series is non-Borel summable. 
 However, the series is  {\it right and left Borel resummable}. These 
 lateral Borel sums are 
 \begin{align} 
{\cal S}_{0^\pm }  {\cal E}  (g^2) & =  \frac{1}{g^2} \int_{C_{\pm}} dt  \; B {\cal E}  (t)  \;  e^{-t/g^2} 
= 
 \Re {\cal S}  {\cal E} (g^2)  \mp i  \frac{8 \xi}{  g^2} e^{-\frac{4 \xi}{g^2}}  \cr
&= \Re  \B_0   \mp  i  \frac{16 \pi}{  g^2N } e^{-\frac{8 \pi}{g^2N}} 
\end{align}
The real part of the lateral Borel sum $ \Re  \B_0 $  is unique and is given by Cauchy's principal part. 
The lateral  resummed energy has a two-fold ambiguous imaginary part, depending on the choice of path. It is important to note that the  imaginary part is {\it not} associated with an {\it  instability}. 
 The spectrum of the Hamiltonian is bounded from below, and  is real.  The interpretation of this result is that  since the lateral resummed  perturbation theory does not have  a smooth limit in the $ \theta \rightarrow 0^{\pm} $ limit, it cannot be used to define the full theory.  In other words, all orders perturbation  theory is not equal to full quantum field theory, or even to full quantum mechanics.
 
The Stokes automorphism (\ref{stokesauto}) connecting different sectorial sums is  defined as the difference between the two lateral Borel sums, and is schematically shown in  Fig. \ref{fig:Hankel}. 
The difference 
\begin{align}
{\cal S}_{0^{+}}  {\cal E} (g^2)  - {\cal S}_{0^{-}}   {\cal E} (g^2)  =  \frac{1}{g^2} \int_{C_{+}}   B {\cal E}  (t)  \;  e^{-t/g^2}  -  \frac{1}{g^2} \int_{C_{-}}   B {\cal E}  (t)  \;  e^{-t/g^2}  = \sum_i  \int_{\gamma_i}  B {\cal E}  (t)  \;  e^{-t/g^2} \qquad
\end{align} 
where  $\{\gamma_1, \gamma_2, \ldots\} $ are the  Hankel contours associated with singularities. 
In our case, the leading factorial growth of the perturbation theory yields
\begin{eqnarray}
{\cal S}_{0^{+}}  {\cal E} (g^2)  - {\cal S}_{0^{-}}   {\cal E} (g^2)  = &&   -
  i  \frac{32 \pi}{  g^2N } e^{-\frac{8 \pi}{g^2N}} 
\label{disc2} 
\end{eqnarray}
Recall from (\ref{firstambiguity}) that the non-perturbative neutral bion amplitude is also two-fold ambiguous:
    \begin{align} 
  [\K_{i}  \bar \K_{i} ]_{\th=0^{\pm}} =& \Re\,[\K_{i}  \bar \K_{i} ] + i \,
   \Im\, [\K_{i}  \bar \K_{i} ]_{\th=0^{\pm}}  \cr
= &  \left(  \log   \left(   \frac{\lambda}{8 \pi } \right)  - \gamma \right) 
\frac{16 }{\lambda}
  e^{-2S_0}  \pm  i  \frac{16 \pi}{\lambda}   e^{-\frac{8 \pi}{\l}} 
    \qquad 
    \label{bion-ambiguity}
     \end{align}
     Both of these ambiguities, as already emphasized, are exponentially more important than the 
     2d instanton-anti-instanton ambiguity.  Remarkably, they   cancel each other {\it exactly}:
         \begin{align} 
\Im \Big[   {\cal S}_{\pm }  {\cal E}  (g^2)   +  [\K_{i}  \bar \K_{i} ]_{\th=0^{\pm}}  \Big]= 0  \qquad \qquad  {\rm up\;  to}   \; e^{-4S_0}  = e^{-4S_I/\beta_0}  
\end{align}
leading to a cancellation of ambiguities up to  $e^{-4S_I/\beta_0}$.   This is an explicit realization of how confluence equations work in the small-${\mathbb S^1}$ regime of  a non-trivial QFT, the \cpn model.

\subsection{In the reverse direction: dispersion relations }
\label{sec:ecalle}

Consider a typical  observable in quantum theory, such as the energy eigenspectrum. 
For concreteness, let ${\cal E}_n(\l)$  be the energy of the state with quantum number $n$  ($n$ may be a collective quantum number;  we do not make such a distinction yet) 
 in units of natural frequency $\xi$ 
as defined in (\ref{bw}). Also, assume that we do not have knowledge of the asymptotic expansion for the energy of the state $n$. And in fact, we will  derive this asymptotic expansion by using the confluence equation (\ref{borelbion}), together with dispersion relations.  Let
 \begin{equation}
 {\cal E}_n(\l) = a_{n,0} +  a_{n,1}\l  +  a_{n,2} \l^2 + \ldots   = \sum_{q=0}^{\infty} a_{n,q} \l^{q}
 \end{equation}
We recall the  dispersion relation from Cauchy's theorem, and explore its consequences in connecting perturbative to  non-perturbative physics \cite{Bender:1969si,arkady,zinn-book}. We assume that  ${\cal E}_n(\l)$ is an analytic function in the cut-plane, where the function has a branch-cut along the positive $\l$ axis.  Cauchy's theorem and a contour deformation  relate the discontinuity, the imaginary part and large-orders in perturbation theory.   
As a result of the confluence equation (\ref{borelbion}), we already have knowledge of  the discontinuity in  ${\cal E}_n(\l)$, and how it  is connected to the discontinuity in the neutral bion amplitude.  
Using Cauchy's theorem provides a prediction for the large-order behavior of perturbation theory, which we test against the Bender-Wu method  applied to the  (reduced) quantum mechanics of   \cpn theory.  

For an analytic function  $f(\l)$ in a cut-plane,    Cauchy's theorem and a contour deformation implies
\begin{align}\label{dispersion1}
f(\l) = \frac{1}{2\pi i} \int_0^\infty d\l' 
\frac{\text{Disc}f(\l')}{\l' - \l}  
- \frac{1}{2\pi i} \oint_{C_\infty} \frac{f(\l')}{\l' - \l}
\end{align}
where $C_\infty$ is a loop at infinity, and $\l$ is a point off the positive real axis.   If $f(\l)$ decays sufficiently fast as $|\l| \rightarrow \infty$,   the last term  drops out.   However, this is not the case  for the energy spectrum,  ${\cal E}_n(\l)$.  Formally,  as $\l \rightarrow \infty$  in the low-energy quantum  mechanics,   the potential term serves as a small perturbation. The system becomes a free particle on a circle with  circumference   $\ell \sim \frac{\pi}{\sqrt \l}$. Thus, the energy levels of the theory are $E_n(\l) \sim \frac{1} {\ell^2} \sim  \l$,  including the ground state energy. We can build an auxiliary function  $f_{\rm aux}(\l)$ for which the boundary term drops out of the equation.  Construct
\begin{align}\label{subtraction}
f_{\rm aux}(\l) =  \frac{1}{\lambda^2} \left( E_n(\l) - a_0 -a_1 \lambda   \right)  =  \sum_{q=0}^{\infty} a_{n, q+2} \l^q 
\end{align}
We divided by  $\l^2$ so that  as $|\l | \rightarrow \infty$, $f_{\rm aux}(\l) \sim 1/\l$  and the  $C_\infty$  term vanishes. We subtracted  two  terms so that we do not generate an undesired  poles for $f_{\rm aux}(\l)$ at the origin, see \cite{zinn-book}. As a result, we have  a more useful form of the dispersion relation:
\begin{align}\label{dispersion2}
f_{\rm aux}(\l) = \frac{1}{2\pi i} \int_0^\infty d\l' 
\frac{\text{Disc}f_{\rm aux}(\l')}{\l' - \l}  
\end{align}
Plugging (\ref{subtraction}) into (\ref{dispersion2}), we find, for the ground state energy, 
\begin{equation}\label{dispersion3}
{\cal E}_0(\lambda) = a_{0,0} +a_{0,1} \lambda +  \frac{1}{2\pi i} \sum_{q=0}^{\infty} \l^{ q+2}  \int_0^\infty d\l' 
\frac{\text{Disc} {\cal E}_0 (\l')} {(\l')^{q+3}}  \,, 
\end{equation}
out of which we extract, 
\begin{equation}\label{dispersion32}
a_{0,q} = \frac{1}{2\pi i} \int_0^\infty d\l 
\frac{\text{Disc} {\cal E}_0(\l')} {\l^{q+1}} 
=  \frac{1}{\pi } \int_0^\infty d\l 
\frac{\text{Im} {\cal E}_0(\l')} {\l^{q+1}} 
\qquad \text{for}\qquad q \geq 2,
\end{equation}
Now, let us investigate the implications of the confluence equation  (\ref{borelbion}) 
\begin{align}\label{borelbion3}
\Im\, {\cal E}_0(\l') \equiv \Im\, \B_{0, \th=0^\pm}   = - \Im\, [\cB_{ii}]_{\th=0^\pm}  \qquad ,\quad {\rm up \; to}  \; e^{-4S_0}\end{align}
The imaginary part of the neutral bion amplitude is given in (\ref{firstambiguity}). The pre-factor 
${\cal A}_i=  \sqrt \frac{ 8}{\lambda}$.  (This is given in Ref. \cite{zinn-book}, Eq.(43.67), 
$ g_{\rm there}= \frac{\lambda}{8\pi}$ here.)  Thus, after taking the pre-factors into account, 
\be
\Im[\cB_{ii}]_{\th=0^\pm} =  \Im \left( \K_{i} \bar \K_i   \times 2 \; I(-g^2) \right)  =  \pm \frac{16 \pi } {\l} e^{-8\pi/\l} 
\ee
Using (\ref{dispersion32}),  we obtain, at  leading order, the large-order behavior of the perturbative 
series  $P_0 (\l) $ as 
\begin{align}\label{dispersion7}
& P_0(\l) \sim  \sum_{q}  a_{0, q} \lambda^q =  - \sum_q 
\frac{2}{\pi}  \left( \frac{\l }{8 \pi} \right)^{q} q!  = -\frac{2}{\pi}  \sum_q   \left( \frac{1}{4 \xi } \right)^{q} q! \; g^{2q}, 
\end{align} 
 This is the expected non-alternating Gevrey-1 series.   A few remarks  are in order:
 \begin{itemize}
 \item The late terms obtained in reduced quantum mechanics  of \cpn, by using the  cancellation of the imaginary parts of the left (right)  $[ \K_{i} \bar \K_i ]$ amplitude  and left (right) Borel-sum of perturbation theory  agrees with the earlier work of Stone and Reeve \cite{Stone:1977au}, who applied the Bender-Wu analysis to a periodic potential.\footnote{In a beautiful paper \cite{Stone:1977au}, Stone and Reeve noted the ambiguity associated with the non-Borel-summability of perturbation theory for the QM periodic potential, and stressed that it is not associated with an instability. However, at that time the counterpart of the neutral bion in  our confluence equation (\ref{borelbion}) was not yet understood.}
 
\item For \cpn, our  expansion 
 is  {\it not} in $g^2$, rather in  $\frac{g^2N}{8 \pi} \equiv \frac{\lambda}{8 \pi} = \frac{N}{2S_I} $ where the instanton action is 
 $S_I= \frac{4\pi}{g^2}$.  
   Consequently,  this leads to   the mechanism for the cancellation of the semi-classically realized IR-renormalon ambiguity on  $\R \times \mathbb S^1_L $, which are closer to origin by a factor of $N$ with respect to the 2d instanton-anti-instanton  $[\cI \bar \cI]$ singularities.

 \item   For \cpn on $\R^2$, the renormalon ambiguity is located at  $\frac{2 n S_I}{N}, n\geq 2$.  On $\R^2$, there are no semi-classical topological configuration that these ambiguities may cancel against. 
 The 2d instanton-anti-instanton events are located  far from the origin, at   $2 n S_I, n\geq 1$. 

\item
  On a locally two-dimensional manifold  on small $\R \times \mathbb S^1_L $,   
   for the $N_f=0$ theory,  we find neutral bion events to cancel the ambiguities at  $\frac{2n  S_I}{N}, n\geq 1$. For the $N_f \geq 1$ theories,  we find neutral bion events to cancel the ambiguities at  $\frac{2n  S_I}{N}, n\geq 2$, in exact agreement with  the renormalon singularities on $\R^2$. According to our analysis, there is one more singularity closer to the origin in the bosonic theory.  
 
\item 
The all-important step  is  {\it continuity}, in the sense we made precise (absence of rapid crossovers at finite-$N$, or sharp phase transitions at $N=\infty$),   the ability to connect the  strong coupling non-trivial holonomy  to  the  weak-coupling  {\it non-trivial}  holonomy  for the $\s$-connection (\ref{hol}).\footnote{All earlier  compactifications  of this class of theories land on  the weak coupling trivial holonomy in the small-$\mathbb S^1$ regime   \cite{D'Adda:1978kp,Jevicki:1979db,Affleck:1980mb,Munster:1982sd,Aguado:2010ex}. This is interesting for other reasons, but this regime  is  unrelated to the  semi-classical treatment of the confined regime due to rapid-crossover or phase transition. Our formulation of the problem provides 
a weak coupling regime which appears to be as close as one can get to the strong coupling regime. }
  \end{itemize}

\subsection{Graded resurgence and resurgent sectors  in the path integral formalism}
\label{sec:graded}
In order to understand the general form of an observable in general QFT or QM with a topological $\Theta$ angle,  we introduce the concept of  ``graded resurgence''.  The main idea of graded resurgence follows from  the following simple observations: 
 \begin{itemize}
\item[\bf 1)] Perturbation theory  is independent of the topological  $\Theta$-angle. Therefore, the ambiguity due to non-Borel summability of perturbation theory is also independent of    $\Theta$. 
  
 \item[\bf 2a)] The amplitude of topological configurations which carry non-vanishing topological charge {\it do depend} on the  $\Theta$-angle. Examples in \cpn are  kink-instantons, 2d instantons.  
 
  \item[\bf 2b)] The amplitude of (molecular/correlated)  topological configurations which carry zero  topological charge  {\it do not depend} on the    $\Theta$-angle. Examples in \cpn are  neutral bions, bion-anti-bions. 

   \item[\bf 3)] Therefore,  the non-Borel summability of the large orders in perturbation theory can {\it never} be cancelled by configurations which carry  non-vanishing topological charge. Rather it can {\it only} be cancelled  by topological configurations  with  zero topological charge, or  equivalently,  without any  $\Theta$-angle  dependence. 
\end{itemize}
This structure leads to  a sectorial mechanism of cancellation, which we call  ``graded resurgence". 
 To apply these ideas to \cpn,  define a   ``cell"  $[n,m]$ as follows: 
 \be
 n= n_{\rm kink} +  n_{\rm anti-kink}  , \qquad  m= n_{\rm kink} -  n_{\rm anti-kink} 
 \ee 
Here $n \frac{S_I}{N}=  n \frac{4\pi}{g^2N}$ is the action  and $\frac{m}{N}$  denotes the topological charge.  The $[n,m]$ sector is composed of  $n_{\rm kink} +  n_{\rm anti-kink}$ {\it correlated}  kink-instanton events. For example, a single  kink event belongs to  $\K_j \in [1,1]$. The proliferation of single-kink events in the Euclidean vacuum is the leading $\Theta$ dependent  contribution to any observable. Neutral and charged   bions belong to $\cB \in [2,0]$, and  their  proliferation generates various physical effects, such as the non-perturbative  mass gap for  the $N_f \geq 1$ theories. 
 
The general form of the contribution of the events in the $[n,m]$ cell  to an observable is given by  
\be 
[n,m] \Longrightarrow \A_{[n,m]}e^{-n \frac{4\pi}{\lambda}  + i m \frac{\Theta +  2 \pi k}{N}} P_{[n,m]}(\lambda)
\ee  
 where $\A_{[n,m]}$ is the pre-factor of the associated  $[n,m]$-defect amplitude, and $P_{[n,m]}(\lambda)$ denotes the formal perturbative fluctuation series  around the  $[n,m]$-defect.   The appearance of $k= 0, \ldots, N-1$ along with $\Theta$ is tied with the multi-branched structure of physical observables in bosonic  \cpn, either on $\R^2$ or $\R \times \mathbb S^1_L$. 
 
{\bf Definition 5: 
\; Graded resurgence triangle:} 
 The sectors in the \cpn model  form a structure that we refer to as the {\it graded resurgence triangle} (\ref{triangle2}), where the rows are at fixed-$n$ (fixed-action).   As one moves downward in the triangle, the action of the whole row  increases by one-unit (in kink-instanton action), namely 
 $n=0,1, 2, \ldots$. 
   \begin{eqnarray}
&[0,0] & \nonumber \\ \cr
 [1, 1]  
&& 
[1, -1] \nonumber\\ \cr
[2, 2]
\qquad \qquad & 
[2, 0] &
\qquad \qquad
[2, -2]
\nonumber\\   \cr
[3,3]
\qquad \qquad
[3,1]
&& 
[3,-1]
\qquad \qquad
[3, -3] \nonumber\\ \cr
[4,4]
\qquad \qquad
[4,2] \qquad \qquad 
&[4,0] 
  & \qquad \qquad 
[4,-2]
\qquad \qquad
[4, -4] \nonumber\\ \cr
\iddots \qquad \qquad \qquad \qquad\qquad &\vdots & \qquad \qquad\qquad \qquad\qquad  \ddots 
\label{triangle2}
\end{eqnarray}
 The row  labelled with $n$ has $n+1$ "cells", these are $m=n, n-2, \ldots, -n+2, -n$. 
  Columns are  fixed-$m$ (fixed topological charge)  sectors. The graded structure inherent to QFTs and QM  with $\Theta$ angle is shown  in (\ref{triangle2}).
For {\it general} QFTs, all cells in the graded resurgence triangle are ambiguous for one of the  two reasons: 
\begin{list}{$\bullet$}{\itemsep=0pt \parsep=0pt \topsep=2pt}
\item {\bf Ambiguities in the Borel resummation of perturbation theory}, ${\cal S}_{\pm} P_{[n,m]} \equiv \B_{[n,m]\pm} $ either  around the perturbative vacuum or in the background of an $[n,m]$-defect. 
\item {\bf   Ambiguities  in the  definition of the non-perturbative  amplitudes} associated with neutral  topological molecules, or molecules which include neutral sub-components,  such as 
$ [(\K)^{n_{\rm k}}(\bar \K)^{n_{\rm ak}}]_{\pm} $.
\end{list}
As discussed earlier,  there is no unique meaning to perturbation theory in the $[0,0]$ cell: the corresponding series is  typically  non-Borel summable,  but left (right)  Borel summable   $\B_{[0,0], \th=0^\pm}$, and   is two-fold ambiguous.  This is to say, usual perturbation theory in  QFT and Rayleigh-Schr\"odinger perturbation theory in quantum mechanics  cannot be used to define the theory.  Analogously,  the $[1, 1]$ sector, which we can write as     $[\K_i] \B_{[1,0], \th=0^\pm}$ is also equally ambiguous, due to the large order behavior of perturbation theory around the kink-instanton background,  whereas   $\K_i$  itself is unambiguous. 

 At the next level  and thereafter, there is a new type of ambiguity. Consider the $[2, 0]$ cell, which has elements like  $[\cB_{ij}] \B_{[2,0], \th=0^\pm}$, as well as  $[\cB_{ii}]_{\th=0^\pm}  \B_{[2,0], \th=0^\pm}$, where $\pm$ is used to indicate the presence of an ambiguity and left-right definitions of the corresponding objects.  In the $[4, 0]$ cell, the events which are ambiguous are $[\cB_{ij} \cB_{ji}]_{\th=0^\pm}$.     

In general QFT, our claim  is that all these ambiguities are interconnected, and once we calculate a physical observable, say the vacuum energy density, mass gap or whatever we want,  the ambiguities cancel to yield a unique unambiguous answer. In the previous section, we have explicitly  demonstrated  this mechanism, recall (\ref{borelbion}). 
This type of cancellation is at the very heart of the Borel-\'Ecalle resummation; if it indeed continues to all non-perturbative orders then it could potentially provide a fully consistent non-perturbative continuum definition of QFT. 
  
\subsection{Confluence equations} 
\label{confluence}
In the graded resurgence triangle, there are certain selection rules, which dictate the possible communications and cancellations between the non-perturbative ambiguities in different  cells.

{ \bf Permitted (and necessary)  communications:} The elements of a fixed-$m$ sector (i.e., a column) in the triangle (\ref{triangle2}) {\it can}  and typically do    talk with each other. These are sectors whose action differ by   two units of (minimal) kink-instanton  action:
  \be 
[n,q] \Longleftrightarrow [n+ 2n', q]
\ee  

{\bf Forbidden communications: } Two different columns can and do contribute to a given observable. However, the ambiguities in a cell of a given column can never be cured by any cell in a different column. In this sense, the communications between 
  \be 
[n,q] \Longleftrightarrow [n', q'], \qquad   q \neq q' , \qquad {n, n' \;\; \rm arbitrary}
\ee  
are forbidden.  This is a simple consequence of the fact that perturbation theory does not depend 
on the $\Theta$-angle.  \footnote{There is actually a far more refined structure in the resurgence triangle compatible with resurgence. Each cell has a sub-structure dictated by the Lie algebra data of 
$SU(N)$. Recall that kinks are associated with the co-roots $\a_i \in \G^r_\v$, whereas the the elements of the $[2,0]$ cell, charged and neutral bions are in one-to-one correspondence with the Cartan matrix entries. For cells associated with higher action, more elaborate structure arises. We will refer to this as ramification of 
graded resurgence triangle. This structure will be explored in more detail in future work. }

In theories such as \cpn, typically, none of the cells exist as  a self-consistent object  on its own. Each cell  is in need of other cells that it communicates to cure its  diseases (ambiguities).  In  special QFTs, 
  cells can exist  self-consistently, this is synonymous with Borel summability. Later, we give evidence 
that extended supersymmetric theories are of this type.

For example,   the ambiguity in ordinary perturbation theory, i.e., in the $[0,0]$ sector can only be cured by the ambiguity in the various neutral bion events in $[2,0], \; [4, 0],\;  [6, 0], \ldots$ cells.  We call this relation 
 {\it  perturbative-non-perturbative confluence equations} 
or  {\it   confluence equations}  for short. 
For the $m=0$ column, the confluence equation is 
\begin{align}\label{rBZJ}
0 = \Im\Bigl(\B_{[0,0], \th=0^\pm}
+ \B_{[2,0],\th=0^\pm} [\cB_{ii}]_{\th=0^\pm}
+ \B_{[4,0],\th=0^\pm} [\cB_{ij}\cB_{ji}]_{\th=0^\pm}
+ \B_{[6,0] \th=0^\pm} [\cB_{ij}\cB_{jk}\cB_{ki}]_{\th=0^\pm} 
+ \ldots \Bigr) .
\end{align} 
Since the ambiguities of $\B_{[0,0]}$ form a sum of  terms of the form $\left\{ \pm i e^{-2S_0}, \pm i e^{-4S_0}, \ldots, \right\}$ and  since  the imaginary (ambiguous) parts of the neutral topological molecules are of the form $\Im [\cB_{ii}]_{\th=0^\pm}  \sim  i e^{-2S_0}  $ , $ \Im [\cB_{ij}\cB_{ji}]_{\th=0^\pm} \sim   \pm i e^{-4S_0} $, $\Im [\cB^n]_{\th=0^\pm} \sim   \pm i e^{-2nS_0} $, 
(\ref{rBZJ}) implies  
a hierarchy of cancellation at each order of ambiguities.  These are given by 
\begin{align}\label{rBZJ2}
 0& = \Im\B_{[0,0] \pm}
+ \Re \B_{[2,0]} \Im [\cB_{ii}]_\pm \;,  \qquad  ( {\rm up\;  to}  \; e^{-4S_0})   \cr 
 0 & = \Im\B_{[0,0] \pm}
+ \Re \B_{[2,0]} \Im [\cB_{ii}]_\pm  +  
 \Im \B_{[2,0]\pm} \Re [\cB_{ii}]  
 +   \Re \B_{[4,0]} \Im [\cB_{ij}\cB_{ji}]_\pm   \qquad  ( {\rm up\;  to}  \; e^{-6S_0})   \cr 
0&=  \ldots 
\end{align} 
where in the first relation, only the ambiguities at order $e^{-2S_0}$ cancel, and in the second  relation, the ambiguities at order $e^{-2S_0}$ and $e^{-4S_0}$  cancel, and so forth. Provided that these confluence equations hold, then a $\Theta$ independent contribution to a general observable will be given by 
\begin{equation}
\label{obser2}
O(g^2) =  \Re\, \B_{[0,0]} +   \Re \B_{[2,0]} \Re [\cB_{ii}]   +  \Im \B_{[2,0]\pm}  \Im [\cB_{ii}]_\pm + 
 \Re \B_{[4,0]} \Re[\cB_{ij}\cB_{ji}] 
\; , \qquad  {\rm up \; to} \;  O( e^{-6S_I})
\end{equation}
which is unambiguous and unique up to order  $O( e^{-6S_I})$. In this relation, $ \Re\, \B_{[2n,0]}$ is $O(1)$, the second term is $O(e^{-2S_0})$, and the third and fourth term are $O(e^{-2S_0})$.

On the other hand, the existence of the kink-instanton sector $[1,1]$,  presents its own set of cancellations, leading to the confluence equations:
\begin{align}\label{rBZJ3}
0 = \Im\Bigl(\B_{[1,1], \th=0^\pm}  [\K_{i}] 
+ \B_{[3,1],\th=0^\pm} [\K_{i} \cB_{jj}]_{\th=0^\pm}
+ \B_{[5,1],\th=0^\pm} [ \K_i \cB_{jk}\cB_{kj}]_{\th=0^\pm} 
+ \ldots \Bigr) .
\end{align}
 This  implies, in hierarchical form 
\begin{align}\label{rBZJ4}
 0& = \Im\B_{[1,1] \pm}  [\K_{i}]  
+ \Re \B_{[3,1]} \Im [ \K_i \cB_{jj}]_\pm \;,  \qquad  ( {\rm  to}  \; e^{-5S_0})   \cr 
 0 & = \Im\B_{[1,1] \pm}  [\K_{i}]  
+ \Re \B_{[3,1]} \Im [ \K_i \cB_{jj}]_\pm  + 
 \Im \B_{[3,1]\pm} \Re [ \K_i \cB_{jj}]  
 +   \Re \B_{[5,1]} \Im [ \K_i \cB_{jk}\cB_{kj}]_\pm   \quad  ( {\rm  to}  \; e^{-7S_0})   \cr 
0&=  \ldots 
\end{align} 
One may be tempted to divide (\ref{rBZJ3}) and (\ref{rBZJ4}) by the kink amplitude $[\K_{i}]$, 
and write an expression virtually identical in form to (\ref{rBZJ2}). This is not quite true,  because  the pre-factor of the $[ \K_i \cB_{jj}] $  amplitude is not obtained through a simple product, but rather a convolution,  an integral over the quasi-zero mode. 
Nevertheless,   it is still true that the large-order asymptotics of  $P_{[0, 0]}(\l) $ and $P_{[1,1]}(\l) $ 
have {\it universal} late terms that can be extracted from the dispersion relations through the formula: 
\begin{align}\label{leadingDisc}
\text{Disc}\, \B_{[0,0]} &= - 2\pi i \l^{-r_2} P_{[2,0]} e^{-2A/\l} 
+ \cO(e^{-4A/\l}) ,
\nonumber\\
\text{Disc}\, \B_{[1,1]}  &= -2\pi i \l^{-r_3+r_1} P_{[3,1]} e^{-2A/\l} 
+ \cO(e^{-4A/\l}) .
\end{align}
Using  (\ref{leadingDisc}) in the dispersion relation,  we obtain 
\begin{align}\label{dispersion4}
a_{[0,0],q} 
& =  \sum_{q'=0}^\infty a_{[2,0],q'}  
\frac{\G(q+r_2-q')}{(2A)^{q+r_2-q'}}  + O \left( \left( \frac{1}{4A}\right)^q \right) \cr
&= \frac{\G(q+r_2-q')}{(2A)^{q+r_2}} \left[ 
a_{[2,0],0}  +  \frac{2A}{ (q+ r_2 -1)} a_{[2,0],1} +  \frac{(2A)^2}{ (q+ r_2 -1) (q+ r_2 -2)}  a_{[2,0],2}+ \ldots \right]   
\cr 
&+ O \left( \left( \frac{1}{4A}\right)^q \right) 
\qquad \cr 
a_{[1,1],q}
& = \sum_{q'=0}^\infty a_{[3,1], q'}  
\frac{\G(q+r_3-r_1-q')}{(2A)^{q+r_3-r_1-q'}}   + O \left( \left( \frac{1}{4A}\right)^q \right) \cr
&= \frac{\G(q+r_3-r_1-q')}{(2A)^{q+r_3-r_1}} \left[ 
a_{[3,1],0}  +  \frac{2A}{ (q+ r_3-r_1 -1)} a_{[3,1],1} \right.\cr
&\hskip 1cm \left. +  \frac{(2A)^2}{ (q+ r_3 -r_1 -1) (q+ r_3-r_1 -2)}  a_{[3,1],2}+ \ldots \right]  
+ O \left( \left( \frac{1}{4A}\right)^q \right) 
\qquad 
\end{align} 
These  equations describe multiple  manifestations of resurgence: 
\begin{itemize} 
\item The large order behavior  of $a_{[0,0],q} $ for large-$q$  is determined  by the first few  $a_{[2,0],q'}$,  mainly by   $a_{[2,0],0}$. The subsequent  terms, associated with the one-loop perturbative fluctuations around $[\cB_{ii}]$  are suppressed by power law corrections in $q$, which are  extra factors of $1/q$.
Analogously, the large-order behavior of perturbation theory around a kink-instanton $a_{[1,1],q}$ is determined by the low orders in perturbation theory around the $[\K \K \bar \K]$ sector, mainly by $a_{[3,1],0}$. 
\item
The one-loop fluctuations around, respectively,  $[\cB \bar\cB]$  and $[\cB \bar\cB \K]$ saddle points determine the sub-series exponentially suppressed by a factor of $2^{-q}$. 

\item The expansion in both cases is dictated by the nearest singularity in the Borel plane, as a  consequence of Darboux's theorem \cite{Dingle:1973,BerryHowls}.  
The   leading large-order behaviors for the $[0,0]$ and $[1,1]$ sectors are given by
\begin{align}\label{dispersion72}
& P_{[0,0]}(\l) \sim \frac{a_{[2,0],0}}{(2A)^{r_2}}
\sum_{q=0}^\infty (q+r_2-1)!  
\left( \frac{\l}{2A} \right)^{q} , 
\nonumber\\ 
& P_{[1,1]}(\l) \sim \frac{a_{[3,1],0}}{(2A)^{r_3-r_1}}
\sum_{q=0}^\infty (q+r_3-r_1-1)!
\left( \frac{\l}{2A}\right)^{q} .
\end{align}
Despite the different  backgrounds, the asymptotics of the perturbative expansions around their respective sectors have a \emph{universal} behavior, determined by the nearest singularity in the Borel plane.  
\end{itemize}
The relations (\ref{dispersion4}) have their  counterparts in matrix models and topological string theory \cite{schiappa,marino}  and 4d gauge theory compactified on $\R^3 \times \mathbb S^1$  \cite{Argyres:2012ka}.

Ignoring order one numerical  factors and other (not so major) factors momentarily,  (\ref{dispersion72}) assumes the form 
\begin{align}
\label{dispersion8}
& P_{[0,0]}(\l) \sim  P_{[1,1]}(\l)  \sim \sum_{q=0}^\infty  \frac{q!}   
{\left( S_{ \K \bar \K} \right)^{q} }  \sim \sum_{q=0}^\infty  \frac{q!}   
{\left( 2S_{ \K } \right)^{q} } 
\end{align}
making it clear that it is the neutral bion  ${ \K \bar \K}$ configuration that controls the large order growth of the series  $P_{[n, m]}(\l)$. Comparing with the ordinary integrals as discussed in 
Section \ref{subsec:examples},   we observe that the counterpart of the singulant  (\ref{singulant}) of ordinary integrals  is the kink-instanton--anti-kink-instanton configuration (and not the kink-instanton configuration itself)   in the compactified  \cpn model.

 \subsection{Extended supersymmetric \cpn and Borel summability}
 \label{extended}
 Consider the extended    supersymmetric theory, for example, ${\cal N}=(4,4)$ \cpn model,  compactified on $\R \times \mathbb S^1$. 
   In this class of theories,  despite the fact that kink-instnatons are present, 
   a superpotential is not permitted because of the number of the fermonic zero modes or large amount of supersymmetry.  
   This is similar to 4d gauge theories with ${\cal N} \geq 2$ supersymmetry compactified on $\R^3 \times \mathbb S^1$.   Since a superpotential is not generated, the counterpart of the neutral bion and charged bions do not exist, and $[2,0]$-cell  is an  empty set. 
 In the extended supersymmetric cases, in fact, most  cells in the resurgence triangle are empty, and  the resurgence triangle (\ref{triangle2}) simplifies to  that shown in (\ref{triangleextendedsusy}).
 \newpage
   \begin{eqnarray}
&[0,0] & \nonumber \\ \cr
 [1, 1]  
&& 
[1, -1] \nonumber\\ \cr
[2, 2]
\qquad \qquad & 
\varnothing &
\qquad \qquad
[2, -2]
\nonumber\\   \cr
[3,3]
\qquad \qquad
\;\;\varnothing\;\; 
&& 
\;\;\varnothing\;\; 
\qquad \qquad
[3, -3] \nonumber\\ \cr
[4,4]
\qquad \qquad
\;\;\varnothing\;\;  \qquad \qquad 
&\;\;\varnothing\;\; 
  & \qquad \qquad 
\;\;\varnothing\;\; 
\qquad \qquad
[4, -4] \nonumber\\ \cr
\iddots \qquad \qquad \qquad \qquad\qquad &\vdots & \qquad \qquad\qquad \qquad\qquad  \ddots 
\label{triangleextendedsusy}
\end{eqnarray}
Since there are no neutral bion configurations, the confluence equation   (\ref{rBZJ}) and   (\ref{rBZJ3})  simplify into 
\begin{align}\label{rBZJsusy}
0 = \Im\Bigl(\B_{[0,0], \th=0^\pm} \Bigr), \qquad  0 = \Im\Bigl(\B_{[1, 1], \th=0^\pm} \Bigr) 
\end{align}
meaning that there is no imaginary ambiguity in the Borel sum of  ordinary perturbation theory, 
as well as in perturbation theory around the instantons.  In other words, the cells  $[n,\pm n]$, 
$n=0,1,2, \ldots$   must be Borel summable, or equivalently, there are no singularities in the Borel plane  along $ \R^{+}$ for extended supersymmetric theories. 
This is the major difference between  the bosonic theory and extended supersymmetric theory. 

It should be noted that the existence of instantons implies that perturbation theory is a divergent asymptotic series. However, whether such a series is Borel summable (alternating, Gevrey-one) or non-Borel summable (non-alternating, Gevrey-one) is a more refined question, which is tied with the existence of singularities on the Borel  complex-$t$ plane along the   $ \R^{+}$ ray. These singularities, in the semi-classical regime, would be  associated with neutral topological  events as opposed to  single instanton events. Consequently, the absence of such neutral molecules in the semi-classical regime of a given theory is the  same as Borel summability. 

Our argument for the Borel summability of the extended supersymmetric theory is for the semi-classical regime.  In these theories, it is believed that there are no phase transition as the holomorphic parameters are varied. Therefore, if this is true, then as the theory moves from the semi-classical regime to the regime of strong coupling, the Borel summability must still hold. This implies the  Borel summability of the extended supersymmetric quantum theory on $\R^2$. 

\subsection{$\Theta$-dependence of vacuum energy density and topological susceptibility}


Once the cancellation of the ambiguous imaginary parts is assured,  we obtain  finite and physical results for observables, such as vacuum energy density,  topological susceptibility, mass gap of the theory.  The result obtained in this manner is  an approximation to the physical result, that can be compared with lattice gauge theory. In  $N_f= 0$   deformed-\cpn model, we find, for example, the $\Theta$-angle dependence of  vacuum energy density as  
 \begin{eqnarray}
{\cal E} (\Theta) -  {\cal E} (0)  
= {\rm Min}_{k = 1 }^N    \left[   -\frac{N}{ \sqrt \lambda } e^{-\frac{ 4\pi}{\l}} \;  \Re \B_{[1,1]}     \cos \left(\frac{\Theta + 2 \pi k }{N} \right)     + \ldots   \right]
\end{eqnarray}
where  $\Re \B_{[1,1]}$  is the unambiguous real part of the Borel sum associated with the perturbative fluctuations in the  1-kink sector.  The factor of $N$  is present because there are $N$ types of kink events $\K_j$ contributing to the vacuum energy density at leading order. There are subleading $O(e^{-\frac{ 8\pi}{\l}})  $ corrections, which we ignore in the weak-coupling regime at this order.  

When $LN\Lambda\gtrsim 1 $, the theory moves to the strongly coupled volume independence domain, where semi-classics is no longer reliable. By  continuity,  and by the evidence provided by the analysis of 
renormalons on $\R^2$ versus semi-classical renormalons (bions etc.) on   $\R \times \mathbb S^1$, 
the semi-classical regime   $LN\Lambda\lesssim 1 $ regime is rather  similar to the  strongly coupled domain. Using the  dimensional transmutation (\ref{dimtrans}), we may therefore write the   topological susceptibility as  
\begin{equation}
\frac{ \partial^2 {\cal E}}{ \partial \Theta^2}  \Big|_{\Theta=0} =    a_1 \frac{ \Lambda^2 }{N} 
\end{equation}
where $a_1$ is a numerical factor.   This result is in qualitative agreement with the large-$N$ result, for which $a_1= 3/ \pi$. If we set this as boundary value at $N=\infty$ for the topological susceptibility, our result provides a prediction for finite values of $N$, which seems to be in accord with numerical lattice simulations  \cite{Vicari:2008jw}. This result implies that in the semi-classical regime, it is the kink-instanton events that are responsible for the $O(1/N)$ topological susceptibility.

\subsection{Mass gap on $\R \times \mathbb S^1$ }
The mass gap of the theory  is the energy required to excite the system from the ground state to the 
 first excited state.  In the $g^2 \rightarrow 0$ limit, i.e., the   weak coupling regime of the deformed-bosonic theory,  the gap is purely non-perturbative.  

 \begin{figure}[htb]
\centering{
\includegraphics[scale=0.4]{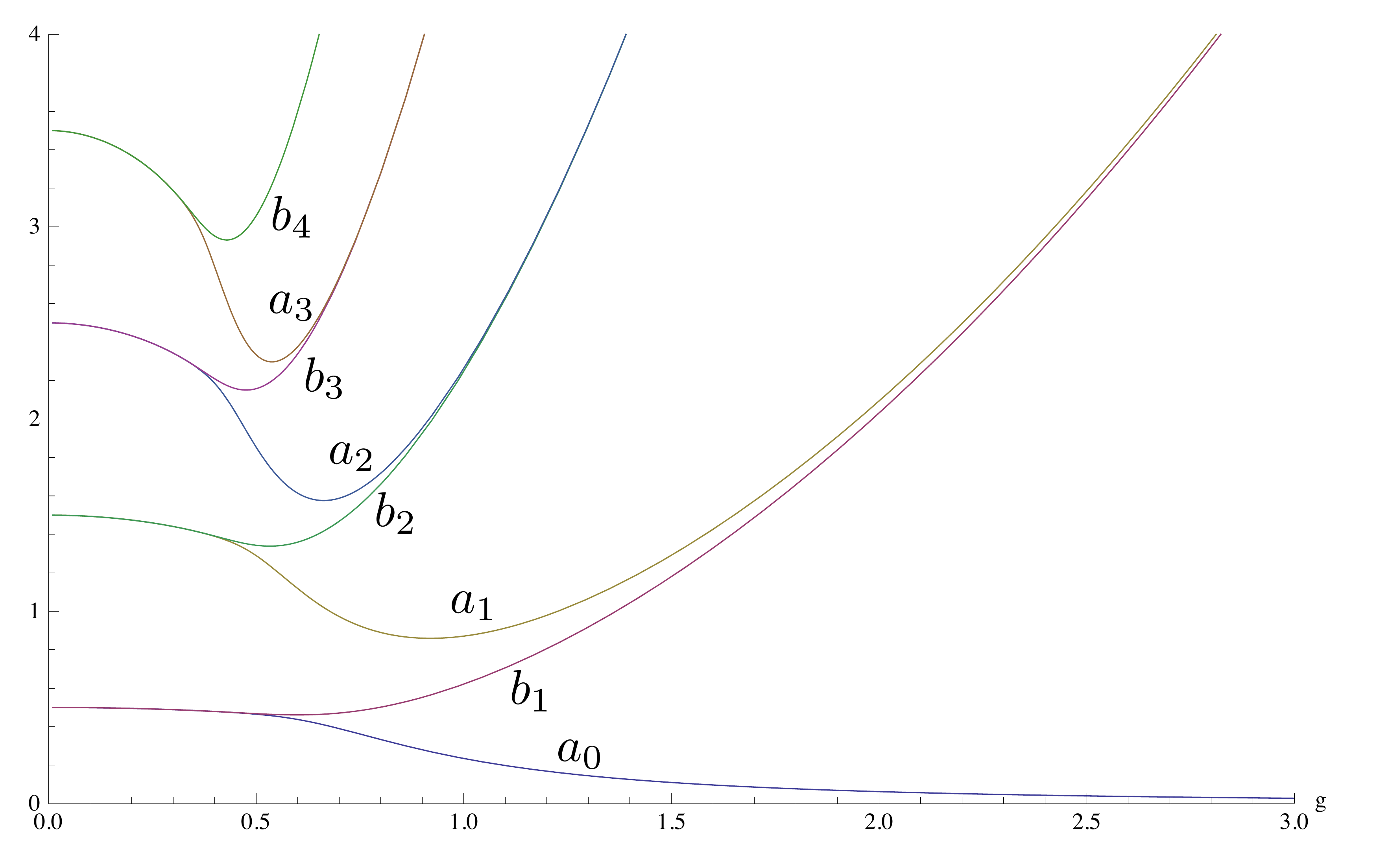}}
\caption{ 
The energy eigenspectrum of the  Mathieu equation as a function of
$g$. For large-$g^2$, it describes a particle with small moment of
inertia $I= 1/g^2$.  The spectrum asymptotes to
$E_{n} = \frac{g^2}{2} n^2$. The small-$g^2$ regime is related to the 
Hamiltonian for  \cpone\,  in the Born-Oppenheimer approximation.  The
curves represent  $a_0$, $b_1$, $a_1$, $b_2$, $a_2$, ..., from
(\ref{spectrum}), starting with the lowest curve. The  mass gap
$E(b_1) - E(a_0)$ is a purely non-perturbative  kink-instanton effect.
} 
\label{fig:spec}
\end{figure}

The study of the spectrum in the Hamiltonian (\ref{zero3b}) in the Born-Oppenheimer approximation, in the case of \cpone, reduces to the the study  of the asymptotics in the Mathieu equation \cite{nist}. 
Let  
\begin{align}
{\cal O}(q) =  - \frac{2 H}{g^2}  = \frac{d^2}{d\theta^2} -2 q \cos 2 \theta, \qquad q= \frac{\xi^2}{4g^4} 
\label{Ham}
\end{align}
denote a second order differential operator.  Then, the eigenstates of the Hamiltonian\footnote{We shifted $\theta$ by $\pi/2$ to match to the standard form of the Mathieu equation, 
$w^{{\prime\prime}}+(a-2q\mathop{\cos\/}\nolimits\!\left(2\theta\right))w=0.$}
   are the eigenfunctions of the Mathieu equation which obey
\begin{align}
& {\cal O}(q) \;    {\rm ce}_n (\theta, q) =  - a_n(q) \;   {\rm ce}_n (\theta, q)  \qquad n=0, 1,2, \ldots  \\
& {\cal O}(q) \;    {\rm se}_n (\theta, q) =  - b_n(q) \;   {\rm se}_n (\theta, q)  \qquad n=1,2, \ldots  
\end{align}
where 
\begin{align}
  {\rm ce}_n (\theta, q) = \langle \theta|  a_n(q) \rangle, \qquad {\rm se}_n (\theta, q) = \langle \theta|  b_n(q) \rangle, 
\end{align}
are the real space wave functions. 

Let P denote the parity operator acting as  $ \theta \rightarrow  - \theta $ 
 and  $T_\pi$ denote translation by $\pi$, $ \theta \rightarrow  \theta + \pi$. The eigenfunctions are also simultaneous eigenstates of  $P$ and $T_\pi$,  transforming as 
\begin{align}
&  P {\rm ce}_n (\theta, q)=    +  {\rm ce}_n (\theta, q),   \qquad  T_\pi {\rm ce}_n (\theta, q)=    (-1)^n  {\rm ce}_n (\theta, q)     \\
& P {\rm se}_n (\theta, q)=    -  {\rm se}_n (\theta, q),  \qquad    T_\pi {\rm se}_n (\theta, q)=    (-1)^n  {\rm se}_n (\theta, q)       
\end{align}
The energy eigenvalues of the Hamiltonian as a function of $g^2$ for low lying states are shown in Fig. \ref{fig:spec}.   For any finite $g^2$, the eigenvalues  obey 
\begin{align}
a_0 <b_1 <  a_1 < b_2 < a_2< \ldots  
\label{spectrum} 
\end{align} 
The large-$g^2  $ limit  (which is not interesting for  our purpose here )\footnote{This limit is sensible in purely quantum mechanical system defined by the Hamiltonian (\ref{Ham}). In our case, $g^2$ is fixed by the asymptotic freedom of the UV theory, and the value of $g^2$ in the long distance Hamiltonian is $g^2(1/L)$ for the \cpone\, theory. We can take the formal large-$g^2$ limit, but this has nothing to do with the continuum  \cpn model. It is, in a sense,  the counter-part of the lattice 
strong coupling expansion.}
 corresponds to a  particle  with a small  moment of inertia $I=1/g^2 \rightarrow 0$.   The Hamiltonian reduces to $H=  \frac{P_\theta^2}{2I}  $ and  
\begin{align} 
H   \;  e^{ \pm i n \theta}  = \frac{P_\theta^2}{2 I}  \;  e^{ \pm i n \theta} =  \frac{g^2}{2}  n^2  \; e^{i n \theta} 
 \end{align}
 The relation between the angular momentum eigenstates  $| \pm n \rangle$ and Mathieu functions in the infinite coupling limit is given by
\begin{align} 
| a_n(q=0) \rangle \pm i   | b_n(q=0) \rangle =  | \pm n \rangle, \qquad  n=1,2, \ldots 
\qquad | a_0(q=0) \rangle=  |  n=0 \rangle\;.
\end{align}
Since the energy is quantized in units of $g^2$,   the splitting between 
the rotational  energy levels become arbitrarily  large.   
Note that in this regime, 
\begin{align}
g^2 =\infty:  \qquad  a_0 =0 , \qquad   a_n=b_n= n^2, \;\;\;  n=1, 2, \ldots
\end{align}
However, this $g^2\to \infty$ limit is not relevant for the weak coupling regime of the \cpn models.

The small-$S^1$ regime of the \cpn  model is  related with the Hamiltonian 
(\ref{Ham}) within the Born-Oppenheimer approximation, as discussed in Section \ref{sec:atwork}.
In the $g^2  \rightarrow 0 $ limit (or $q\to\infty$) limit, 
the pair of  states  
\begin{align} 
| a_n(q \rightarrow \infty) \rangle   \leftrightarrow   | b_{n+1}(q \rightarrow \infty) \rangle, \qquad   n=0, 1, 2, \ldots
\end{align}
or in configuration space 
 ${\rm ce}_n (\theta, q)$  and    ${\rm se}_{n+1} (\theta, q)$  (not $n$, see the figure) 
become  degenerate to all orders in perturbation theory.
 At $g^2 =0$, their eigenenergy is $E (b_{n+1})  =  E (a_{n}) \sim    (n + \frac{1}{2})$, the one of simple harmonic oscillator.   The splitting  
$E(b_{n+1}) - E(a_{n}) $  is purely non-perturbative and is given by (here $h=\frac{\xi}{2g^2} =  \frac{2 \pi } {N (2 g^2) } \equiv \frac{\pi}{\lambda} $):
\begin{eqnarray}
\Delta E_n &=& \frac{g^2}{2} (
\mathop{b_{{n+1}}\/}\nolimits\!\left(h^{2}\right)-\mathop{a_{{n}}\/}\nolimits\!\left(h^{2}\right)) \nonumber\\
&=&  \frac{g^2}{2}\left( \frac{2^{{4n+5}}}{n!}\left(\frac{2}{\pi}\right)^{{\frac{1}{2}}}h^{{n+\frac{3}{2}}}e^{{-4h}}\*{\left(1-\frac{6n^{2}+14n+7}{32h}+\mathop{O\/}\nolimits\!\left(\frac{1}{h^{{2}}}\right)\right)} \right). 
\end{eqnarray}
The mass gap $m_g$  of  deformed-\cpone  \;  in the small-$S^1_L$ regime is given by 
\begin{eqnarray}
\label{massgap1}
m_g = \Delta E_{n=0} =  \frac{8 \pi }{g}   \left(1 - \frac{7 g^2 }{16 \pi}  + O(g^4) \right) e^{-\frac{2 \pi}{g^2}}   \sim    e^{-S_I/2} \qquad   {\rm for}  \;  \C\P^{\rm 1}
\end{eqnarray}
whereas for  the \cpn,  by generalizing the deformed-\cpone  discussion,   it is given by  
\begin{eqnarray}
\label{massgap2}
m_g = \Delta E_{n=0} =  \frac{C }{\lambda}   \left(1 - \frac{7 \lambda }{32 \pi}  + O(\lambda^2) \right) e^{-\frac{4 \pi}{\lambda}}   \sim e^{-S_I/N} \qquad   {\rm for}  \;  \C\P^{\rm N-1}
\end{eqnarray}  
Both  (\ref{massgap1}) and (\ref{massgap2}) are  remarkable non-perturbative consequences of the formalism, and they  deserve multiple comments: 

\begin{list}{$\bullet$}{\itemsep=0pt \parsep=0pt \topsep=2pt}
\item The mass gap is zero to all orders in perturbation theory.
\item The mass gap  in the weak coupling regime is a purely non-perturbative factor  proportional to $e^{-S_I/N}$. This is the first derivation of the all-important non-perturbative  factor  $e^{-S_I/N}$ from microscopic considerations. The mass gap   at small-$L$ may be considered as {\it the  germ of the mass gap} for the theory on $\R^2$. 
    
\item A physicist who is looking for the microscopic origin of the mass gap would be quite happy with this result, whereas a mathematician may feel disappointed.  There is a  series multiplying the kink-instanton, 
$P_{[1,1])} (\l)$ and it is  a divergent  asymptotic   Gevrey-one series. So, is there  a well-defined meaning to the mass gap that we obtained?  Since  $P_{[1,1])} (\l)$  is not even Borel summable,  what does the germ of the mass gap that we obtained really mean?   
\item The rather deep and provocative answer to the problem in the previous question is that the resurgence theory  cures  all the ambiguities, canceling  the ambiguities in the $[1,1]$ cell  
with the ambiguities in the $[3,1]$ sector,  as we have explicitly shown for the vacuum energy. These are encoded into our confluence equations,  (\ref{rBZJ2}) and (\ref{rBZJ4}). 
\item We claim the result is physical and meaningful 
\begin{eqnarray}
m_g = \Delta E_{m=0} =  \frac{C }{\lambda}  \Re \B_{[1,1]}  
e^{-\frac{4 \pi}{\lambda}}   \sim e^{-S_I/N} 
\end{eqnarray}
up to ambiguities of order $O( e^{-4S_I/N})$, where $\Re \B_{[1,1]}$ is the ambiguity-free real part of the 
(left or right) Borel sum. Although we have not shown this fully (only a partial construction is given here),  we anticipate that the mass gap, as well as other non-perturbative observables, in this theory are resurgent functions in the sense of \'Ecalle, and resurgence theory takes care of all the ambiguities. We believe this statement can be proven 
along the lines of \cite{ddp,delabaere-periodic}. 
\item The Born-Oppenheimer approximation  in the weak coupling limit is justified because of the hierarchy of the energy scales, 
\begin{align}
\Delta E_{m=0} \ll  E(a_1) - E(b_1) \sim \Delta E_{\phi}
\end{align}
where $ \Delta E_{\phi}$ is the $\phi$-sector discussed around (\ref{zero3}). 

\item  Another remarkable result is that   the mass gap obtained in the semi-classical regime coincides with the result (\ref{gap-largeN}) obtained by large-$N$ consideration, although the nature of the two semi-classical limits is  completely different, weak in $g^2$ in terms of original degrees of freedom vs.  weak in the coupling of the confined states where interactions are  $1/N$.  This item deserves more consideration, 
perhaps by incorporating the volume independence property (\ref{section-oneloop-c}).
\end{list}

The resurgent analysis of (\ref{Ham}) from the Hamiltonian perspective, along with some newly developed mathematical methods,    will be  discussed in detail in \cite{sg}.

\section{Conclusion: Towards a non-perturbative continuum  definition of QFT}

This work provides some steps  towards the  construction of  a non-perturbative continuum definition of QFT, in particular, the 
two-dimensional \cpn model. Our goal with such a construction, apart from providing 
a rigorous foundation to QFT, is to have a continuum formulation of practical value.  We feel obliged to emphasize that this is not a formal problem. 
In our view, the lack of such a formulation is the root-cause  underlying our    rather insufficient   understanding  of these theories.  In this work, we hope that we have made  progress in this direction.  There are 
two key elements, one from mathematics and one from physics: 
\begin{itemize}
\item \'Ecalle's theory of resurgent functions
\item Continuity:  the absence of phase transitions 
or rapid-crossovers 
upon spatial compactification of QFTs. 
\end{itemize} 

There are at present rigorous results using resurgent functions in quantum mechanics \cite{ddp,delabaere-periodic}. These authors,  using \'Ecalle's theory,  prove  that the semi-classical trans-series  expansion for some of these theories is resummable to finite, {\it exact} and unambiguous results. 

Continuity and spatial compactification provide  the new physical inputs necessary to extend these QM rsults to non-trivial QFTs such as \cpn or QCD.\footnote{There is an old idea of Bjorken: the femto-universe both in QCD and other theories. To move from the femto-universe to the large-scale-universe, there is either a phase transition or a rapid crossover. Our construction and idea of continuity relates the quantum mechanics in the small-circle (or torus limit) in an ``as smooth as possible'' manner passage to the large-volume QFT, and preserves a substantial amount of information about the QFT. This paper presents evidence for the existence of this type of passage, and another example was given in \cite{Argyres:2012ka}.} 
These are  used  in a new way 
to  reduce a  non-trivial QFT, in its  low energy limit, 
to quantum mechanical systems that can be studied through resurgence. Although we did this for 
the 2d \cpn model, it can be done for even more interesting theories, primarily QCD(adj) and (deformed) Yang-Mills theory. A quantum mechanical version of these theories exists, in which  rigorous results can be proven along the lines of  \cite{ddp,delabaere-periodic}.  It is apparent that there is  ample opportunity here to improve  significantly our understanding of QFT. In the next subsection we list some of the problems in which we feel  progress can be made: 
\\

To summarize, the main results of this paper are: 

\begin{itemize}

\item{We introduced a new parametrization of  the \cpn manifold  (\ref{chs}) which makes the analysis of the QFT simpler. This parametrization immediately yields a new order parameter, that we referred to as the $\s$-connection holonomy (\ref{hol}), a matrix valued  operator which  is the counter-part of the Wilson line in non-abelian $SU(N)$ gauge theory, and which carries more information than the regular $U(1)$ Wilson line in the \cpn theory. }

\item{The classical background of the $\s$-connection holonomy (\ref{hol}) is equivalent to the twisted boundary conditions for the \cpn fields. The quantum mechanical stability (instability) of the $\Z_N$-symmetric background follows from spatial (thermal) compactification.  The former  (but not the latter) admits  a semi-classical weak coupling study of the confined phase for all $N_f\geq 1$, different from {\it all} earlier studies of the \cpn theories.  In the $N_f=0$ bosonic theory, we use a deformation to stabilize the $\Z_N$-symmetric background. This can also be achieved by integrating out heavy fermions.}

\item{
In the  weak coupling regime   ($LN \Lambda \lesssim 1$),  we have shown that the leading finite action topological configurations are not 2d instantons, rather kinks with action $S_I/N$. }

\item{ 
In the same regime, we have shown that the non-perturbative ambiguities in perturbation theory are  cancelled against the  ambiguities in neutral bions, or bion-anti-bion  configurations with action  $2S_I/N$, $4S_I/N, \ldots$.  This is the content of our confluence equations (\ref{rBZJ}) in QFT. 
This shows that  the Bogomolny-Zinn-Justin mechanism of cancellation of ambiguities of perturbation theory against  ambiguities in non-perturbative semiclassical configurations  in quantum mechanics  can be successfully applied to an asymptotically free QFT such as the \cpn model. 
 } 

\item  {The standard renormalon analysis on $\R^2$  suggests  singularities in the Borel plane located at  $t_n = \frac{2S_I}{\beta_0} n, n=2,3, \ldots$  \cite{'tHooft:1977am}.  There is no known (semi-classical) configuration that these ambiguities of perturbation theory may cancel against. The 2d instanton-anti-instanton cancels a much suppressed and unimportant ambiguity of the perturbation theory located at  $t^{I\bar I} = n(2S_I), n=1,2, \ldots $ 
in the Borel plane.  Our findings imply that the  neutral  neutral bions {\it etc.}   are the semi-classical realization of IR renormalons. }

\item The mass gap  in the weak coupling regime of the bosonic theory is purely non-perturbative,  proportional to $e^{-S_I/N} = e^{-4 \pi/\l} $. This is the first  derivation of  the mass gap from  microscopic considerations, and is sourced by the proliferation of kink-instanton events.  This result, valid for arbitrary $N$, also  matches with the  large-$N$ result.  We conjecture that the mass gap is a resurgent function, at least in the small-${\mathbb S^1}$ regime. 

\item The 2d-instantons, in the large-$N$ limit, scale as $e^{- 4\pi/g^2} \sim e^{-N}$. In the semi-classical domain of the theory, the kink-instantons scale as $e^{- 4\pi/(g^2N)} \sim e^{-1}$ and are unsuppressed in the large-$N$ limit. If indeed, our claim that the renormalons and our semi-classical  neutral molecules being continuously connected is correct (we provided evidence that it is so), then the continuation of kinks (or charged bions) to the strong coupling domain are the resolution of the large-$N$ vs. instanton puzzles. At least in the weak coupling domain, they are indeed the resolution of the puzzles noted in \cite{Witten:1978bc}. 
  
\item For QFT (and QM) with degenerate vacua, and an associated topological charge and $\Theta$-parameter, we have introduced the notion of a ``graded resurgence triangle'' to explain how such cancellations can be categorized according to the $\Theta$-dependence, based on the simple observation that perturbation theory is insensitive to the $\Theta$-parameter. Thus, non-Borel summability of the large orders of perturbation theory can never be cured by configurations with non-vanishing topological charge, but only by certain {\it neutral} bion molecules. In the full resurgent framework, such cancellations should proceed to all non-perturbative orders, suggesting that this approach could provide a fully consistent non-perturbative definition of QFT in the continuum.

\end{itemize}

In the small-${\mathbb S^1}$ regime, at length scales larger than $\xi^{-1} = \frac{LN}{2\pi}$,  the 2d instanton should be viewed as a composite, and kink-instantons are elementary. This, combined with 
 large-$N$, matching of the mass gap between small-$\l$ semi-classics and large-$N$ ($1/N$ expansion), and  the semi-classical description of IR renormalons suggests that  the 
 2d-instanton on $\R^2$   should be viewed as an object with sub-structure. We claim that 
  the ``fractionalization  scale" of a 2d instanton is 
 \be
L^*=   \min ( LN, \Lambda^{-1}) 
\label{claim}
\ee
This seems to be   the  only reasonable   possibility in order to merge the  semi-classical domain  $LN \Lambda/(2\pi)  \lesssim $  with the volume independence domain $LN \Lambda/(2\pi) \gg  1$.  
When $LN < \Lambda^{-1}$, this is in fact true as we have shown already. If $LN \gg \Lambda^{-1}$,  then, the theory is the same as the theory on $\R^2$ as per volume independence, for which the only relevant length scale is the strong scale $\Lambda^{-1}$.  
 Therefore, the scale (\ref{claim}) seems to be the most reasonable guess for the fractionalization scale. 

\subsection{Prospects and open problems}
~

{\bf 1: Borel-\'Ecalle summability at higher orders or proof of all-orders confluence equations:}  We showed the leading cancellation of ambiguities in the bosonic theory. The demonstration of the whole set of  confluence equations would be  a major step, plausibly equivalent to the demonstration of the existence of QFT in continuum.

 {\bf 2: Which QFTs are Borel resummable? } The general classification of which QFTs  are Borel summable, and which are not,  is a meaningful and important question that can  partially be addressed using continuity, weak-coupling methods, and the graded resurgence triangle. For example, all extended supersymmetric theories seem to be Borel summable according to our criteria here and \cite{Argyres:2012ka}. 
 
   {\bf 3: O(N) models:} The techniques of this work easily generalize to the $O(N)$ model. It is often asserted, based on homotopy considerations, that there are no stable instantons for $N \geq 4$. However, our preliminary investigation  shows that with judiciously chosen boundary conditions, there are kink configurations and bion configurations. The classification of these seems to be identical to $SO(N)$ gauge theory on $\R^3 \times S^1$ \cite{Argyres:2012ka}.
  
    {\bf 4: Grassmannian models:} The techniques of this work are also suitable for Grassmannian models, with or without fermions. 
The ${\mathbb C}{\mathbb P}^{N-1}$ model  is a rank-one 
  Grassmannian model and is intimately connected to $SU(N)$ gauge theory on $\R^4$. It would be interesting to find the theory associated with higher rank Grassmannians.

 {\bf 5: Rigorous study of the 4d gauge theory-2d sigma model connection: }  The relation between 4d gauge theory and 2d ${\mathbb C}{\mathbb P}^{N-1}$ can be made non-perturbatively rigorous by compactifying the former on asymmetric $T^2 \times S^1_L \times \R$, and the latter on 
$S^1_L \times \R$. It is transparent from our analysis and the one of the \cite{Argyres:2012ka}  that the Lie algebraic classification of the topological configurations is actually identical. 

{\bf 6: Three-dimensional models:} The fractionalization of instantons has been studied in three-dimensional \cpn models \cite{Collie:2009iz} from a very different perspective, and it would be interesting to see if any of our ideas could be usefully applied therein.
   
      {\bf 7: $\Theta$ angle dependence: } There are conjectures for the behavior of the  ${\mathbb C}{\mathbb P}^{1}$ model at $\Theta=\pi$. These conjectures may be tested in our framework. The $\Theta$ dependence of general  ${\mathbb C}{\mathbb P}^{N-1}$ theory can also be studied. 
 
     {\bf 8: Singularities in the Borel plane vs. phase transitions:} Depending on whether the background on small $S^1$ regime is center-asymmetric (degenerate eigenvalue for  the holonomy matrix $^L\Omega$)
     vs. $\Z_N$  center-symmetric (maximally non-degenerate eigenvalue distribution) the Borel plane structure of singularities is drastically different. The singularity structure for the latter is almost identical to the theory on $\R^2$, perhaps only a small deformation thereof. The singularity structure for the former can be deduced from the combination of our work and Kontsevich's work on resurgence in quantum mechanics \cite{Kontsevich}, see footnote \ref{fn:k}.  It is evident that the drastic changes in the singularity structure in the Borel plane is associated either with a sharp phase transition or a rapid crossover. 
     
{\bf 9: Index theorem:}   The index theorem (\ref{index}) for the     Fredholm-type  Dirac operator on  $\R \times \mathbb S^1_L$    is stated without derivation. 
It can be derived along the same lines as in the joint work with E. Poppitz \cite{Poppitz:2008hr} or  \cite{Nye:2000eg}.  For the $O(N)$  and  Grassmannians sigma models there exists a corresponding  index theorem.

  {\bf 10: Chiral symmetry and sectors in quantum mechanics:} The low energy limit of multi-flavor 
  ${\mathbb C}{\mathbb P}^{N-1}$ is described by  quantum mechanics systems with $N_f$ types of fermions. This amounts to considering, in the quantum mechanical context, particles with spin (where the spin is determined by $N_f$) instead of spin-zero particles. There are crucial changes in the observables due to fermion zero modes. In particular, the number of sectors of associated quantum mechanics will be related to the number of discrete chiral symmetry breaking vacua in the field theory on $\R^2$. This deserves a detailed study of its own. 
  
    {\bf 11: Duality in quantum mechanics versus mirror symmetry:} For the ${\cal N}=(2,2)$ theory, the  dual theory obtained in  quantum mechanics  is equivalent to  the well-known mirror symmetry dual in the stringy framework of the very same theory. This also deserves further deliberations. 
    
{    \bf 12: Relation with the $\Z_N$ twisted mass deformed \cpn model:} A calculable deformation of the \cpn model on $\R^2$ is the   twisted mass deformed \cpn model: see for example,  
\cite{Ogilvie:1981yw, Hanany:1997vm, Dorey:1998yh, Gorsky:2005ac, Cui:2010si, Bolokhov:2011mp}. We believe that this model can be studied at arbitrary size ${\mathbb S^1_L}$ by the methods of this work. It would also be useful to understand  more precisely 
the relation between the topological configurations on $\R^2$ and in its compactified version.   

\section*{Acknowledgement} 
\hspace{0.5cm}
The authors would like to thank P. Argyres,  E. Poppitz, M. Shifman and A. Wipf for discussions and comments. This work was supported by the US DOE under grants DE-FG02-92ER40716 (GD) and  DE-FG02-12ER41806 (MU).

\end{document}